\documentclass[a4paper,11pt]{article}
\pdfoutput=1 

\usepackage{jcappub} 
\usepackage[dvipsnames]{xcolor}
\usepackage[T1]{fontenc} 
\def\bea{\begin{eqnarray}}
\def\eea{\end{eqnarray}}

\def\ba{\begin{eqnarray}}
\def\ea{\end{eqnarray}}

\usepackage{soul}
\usepackage{amsmath,mleftright}
\usepackage{xparse}
\usepackage{tabularx}
\newcolumntype{C}[1]{>{\centering\arraybackslash}m{#1}}

\def\mnras{MNRAS}

\newcommand{\Dk}[1]{\frac{d^3#1}{(2\pi)^3}}

\newcommand{\ve}[1]{{\text{\bf #1}}} 
\newcommand{\vk}{\ve k}
\newcommand{\vp}{\ve p}
\newcommand{\vq}{\ve q}

\newcommand{\mA}{\mathcal{A}}
\newcommand{\mB}{\mathcal{B}}

\newcommand{\dD}{\delta_\text{D}}

\usepackage{multirow}
\title{\boldmath An accurate perturbative approach to redshift space clustering of biased tracers in modified gravity}

\author[a]{Georgios Valogiannis,}
\author[a]{Rachel Bean}
\author[b,c]{and Alejandro Aviles}

\affiliation[a]{Department of Astronomy, Cornell University, Ithaca, NY, 14853, USA,}
\affiliation[b]{Consejo Nacional de Ciencia y Tecnologia, Av. Insurgentes Sur 1582, Colonia Credito, \\Constructor, Del. Benito Jurez, 03940, Ciudad de Mexico, Mexico,}
\affiliation[c]{Departamento de Fisica, Instituto Nacional de Investigaciones Nucleares, Apartado Postal \\18-1027, Col. Escandon, Ciudad de Mexico,11801, Mexico.}

\emailAdd{gv89@cornell.edu}
\emailAdd{rbean@astro.cornell.edu}
\emailAdd{avilescervantes@gmail.com}

\abstract{We extend the scale-dependent Gaussian Streaming Model (GSM) to produce analytical predictions for the anisotropic redshift-space correlation function for biased tracers in modified gravity models.  

Employing the Convolution Lagrangian Perturbation Theory (CLPT) re-summation scheme,  with a local Lagrangian bias schema provided by the peak-background split formalism, 
we predict the necessary ingredients that enter the GSM, the real-space halo pairwise velocity and the pairwise velocity dispersion. 
We further consider effective field theory contributions to the pairwise velocity dispersion in order to model correctly its large scale behavior.
We apply our method on two widely-considered 
modified gravity models, the chameleon-screened $f(R)$ Hu-Sawicki model and the $n$DGP Vainshtein model and compare our predictions against state-of-the-art N-body simulations for these models. 

We demonstrate that the GSM approach to predict the monopole and the quadrupole of the redshift-space correlation function for halos, gives very good agreement with the 
simulation data, for a wide range of screening mechanisms, levels of screening and halo masses at $z=0.5$ and $z=1$. Our work shows the applicability of the GSM, for cosmologies beyond GR, demonstrating that it can serve as a powerful predictive tool for the next stage of cosmological surveys like DESI, Euclid, LSST and WFIRST.}

\begin{document}
\maketitle
\flushbottom

\section{Introduction}
\label{sec:intro}

Based on the widely considered standard $\Lambda$-Cold Dark Matter ($\Lambda$CDM) cosmological model, a broad set of cosmological observations can be successfully explained \citep{Eisenstein:2005su, Percival:2007yw, Percival:2009xn, Kazin:2014qga, Spergel:2013tha, Ade:2013zuv, Ade:2015xua}, provided that two components are added to the cosmic energy budget in addition to those in the Standard Model of particle physics.
The first of these two ``dark'' components, cold dark matter, interacts merely gravitationally and plays a vital role in the formation of the inhomogeneous Large-Scale Structure (LSS) of the universe, through the process of gravitational instability.  The latter, the cosmological constant $\Lambda$, serves to drive the expansion of the universe to accelerate, as first observed two decades ago \cite{Perlmutter:1998np, Riess:2004nr}.

The mismatch between the theoretically predicted value for $\Lambda$, from high energy physics, and the observationally determined values, the cosmological constant problem \citep{Weinberg:1988cp}, has cast doubt on the foundations of the concordance cosmology. As a consequence, a plethora of theoretical alternatives has been proposed, with the most common one being a minimally-coupled ``dark energy'' field with negative pressure that evolves over time, but suffers from equally undesirable fine-tuning problems.

Another theoretical avenue towards the explanation of cosmic acceleration, lies in the fact that the above picture assumes a gravitational evolution governed by General Relativity (GR) and entertains the possibility that the observed acceleration is a manifestation of a change in the behavior of gravity at large scales, rather than a new energy component; these are the Modified Gravity (MG) proposals \citep{Koyama:2015vza, Ishak:2018his,Ferreira:2019xrr}. GR is a very successful, well-tested theory \citep{Will:2005va}, however, which poses strict restrictions on the magnitude of possible deviations introduced by the non-minimally coupled MG fields. The LIGO/Virgo collaboration has recently observed gravitational waves and electromagnetic counterparts arriving from the same source, almost simultaneously \citep{TheLIGOScientific:2017qsa,Goldstein:2017mmi,Savchenko:2017ffs,Monitor:2017mdv,GBM:2017lvd}, which has imposed additional limitations \citep{Lombriser:2015sxa,Lombriser:2016yzn,Sakstein:2017xjx,Ezquiaga:2017ekz,Creminelli:2017sry,Baker:2017hug} on all the possible ways through which MG fields can be combined into a Lagrangian that generates second order equations of motion, the Horndeski Lagrangian \citep{Horndeski1974,PhysRevD.84.064039}. 

Subject to these tight restrictions, interesting MG candidates typically employ a restoring, ``screening'' mechanism \citep{Khoury:2010xi,Khoury:2013tda}, that weakens the magnitude of the deviations in the high-density regime, through self-interaction terms and guarantees their phenomenological viability. The best-studied classes of screening, are arguably the Vainshtein mechanism \citep{VAINSHTEIN1972393,Babichev:2013usa} and the ``chameleons'' \citep{PhysRevD.69.044026,PhysRevLett.93.171104}. In the former type, the self-interactions grow large in the high-density regions, effectively weakening the fifth-force couplings to matter, causing deviations to be highly suppressed away from a source. In the latter class, the chameleons, screening is achieved thanks to a heavy Yukawa suppression of the fifth-force mediated by the scalar fields, which grow very massive in regions of high gravitational potential. Unfortunately, a ``no-go'' theorem prevents chameleons from self-accelerating \citep{Wang:2012kj}, but their very interesting phenomenology renders them perfect candidates to explore cosmological tests of gravity \citep{Burrage:2017qrf} and so we consider them as well in this work. Other types of screening, that we don't consider in this paper, are the symmetrons \citep{PhysRevLett.104.231301,Olive:2007aj}, which are phenomenologically similar to the chameleons, or the K-Mouflage fields \citep{Dvali:2010jz,Babichev:2013usa}.

We are currently in an era of ``precision'' cosmology that will continue to be refined with a plethora of upcoming photometric and spectroscopic surveys like e.g. DESI \citep{Levi:2013gra}, the LSST \citep{Abell:2009aa}, WFIRST \citep{Spergel:2013tha} and Euclid \citep{Laureijs:2011gra}. The LSS of the universe will be mapped out with unparalleled accuracy, allowing the prospect to place tight constraints on the various models of cosmic acceleration. Given, in particular, that the observed lumpy pattern of galaxies has emerged from the primordial density field, under the influence of non-linear gravitational evolution and is thus sensitive to the properties of the underlying gravitational law, the opportunity to test the MG models with modern cosmological surveys is unique \citep{Ishak:2019aay}. Maximizing the scientific return of such endeavors is a great challenge for the community that manifests itself in both the theoretical and experimental demands.

In the hierarchical picture of structure formation, the tiny perturbations in the primordial dark matter density field grow, under the influence of non-linear gravitational collapse, partly opposed by cosmic expansion, to give rise to the rich cosmic pattern observed today. On the large, linear scales and when GR is assumed, dark matter over-densities evolve as a simple function of time, for all scales, whereas on smaller, non-linear scales, computationally expensive N-body simulations are inevitable. This picture is further complicated by the fact that the galaxies observed by surveys of the LSS, do not perfectly trace the underlying dark matter density field, but are biased tracers of it \citep{1984ApJ...284L...9K} and, are observed in redshift space \citep{10.1093/mnras/227.1.1,1988MNRAS.235..715E}, which introduces redshift-space distortions (RSD) to the observed clustering statistics. In the case of MG models, another layer of complexity is added -- one needs to account for the presence of the additional degree of freedom that enhances structure formation and interferes with the evolution of dark matter and biased tracers in a non-linear manner. In the intermediate, quasi-linear scales, higher order Perturbation Theory (PT) \citep{Bernardeau:2001qr,2009PhRvD..80d3531C} approaches or hybrid methods \citep{Tassev:2013pn,Valogiannis:2016ane}, integrating both analytic and numerical simulation approaches, are of great benefit. 

RSD effects are induced by the peculiar velocity field of galaxies about the Hubble flow, which breaks the isotropy of the two-point correlation function of galaxies detected through spectroscopic means. At large scales, RSD lead to an enhancement of the amplitude of the correlation function, the ``Kaiser boost'' \citep{10.1093/mnras/227.1.1}, that can be modeled analytically, while on the opposite end, the non-linear regime, the Fingers-Of-God (FOG) effect suppresses the correlation function, an effect that is frequently captured through phenomenological ``streaming'' models \citep{1983ApJ...267..465D,10.1093/mnras/258.3.581}. In \citep{doi:10.1111/j.1365-2966.2011.19379.x}, the Gaussian Streaming Model (GSM) was introduced, to model the RSD correlation function in the quasi-linear scales. It used a non-perturbative resummation of the linear treatment by \citep{1995ApJ...448..494F} that convolves the real-space correlation function of biased tracers with a Gaussian pairwise velocity distribution function \citep{PhysRevD.70.083007}. The accuracy of the original approach, that used Eulerian Standard PT (SPT) to model the velocity moments, was further improved in \citep{Wang:2013hwa}, using the Lagrangian Perturbation Theory (LPT) approach to structure formation \citep{Zeldovich:1969sb,1989A&A...223....9B,Bouchet:1994xp,Hivon:1994qb,Taylor:1996ne,Matsubara:2007wj,Matsubara:2008wx,2013MNRAS.429.1674C,Matsubara:2015ipa} with a resummation scheme called Convolution LPT (CLPT) \citep{2013MNRAS.429.1674C} in which the effects of the bulk flows are not expanded in perturbative order. Further advancements included adding higher order velocity moments \citep{Uhlemann:2015hqa,Bianchi:2016qen} or small-scale physics effects through corrections from Effective Field Theory (EFT) \citep{Vlah:2015sea,Vlah:2016bcl}.

While halo bias and RSD have been studied in tandem for modified gravity in the context of N-body simulations, for example \citep{10.1093/mnras/stx196,PhysRevD.94.084022,Hernandez-Aguayo:2018oxg}, they have only been studied separately, to date, for perturbative approaches to the clustering statistics \citep{PhysRevD.79.123512,Taruya:2013quf,PhysRevD.88.023527,PhysRevD.90.123515,PhysRevD.92.063522,Fasiello:2017bot,Bose:2017dtl,Bose:2016qun,Bose:2018zpk,Aviles:2018saf,Aviles:2018qot,Valogiannis:2019xed}. In \citep{Aviles:2018saf,Valogiannis:2019xed}, CLPT was extended to predict the two-point statistics for biased tracers in MG, based on the LPT framework for MG developed in \citep{Aviles:2017aor} and also an analytical model for the prediction of the Lagrangian bias factors in MG \citep{Valogiannis:2019xed}, extending the Peak-Background Split formalism (PBS) \citep{1984ApJ...284L...9K,1986ApJ...304...15B,PhysRevD.88.023515,Mo:1996cn}. Applied on the f(R) Hu-Sawicki \citep{Hu:2007nk} and the nDGP \citep{Dvali:2000hr} models, it was shown to perform very well against results obtained by N-body simulations across a wide variety of screening levels and cosmological redshifts. In the work of \citep{Bose:2017dtl}, the GSM model was employed to model the RSD correlation function in MG models, shown to work very well against data obtained by N-body simulations, but only quantified in the context of pure dark matter considerations and with a local linear bias. Furthermore, \citep{Bose:2017dtl} used RegPT and also the SPT scheme previously used by \citep{doi:10.1111/j.1365-2966.2011.19379.x}, for the perturbative representations of the GSM ingredients, but not LPT, which was used in the GSM implementation by \citep{Wang:2013hwa} and will be the focus of this work. 

Building upon our previous work \citep{Aviles:2018saf,Valogiannis:2019xed}, in this paper we move forward to expand the scale-dependent GSM, in particular as presented in \citep{Wang:2013hwa,Vlah:2016bcl}, 
so as to make analytical predictions for the anisotropic redshift-space correlation function for biased tracers in MG theories. The underlying density field is evolved using the LPT for scalar-tensor theories 
presented in \citep{Aviles:2017aor}, while the effect of bias is captured through a local Lagrangian bias, up to second order, with the corresponding bias values predicted by the ST model for MG that was presented, 
and found to work well, in \citep{Valogiannis:2019xed}. We apply this framework on two widely-considered MG models, the chameleon f(R) Hu-Sawicki \citep{Hu:2007nk} and the Vainshtein-screened nDGP \citep{Dvali:2000hr} 
braneworld model and compare our results against state-of-the-art N-body simulations. We first make sure that our CLPT predictions for the remaining GSM ingredients, the pairwise velocity and the scale-dependent velocity 
dispersion, match the simulations sufficiently well, as already done for the real-space $2$-point correlation function in \citep{Valogiannis:2019xed}, before proceeding to cross-check the predictions for the monopole and 
the quadrupole of the RSD 2-point correlation function against the corresponding ones from the simulations. This last step is crucial for confirming the robustness of our analytical predictions, as well as the level of 
their accuracy, as we enter the era of precision cosmology. Our analytical approach is the first one, to the best of our knowledge, that captures both the effects of halo bias and RSD in the context of MG.

Our paper is structured as follows: in Sec. \ref{MGmodels} we introduce the MG scenarios on which we focus and also introduce the N-body simulations  used to cross-validate our analytical results.  In Sec. \ref{RSDMG} we show how the GSM is implemented in scenarios in which gravity deviates from  GR. Then, in Sec. \ref{results}, we discuss the accuracy of our results through the comparison against the N-body simulations, before concluding in Sec. \ref{Conclusions}. The details of the various derivations are laid out in Appendices \ref{AppendixGSM} and \ref{AppendixDirect}.

\section{Modified Gravity Scenarios and Simulation Tools}\label{MGmodels}

In this section, we briefly introduce the MG models we consider and also present the N-body simulations we used to cross-check our model's validity. 
\subsection{Modified Gravity Scenarios}
One of the oldest ways to depart from GR in the literature, proposes adding a function of the Ricci scalar to the standard form of the Einstein-Hilbert action. In particular, if $R$ is the Ricci scalar, these models, the ``f(R)'' class of theories \citep{DeFelice:2010aj}, are described by an action $S$ of the form:
\begin{equation}\label{actFr}
S=\int d^4x \sqrt{-g} \left[\frac{R+f(R)}{16 \pi G} + \mathcal{L}_m \right],
\end{equation}
where in the above expression, modifications to gravity manifest themselves through the nonlinear function $f(R)$. In (\ref{actFr}) we use $\mathcal{L}_m$ for the matter sector Lagrangian and G for the gravitational constant. The renewed interest in modifications of this type, is motivated by the possibility that such models can be responsible for the observed accelerated expansion of the universe \citep{Carroll:2003wy}. In what is probably the best-studied candidate of this class, the $f(R)$ Hu-Sawicki model \citep{Hu:2007nk}, the modifying function is of the form:
\begin{equation}\label{fRHu}
f(R)=-m^2\frac{c_1\left(R/m^2\right)^n}{c_2\left(R/m^2\right)^n+1},
\end{equation}
with $m=H_0\sqrt{\Omega_{m0}}$, $H_0$ being the Hubble constant, $\Omega_{m0}$ the fractional matter density evaluated today and $c_1, c_2$ and $n$ the free parameters of the model. The number of free parameters is further reduced by imposing a background expansion that matches the $\Lambda$CDM one in the high curvature limit ($R\gg m^2$), which gives:
\begin{equation}\label{fr0}
\bar{f}_{R_0} =-n\frac{c_1}{c_2^2}\left(\frac{\Omega_{m0}}{3(\Omega_{m0}+\Omega_{\Lambda0})}\right)^{n+1},
\end{equation}
where we defined the scalaron, $f_{R}=\frac{df(R)}{dR}$, that is evaluated today in equation (\ref{fr0}). Thanks to this expression, this model is usually parametrized with $|f_{R_0}|$ and $n$. Its popularity lies in the fact that it realizes the interesting phenomenology of the chameleon screening mechanism \citep{PhysRevD.69.044026,PhysRevLett.93.171104}, as can be shown through a conformal transformation \citep{Brax:2008hh}. As $|f_{R_0}|\rightarrow0$ and/or $n\rightarrow \infty$, the deviations are suppressed and GR is recovered. In our analysis, we always fix $n=1$ and consider three variations of $|\bar{f}_{R_0}|=\{10^{-6},10^{-5}, 10^{-4}\}$, that will be referred to from now on as F6, F5 and F4, respectively.

In the case of Vainshtein screening, there is a characteristic scale away from a massive source, the Vainshtein radius, below which fifth forces are strongly suppressed, due to the existence of large second derivatives of the gravitational potential. A MG model that exhibits this behavior is the Dvali-Gabadadze-Porrati (DGP) model \citep{Dvali:2000hr}, in which spacetime is actually 5-dimensional (5D), with an action of the form:
\begin{equation}\label{actDGP}
S=\int d^4x \sqrt{-g} \left[\frac{R}{16 \pi G} + \mathcal{L}_m \right] + \int d^5x \sqrt{-g_5} \left(\frac{R_5}{16 \pi G r_c}\right),
\end{equation}
where by $R_5$ and $g_5$ we label the 5D equivalent versions of the Ricci scalar and the metric determinant. Gravity does become 4-dimensional, however, below a characteristic scale $r_c$ and the usual 4D spacetime corresponds to a brane, on which the Standard Model fields are confined. The DGP model contains a self-accelerating branch (sDGP), which unfortunately has been shown to exhibit undesirable "ghosts" that make it unstable \citep{Koyama:2007zz}. For this reason, we consider the ``normal'' branch instead, called the $n$DGP, that is assumed to co-exist with a dark energy component, so that a $\Lambda$CDM homogeneous evolution is matched. We study two instances of the $n$DGP model, those with $n\equiv H_0 r_c=1$ and $n=5$, that we label and, from now on call, N1 and N5, correspondingly. 

\subsection{N-body Simulations}\label{Nbody}

In this section, we briefly introduce the N-body simulations we will use to assess the performance of our analytical model, which is a crucial step for our analysis. 

The first set of simulations, that we will refer to from now on as Group I simulations, are the Extended LEnsing PHysics using ANalaytic ray Tracing ({\sc ELEPHANT}) simulations \citep{Cautun:2017tkc}, that were performed with two modified versions of the GR code ({\sc RAMSES}): the {\sc ECOSMOG} module \citep{1475-7516-2012-01-051,Bose:2016wms} produced snapshots for the F6, F5 and F4 cases at a cosmological redshift of $z=0.5$, while {\sc ECOSMOG-V} \citep{Li:2013nua,Barreira:2015xvp,Hellwing:2017pmj} was used to produce the $n$DGP N1 and N5 realizations, also at $z=0.5$. $1024^3$ dark matter particles were evolved, in a simulation box with a side $L_{box}=1024 Mpc/h$ and a cosmology specified by the following parameters:
\begin{equation}
\{\Omega_{m0},\Omega_{\Lambda0},h,n_s,\sigma_8, \Omega_b\}=\{0.281,0.719,0.7,0.971,0.82, 0.046\}.
\end{equation}
So as to reduce the effects of cosmic variance, each model is run using 5 different random realizations. Finally, the dark matter halos in each snapshot are identified through the  \verb|ROCKSTAR| halo finder \citep{2013ApJ...762..109B}.

The second group of simulations available, that we will call Group II, come from the MG lightcone simulation project \citep{Arnold:2018nmv}, that employed the MG code {\sc MG-GADGET} \citep{doi:10.1093/mnras/stt1575}, to simulate GR and F5 cosmologies at a variety of redshifts; in our work we focus on the $z=1$ snapshot. Using $2048^3$ dark matter particles in a cubic box with side $L_{box}=1536 Mpc/h$, they are the largest-volume MG simulations performed up-to-date, which allows us to explore our GSM predictions at the Baryon Acoustic Oscillation (BAO) scales \citep{Eisenstein:2005su}. The $\Lambda$CDM 
cosmology in these simulations is given by:
\begin{equation}
\{\Omega_{m0},\Omega_{\Lambda0},h,n_s,\sigma_8, \Omega_b \}=\{0.3089,0.6911,0.6774,0.9667,0.8159, 0.0486\}.
\end{equation}
The halo catalogues are produced making use of the {\sc SUBFIND} code \citep{2001MNRAS.328..726S} and each model is simulated for only one random seed. 

Finally, to compute the real-space two-point correlation function, the RSD anisotropic correlation function and also the velocity information from the simulations, we utilize the publicly available code \verb|CUTE| \citep{2012arXiv1210.1833A}, using 30 linearly space bins in the range $0-140$ Mpc/h, both in the real and in the redshift space. For the Group I simulations, all error bars are calculated as the standard deviations over the 5 available realizations, whereas in the Group II case, where only 1 realization is available, we use the Jackknife method, splitting the simulation box into 64 sub-volumes.  

As was also done in \citep{Valogiannis:2019xed}, for the Group I  simulations, we analyze a $z=0.5$ snapshot of halos in the mass range $(2-3.5)\ \times 10^{12} M_{\odot}/h$, for all models, using only the main halos identified by \verb|ROCKSTAR|. For the Group II simulations, on the other hand, we focus our predictions on a redshift of $z=1$, considering halos in three separate mass bins: a lower mass bin of $9\times10^{11}-2\times 10^{12}\ M_{\odot}/h$, an intermediate bin of $5\times10^{12}-1\times 10^{13}\ M_{\odot}/h$ and a higher mass bin, $1.1\times10^{13}-9\times 10^{13}\ M_{\odot}/h$. 

\section{Redshift-Space Correlation Function For Biased Tracers In Modified Gravity}\label{RSDMG}
In this section, we present our analytical framework for the redshift-space correlation function of biased tracers in modified gravity cosmologies. Before the topic of RSD is addressed, we briefly summarize the LPT framework for structure formation in MG cosmologies, as well as the analytical treatment of Lagrangian bias for dark matter halos in such scenarios. 

\subsection{Lagrangian Perturbation Theory For Dark Matter In Modified Gravity}\label{LPTMG}

In the Lagrangian Perturbation Theory framework \citep{Zeldovich:1969sb,1989A&A...223....9B,Bouchet:1994xp,Hivon:1994qb,Taylor:1996ne,Matsubara:2007wj,Matsubara:2008wx,2013MNRAS.429.1674C,Matsubara:2015ipa}, the time-dependent growth of dark matter overdensities is traced in a coordinate system that is comoving with matter particles, as they move along their fluid trajectories. In particular, the fundamental element of LPT is a displacement, vector, field $\bold{\Psi}$, which in each moment of interest $t$, maps a particle from an initial ``Lagrangian'' position $\bold{q}$ to its final, Eulerian position vector $\bold{x}(\bold{q},t)$, encoded through the following relationship:
\begin{equation}\label{Lagpos}
\bold{x}(\bold{q},t) = \bold{q} + \bold{\Psi}(\bold{q},t).
\end{equation}
Imposing conservation of matter mass between $\bold{q}$ and $\bold{x}$, one gets
\begin{equation}\label{delJac}
\delta_m(\bold{x},t) = \frac{1-J(\bold{q},t)}{J(\bold{q},t)},
\end{equation}
with $\delta_m(\bold{x},t)=\frac{\rho_m}{\bar{\rho}}-1$ in  (\ref{delJac}) denoting the fractional matter overdensity and $J(\bold{q},t) = det(J_{ij})$, the determinant of the Jacobian of the mapping (\ref{Lagpos}), given by 
\begin{equation}\label{Jacobian}
J_{ij} = \frac{\partial x^i}{\partial q^j} = \delta_{ij} + \frac{\partial \Psi^i}{\partial q^j}.
\end{equation}
Unlike the Eulerian approach, in LPT the expansion parameter is the displacement $\bold{\Psi}$, as 
\begin{equation}\label{eq:psiexp}
\bold{\Psi}(\bold{q},t) = \sum_{n=1}^{\infty}\bold{\Psi}^{(n)}(\bold{q},t) = \bold{\Psi}^{(1)}(\bold{q},t)+\bold{\Psi}^{(2)}(\bold{q},t)+\bold{\Psi}^{(3)}(\bold{q},t)...
\end{equation}
Equations (\ref{Lagpos})-(\ref{Jacobian}) form a closed system that can be solved, order by order, when combined with the coupled pair of the geodesic and Poisson equations 
\begin{equation}
\begin{aligned}\label{eq:geopoisson}
\ddot{\bold{x}} + 2H\dot{\bold{x}} &= -\frac{1}{a^2}\nabla_{\bold{x}}\psi(\bold{x},t), \\
\frac{1}{a^2}\nabla^2_{\bold{x}}\psi(\bold{x},t) &=  4 \pi G\bar{\rho}_m \delta(\bold{x},t),
\end{aligned}
\end{equation}
with $\psi(\bold{x},t)$ representing the scalar metric perturbation. Equations (\ref{eq:geopoisson}) have been derived perturbing about a Friedmann-Robertson-Walker (FRW) background assuming a GR-like evolution. The first order solution to the system (\ref{Lagpos})-(\ref{eq:geopoisson}) is the well-known Zel'dovich approximation \citep{Zeldovich:1969sb}, in which
\begin{equation}\label{eq:ZeldispsolGR}
{\Psi}^{j}(k,t) = \frac{i k^j}{k^2} D^{(1)}(t)\delta^{(1)}(\bold{k},t=0),
\end{equation}
where $\delta^{(1)}(\bold{k},t=0)$ is the linearized overdensity at early times and $D^{(1)}(t)$ the linear growth factor in GR, which is the growing solution of 
\begin{equation}\label{growth1stGR}
\mathcal{\hat{T}}D^{(1)}(t)=A_0 D^{(1)}(t),
\end{equation}
with $A_0=4 \pi G\bar{\rho}_m$ and the differential operator $\mathcal{\hat{T}}=\frac{d^2}{dt^2} + 2H\frac{d}{dt}$, defined in \citep{Matsubara:2007wj}.

In the presence of a modification to gravity, the above picture is complicated by the action of the additional degree of freedom, the impact of which should be taken into account by the LPT framework. In the work of \citep{Valogiannis:2016ane}, LPT was studied for the MG chameleons and symmetrons, up to second order, in the context of the COLA hybrid approach, which was found to recover results from full N-body simulations with high accuracy. LPT was first expanded to capture MG theories up to third order by \citep{Aviles:2017aor} (also see \citep{Aviles:2018qot,Aviles:2018saf,Valogiannis:2019xed}), which is the approach we closely follow and briefly summarize here. For a scalar-tensor theory, equations (\ref{eq:geopoisson}) are replaced by the modified version of the perturbed Einstein equations: 
\begin{equation}
\begin{aligned}\label{eq:modpoisson}
&\nabla_{\bold{x}}\mathcal{\hat{T}}\bold{\Psi} = -\frac{1}{a^2}\nabla^2_{\bold{x}}\psi(\bold{x},t), \\
&\frac{1}{a^2}\nabla^2_{\bold{x}}\psi(\bold{x},t) =  4 \pi G\bar{\rho}_m \delta(\bold{x},t) + \frac{1}{2 a^2}\nabla^2_{\bold{x}}\phi,
\end{aligned}
\end{equation}
combined with the Klein-Gordon (KG) equation 
\begin{equation}\label{BransDicke}
(3 + 2 \omega_{BD})\frac{1}{a^2} k_x^2\phi(\bold{k}_x,t) = 8 \pi G \bar{\rho}_m \delta(\bold{k}_x,t) - \mathcal{I}(\phi).
\end{equation}
Here, $\omega_{BD}$ is a function that depends on the specific theory, and is named in that way because for Brans-Dicke (BD) theories it reduces to the BD parameter. The under-script x is meant to show that the quantities are evaluated in the Eulerian basis. The term $\mathcal{I}(\phi)$ is the perturbative representation of the screening due to the field $\phi$ self-interactions \citep{PhysRevD.79.123512}, given by 
\begin{eqnarray}
\label{interaction}
 \mathcal{I}(\phi) &=& M_1(\bold{k},t)\phi + \frac{1}{2}\int \frac{d^3 k_1 d^3 k_2}{\left(2 \pi\right)^3} \delta_D(\bold{k}-\bold{k}_{12})M_2(\bold{k}_{1},\bold{k}_{2})\phi(\bold{k}_{1})\phi(\bold{k}_{2})\nonumber\nonumber \\
&& + \frac{1}{6}\int \frac{d^3 k_1 d^3 k_2 d^3 k_3}{\left(2 \pi\right)^6} \delta_D(\bold{k}-\bold{k}_{123})\nonumber M_3(\bold{k}_{1},\bold{k}_{2},\bold{k}_{3})\phi(\bold{k}_{1})\phi(\bold{k}_{2})\phi(\bold{k}_{3}),
\end{eqnarray}
with $M_1(\bold{k},t)$, $M_2(\bold{k}_{1},\bold{k}_{2})$ and $M_3(\bold{k}_{1},\bold{k}_{2},\bold{k}_{3})$ being mass terms and where we used the convention $\bold{k}_{ijk}=\bold{k}_{i}+\bold{k}_{j}+\bold{k}_{k}$. The first order solution in MG is 
\begin{equation}\label{eq:ZeldispsolMG}
{\Psi}^{j}(k,t) = \frac{i k^j}{k^2} D^{(1)}(k,t)\delta^{(1)}(\bold{k},t=0),
\end{equation}
where the MG linear growth factor $D^{(1)}(k,t)$ is now obtained through 
\begin{equation}\label{growth1stMG}
\mathcal{\hat{T}}D^{(1)}(k,t)=A(k)D^{(1)}(k,t),
\end{equation}
with 
\begin{equation}
\begin{aligned}\label{eq:sources}
A(k) &=  4 \pi G\bar{\rho}_m\left(1 + \frac{k^2}{a^2 3\Pi(k)}\right), \\
\Pi(k) &= \frac{1}{3a^2}\left[\left(3+2\omega_{BD}\right)k^2 + M_1a^2\right].
\end{aligned}
\end{equation}
A generic feature of many MG models, for which $M_1\neq0$, is that, unlike in the GR case (\ref{growth1stGR}), the linear growth factor is scale-dependent. The $2^{nd}$ and $3^{rd}$ order LPT solutions in MG contain two additional contributions, compared to the simpler GR case: a screening term, due to the field self-interactions and a, geometric in nature, Frame-Lagging component that arises when transforming the KG equation from an Eulerian to a Lagrangian basis. The expressions for the $2^{nd}$ and $3^{rd}$ order solutions, as well as for the perturbative mass terms $M_1-M_3$ in the f(R) and $n$DGP cases can be found in \citep{Aviles:2017aor} and also \citep{Aviles:2018qot,Aviles:2018saf,Valogiannis:2019xed}.

We finally note that a scale dependent linear growth function implies the linear growth rate is also scale dependent,
\begin{equation}
f(k) = \frac{d\ln D^{(1)}(k,a)}{d\ln a}.
\end{equation}
We find it useful to define the large-scale value, $f_0$, as
\begin{equation}
f_0=f(k=0).
\end{equation}
For MG models with $M_1\neq 0$, $f_0$ coincides with the standard growth rate in $\Lambda$CDM. In models with vanishing  mass, 
such as the $n$DGP,  $f(k)=f_0$ is scale independent, but its value does not coincide with that of $\Lambda$CDM.

\subsection{Lagrangian Biased Tracers In Modified Gravity}\label{BiasMG}

Galaxies do not exactly trace the underlying distribution of dark matter, which in principle biases observable quantities \citep{1984ApJ...284L...9K}, like the clustering statistics, extracted from such observations. This effect is taken into account by the perturbative theory of galaxy clustering, which has been greatly explored using a variety of analytical approaches; a comprehensive review of this topic can be found in \citep{Desjacques:2016bnm}. Given an analytical model for the nonlinear evolution of the underlying dark matter overdensities (for which we choose LPT), the effects of halo or galaxy formation are captured through a set of bias parameters. Building upon our LPT formalism laid out in the previous section \ref{LPTMG}, we employ a local Lagrangian bias to study overdensities of halos, within which the observed galaxies form and reside. Before addressing the biasing scheme in MG scenarios, it is worth mentioning that in the context of GR, approaches in the literature have employed a variety of Lagrangian bias schemes, ranging from a local-in-matter-density Lagrangian bias \citep{Matsubara:2008wx,2013MNRAS.429.1674C}, to a local Lagrangian bias including curvature and tidal terms \citep{Vlah:2016bcl}, all the way to extensions with a non-local Lagrangian bias \citep{PhysRevD.83.083518}. 

The two-point statistics of biased tracers in modified gravity models have recently been expressed using a local-in-matter and curvature Lagrangian bias \citep{Aviles:2018saf}, as well as a local-in-matter density bias \citep{Valogiannis:2019xed}. 
While we will focus on predictions using the latter,  we will present the expressions including the curvature bias, which is more general and contains, as we will see below, the local in matter density terms as a limiting case. We make the common assumption that tracers are initially identified in the primordial dark matter density field, at a sufficiently early time $t_0$, through a local function $F$. In particular, if by $\delta_R(\bold{q},t_0)\equiv\delta_R(\bold{q})$ we denote the dark matter density field, smoothed out over a spatial scale $R$, then the initial fractional overdensity of tracers (in our case halos), $\delta_X(\bold{q},t_0)\equiv\delta_X(\bold{q})$, will be given by \citep{Aviles:2018saf} \footnote{In principle, in MG theories with an additional scalar field $\phi$, the bias function F should also depend on the Laplacian, $\nabla^2\phi$, but as noted in \citep{Desjacques:2016bnm} and further developed in \citep{Aviles:2018saf}, expanding the KG equation reveals that this dependence is degenerate with $\nabla^2\delta$ for k-modes smaller than the scalar field mass and can thus be absorbed.}: 
\begin{equation}\label{biasF}
1+ \delta_X(\bold{q})= \frac{\rho_{X}(\bold{q})}{\bar{\rho}_{X} } =F\left[\delta_R(\bold{q}),\nabla^2\delta_R(\bold{q})\right].
\end{equation}
When $F=1$ in  (\ref{biasF}), we get $\delta_X(\bold{q})=0$, recovering thus the dark matter case. Having identified the initially biased tracers through (\ref{biasF}), their subsequent nonlinear evolution is found after applying the continuity equation between $\bold{q}$ and $\bold{x}$:
\begin{equation}\label{delJacX}
1 + \delta_X(\bold{x},t) = \int d^3q F\left[\delta_R(\bold{q}),\nabla^2\delta_R(\bold{q})\right]\delta_{D}\left[\bold{x}-\bold{q}-\bold{\Psi}(\bold{q},t)\right],
\end{equation}
where in the above equation $\delta_R(\bold{q})$ is the extrapolated linear density field evaluated at the observation time.
Our goal is to model the two-point correlation function for halos, defined by 
\begin{equation}\label{xi}
\xi_X(r) = \langle \delta_X(\bold{x})\delta_X(\bold{x+r})\rangle,
\end{equation}
where the angle brackets indicate an ensemble averaging. Plugging the result (\ref{delJacX}) into (\ref{xi}) and after performing a multinomial expansion and several integrations, one gets the two-point correlation function for biased tracers, up to 1-loop order, as \citep{2013MNRAS.429.1674C,Aviles:2018saf,Valogiannis:2019xed}:
\begin{eqnarray}
\label{xiXfinal}
1 + \xi_{X}(r) &=& \int d^3q \frac{e^{-\frac{1}{2} (q_i - r_i) (A^{-1}_{L})_{ij}(q_j - r_j)}} {\left(2\pi\right)^{3/2} |A_{L}|^{1/2}} \times    \Biggl( 1 - \frac{1}{2}G_{ij} A^{loop}_{ij} + \frac{1}{6}\Gamma_{ijk} W_{ijk} \nonumber\\
 && - b_1 \left(2U_i g_i + A^{10}_{ij}G_{ij}\right) -b_2\left(U^{(1)}_iU^{(1)}_jG_{ij}+U^{20}_ig_i\right)\nonumber \\
&&+b_1^2\left(\xi_L-U^{(1)}_iU^{(1)}_jG_{ij}-U^{11}_ig_i\right) +\frac{1}{2}b_2^2\xi_L^2 -2b_1b_2\xi_LU^{(1)}_ig_i\nonumber\\
&& +2\left(1+b_1\right)b_{\nabla^2\delta}\nabla^2\xi_L +b^2_{\nabla^2\delta}\nabla^4\xi_L \Biggr),
\end{eqnarray}
with 
\begin{eqnarray}
\label{determs}
 g_i &\equiv& (A^{-1}_{L})_{ij}(q_j - r_j), \nonumber \\
 G_{ij} &\equiv& (A^{-1}_L)_{ij}-g_ig_j, \nonumber\\
 \Gamma_{ijk} &\equiv& (A^{-1}_{L})_{ij} g_k +(A^{-1}_{L})_{ki} g_j +(A^{-1}_{L})_{jk} g_i - g_i g_j g_k,
\end{eqnarray}
and where we defined
\begin{equation}
\begin{aligned}\label{eq:correl}
\sigma_R^2 &= \langle \delta^2 \rangle_c \\
\xi_L(\vec{q}) &= \langle \delta_1 \delta_2 \rangle_c,  \\
A_{ij}^{m n}(\vec{q}) &= \langle \delta_i^{m} \delta_j^{n} \Delta_i \Delta_j\rangle_c,  \\ 
W_{ijk}^{m n}(\vec{q}) &= \langle \delta_i^{m} \delta_j^{n} \Delta_i \Delta_j \Delta_k\rangle_c,  \\ 
U_{i}^{m n}(\vec{q}) &= \langle \delta_1^{m} \delta_2^{n} \Delta_i \rangle_c.
\end{aligned}
\end{equation}
The Lagrangian correlators (\ref{eq:correl}), as first defined in \citep{2013MNRAS.429.1674C}, are the elementary ingredients of the LPT correlation function (\ref{xiXfinal}) and contain cumulants of the differential LPT displacement field, $\bold{\Delta}=\bold{\Psi}_2-\bold{\Psi}_1$, where we adopted the shorthand notation $\bold{\Psi}(\bold{q}_1)=\bold{\Psi}_1$, etc. In (\ref{xiXfinal}) we also defined $A^{00}_{ij}\equiv A_{ij}$, $W^{000}_{ijk}\equiv W_{ijk}$ and $U^{10}_i\equiv U_i$. These correlators (\ref{eq:correl}) differ in GR \citep{2013MNRAS.429.1674C} and MG \citep{Aviles:2018saf,Valogiannis:2019xed}, because the LPT displacement fields they contain follow a different time-evolution in each of these two cases, as explained in the previous Section \ref{LPTMG} (this difference manifests itself in the ``k-functions'' in \citep{2013MNRAS.429.1674C} and \citep{Aviles:2018saf,Valogiannis:2019xed}). Finally, we note that in (\ref{xiXfinal}) only the linear part of $A_{ij}$ is kept exponentiated following \citep{Vlah:2015sea,Vlah:2016bcl}, a variant of the Convolution Lagrangian Perturbation Theory (CLPT) resummation scheme \citep{2013MNRAS.429.1674C} that maintains also the loop components in the exponential. 

In expression (\ref{xiXfinal}), we identify the local-in-matter-density bias parameters \citep{Matsubara:2008wx,PhysRevD.83.083518}  
\begin{equation}
\begin{aligned}\label{biasfn}
b_n\equiv \int \frac{d \lambda}{2\pi}\tilde{F}e^{-\frac{1}{2}\lambda^2 \sigma^2_R}\left(i \lambda \right)^n, 
\end{aligned}
\end{equation}
where $\tilde{F}$ is the Fourier-space representation of the Lagrangian function $F$. The extension of (\ref{biasfn}) to include the higher-order bias $b_{\nabla^2\delta}$ can be found in \citep{Aviles:2018saf,Aviles:2018thp}.

One approach to evaluate biases is the excursion set approach \citep{1991ApJ...379..440B}. This does not, however, have an analytical solution for generic MG models, due to the fact that the critical overdensity for gravitational collapse is not a constant at a given cosmological time, as it is in GR. One can then perform brownian-walk simulations in that case, as was done in \citep{Aviles:2018saf}. 

In our analysis here, we will evaluate predictions from (\ref{xiXfinal}) $\it{only}$ with a local-in-matter density Lagrangian bias, which simply corresponds to the limit $b_{\nabla^2}=0$ of this relationship. In the total absence of an analytical method to evaluate the bias parameters $b_n$, they can be treated as free parameters to be fitted over the N-body simulations, a method followed by \citep{2013MNRAS.429.1674C,doi:10.1111/j.1365-2966.2011.19379.x,Vlah:2016bcl}, for instance. In \citep{Valogiannis:2019xed}, an analytical model was developed, for the calculation of the bias parameters in MG models, which is the one we adopt in this work. Based on the PBS formalism \citep{1984ApJ...284L...9K}, the Lagrangian bias factors of order $n$ are given by \citep{Mo:1995cs,Mo:1996cn,Sheth:1998xe}:
\begin{equation}\label{biasrig}
b_{n}^L(M)= \frac{1}{\bar{n}_h(M,0)}\frac{d^n\bar{n}_h(M,\Delta)}{d \Delta^n}\Biggr|_{\substack{\Delta=0}}, 
\end{equation}
where $\bar{n}_h(M,0)$ is the halo mass function of halos with mass $M$ and $\bar{n}_h(M,\Delta)$ is its response, in the presence of a long-wavelength density perturbation $\Delta$. By suitably modeling $\bar{n}_h(M,0)$ and its response in MG, using the Sheth-Tormen (ST) model \citep{Sheth:1999mn} with an environment-dependent gravitational collapse, \citep{Valogiannis:2019xed} derived the PBS biases in MG models (relationships (78)-(80) in that work). This approach was shown to agree very well with simulations and thus we adopt it in this work as well; readers interested in more details about this implementation, are referred to \citep{Valogiannis:2019xed}. The halo bias values $b_1$ and $b_2$, used in this paper, are the ones shown in Table I of \citep{Valogiannis:2019xed}, predicted for each gravity model, halo mass range and cosmological time.

\subsection{Direct Lagrangian Approach to RSD in Modified Gravity}\label{DirectLPT}

In the previous section \ref{BiasMG}, we discussed how LPT can be used to robustly model the two-point statistics of halos in both the cases of GR and MG. However, the peculiar velocities of the observed galaxies, sourced by the perturbations in the underlying density field, contribute to the line-of-sight component of the observed recession velocity, contaminating thus the information extracted from spectroscopic means. These ``Redshift-Space Distortions'', in particular, introduce an anisotropy in the observed clustering pattern \citep{10.1093/mnras/227.1.1,1992ApJ...385L...5H,Hamilton:1997zq}. To model their impact on the clustering statistics, and following standard practice, we fix the line-of-sight in the Cartesian $\hat{z}$ direction for all objects, adopting the plane-parallel approximation. This approximation has been shown to work well in the context of modern surveys of the LSS \citep{10.1111/j.1365-2966.2011.20169.x,Yoo:2013zga}. Having adopted this approach, if $\bold{x}$ is the real-space position of a tracer with peculiar velocity $\bold{v}(\bold{x})$, then its observed, ``redshift-space'' position $\bold{s}$ will be:
\begin{equation}\label{rsdpos}
\bold{s} = \bold{x} + \frac{\hat{z}\cdot\bold{v}(\bold{x})}{a H(a)}\hat{z},
\end{equation}
with $H(a)$ the Hubble factor at a given scale-factor $a$. As a consequence, the redshift-space 2-point correlation function for halos
\begin{equation}\label{xiRSD}
\xi_X^s(\bold{r}) = \langle \delta_X(\bold{s})\delta_X(\bold{s+r})\rangle, 
\end{equation}
becomes directionally dependent, unlike the real-space expression given by (\ref{xi}). 

With the advent of precision cosmology, a great deal of theoretical effort has been put into analytically modeling (\ref{xiRSD}), with the various different approaches summarized in \citep{Vlah:2018ygt}. Within the framework of LPT, the most straightforward approach, called ``Direct Lagrangian'' in \citep{Vlah:2018ygt} and considered in \citep{2013MNRAS.429.1674C,White:2014gfa,Vlah:2015sea}, takes advantage of the fact that, in LPT, the displacement field simply transforms as:
\begin{equation}\label{Psimap}
\bold{\Psi}^s = \bold{\Psi} + \frac{\hat{z}\cdot\bold{\dot{\Psi}}(\bold{x})}{H(a)}\hat{z}.
\end{equation}
This can be easily seen if ones combines (\ref{rsdpos}) and (\ref{Lagpos}) with the fact that the peculiar velocity in LPT is given by $\bold{v}=a \bold{\dot{\Psi}}$. (\ref{Psimap}) can then be further simplified if one notices that, up to order $n$, the LPT field evolves as $\bold{\Psi}^{(n)}\propto D^n(a)$, giving $\bold{\dot{\Psi}}^{(n)}= n f_0 H \bold{\Psi}^{(n)}$, with $f_0(a)=\frac{d \ln D}{d \ln a}$ the GR growth rate, thus giving
\begin{equation}\label{PsiRSD2}
\Psi_i^{s(n)}=\left(\delta_{ij}+n f_0\hat{z}_i\hat{z}_j\right)\Psi_j^{(n)}.
\end{equation}
Then it quickly follows that the Lagrangian correlators (\ref{eq:correl}) will also transform accordingly, e.g. 
\begin{equation}\label{CorrelRSD}
U_i^{s(n)}=\left(\delta_{ij}+n f_0\hat{z}_i\hat{z}_j\right)U_j^{(n)}.
\end{equation}
In the Direct Lagrangian approach, the RSD correlation function is calculated through directly mapping  the Lagrangian correlators (\ref{eq:correl}) to redshift-space, as in (\ref{PsiRSD2})-(\ref{CorrelRSD}), and then using (\ref{xiXfinal}) with the shifted correlators. 

In MG, however, the situation is a little more complicated, because, as we saw in (\ref{growth1stMG}), the MG growth factor is scale-dependent. As a result, $\bold{\dot{\Psi}}$ cannot be simplified as in GR and  (\ref{PsiRSD2}) does not apply. Instead, the LPT displacement field will now transform as 
\begin{equation}\label{PsiRSDMG}
\Psi_i^{s(n)}=\Psi_i^{(n)}+\hat{z}_i\hat{z}_j\frac{d\Psi_j^{(n)}}{d \ln a},
\end{equation}
where the added shift has to be evaluated numerically. In the Appendix \ref{AppendixDirect}, we present the details on how the Direct Lagrangian approach is implemented in MG theories with scale-dependence. It is worth emphasizing, at this point, that the fact that the RSD shift depends so sensitively on the underlying gravity model, is exactly what makes it such a powerful cosmological probe.

\subsection{The Gaussian Streaming Model In Modified Gravity}\label{GSMG}

In the previous section, we saw that, despite their success at accurately capturing the real-space clustering statistics for a wide range of models, Lagrangian methods prove to be inadequate at $\it{directly}$ predicting the velocity-induced redshift-space anisotropies. This problem can be overcome by employing the Gaussian Streaming Model (GSM), first proposed in \citep{doi:10.1111/j.1365-2966.2011.19379.x}, inspired by the the work of \citep{1995ApJ...448..494F}. In order to address the discrepancy between the traditional phenomenological ``streaming'' (or dispersion) models \citep{1983ApJ...267..465D,10.1093/mnras/258.3.581} and the linear Kaiser limit \citep{10.1093/mnras/227.1.1}, \citep{1995ApJ...448..494F} adopted a probabilistic approach to relate the distributions of tracers in the real and redshift space. In particular, if $\mathcal{P}$ is the pairwise velocity Probability Density Function (PDF), then the real-space correlation function of tracers, $\xi^r_X(r)$, will be mapped to the redshift-space one as \citep{1995ApJ...448..494F,PhysRevD.70.083007}:
\begin{equation}\label{Fisherxi}
1 + \xi^s_X(s_{\perp},s_{\parallel}) = \int dy [1+ \xi^r_X(r)]\mathcal{P}(y = s_{\parallel}-r_{\parallel}|\bold{r}),
\end{equation}
where $s_{\perp},s_{\parallel}$ are the perpendicular and parallel to the line-of-sight components of the redshift-space separation s, with $s=\sqrt{s_{\perp}^2+s_{\parallel}^2}$ and $r=\sqrt{s_{\perp}^2+y^2}$. By taking the linear limit of  (\ref{Fisherxi}), and allowing for the scale-dependence of the pairwise velocity moments,  \citep{1995ApJ...448..494F} showed that line-of-sight variations of the pairwise velocity and its dispersion drive the correlation function away from isotropy in redshift space. The pairwise velocity PDF $\mathcal{P}$ is in principle not Gaussian, even in the case of a Gaussian density field, but can be well approximated by a Gaussian near its peak \citep{PhysRevD.70.083007}. Using a non-perturbative resummation of the linearized limit of (\ref{Fisherxi}) in \citep{1995ApJ...448..494F}, \citep{doi:10.1111/j.1365-2966.2011.19379.x} proposed the GSM expression:
\begin{equation}\label{xiGSM}
1+\xi^s_X(s_{\perp},s_{\parallel}) =\int_{-\infty}^\infty \frac{d y}{\sqrt{2\pi\sigma_{12}^2(r,\mu)}} 
[1+\xi^r_X(r)] \exp\left[-\frac{\left(s_\parallel-y-\mu v_{12}(r)\right)^2}{2\sigma^2_{12}(r,\mu)}   \right],
\end{equation}
where $\mu=\hat{r}\cdot\hat{z}= \frac{y}{r}$, $\mu v_{12}(r)$ the pairwise velocity and $\sigma^2_{12}(r,\mu)$ the pairwise velocity dispersion along the line-of-sight. Using CLPT to model the ingredients of the GSM, the accuracy of the initial approach was improved in \citep{Wang:2013hwa}, which was able to match the redshift-space halo clustering statistics extracted from N-body simulations at the few $\%$ level. Further improvements included adding tidal bias and EFT corrections \citep{Vlah:2016bcl}, as well as higher moments in the cumulant expansion \citep{Uhlemann:2015hqa,Bianchi:2014kba,Bianchi:2016qen}, while the GSM was also applied to observational data \citep{2012MNRAS.426.2719R,2013MNRAS.429.1514S,2014MNRAS.439.3504S,Alam:2016hwk,Zarrouk:2018vwy}. In \citep{Bose:2017dtl}, the GSM was employed to model the anisotropic correlation function for dark matter in MG, using RegPT and the SPT approach in \citep{doi:10.1111/j.1365-2966.2011.19379.x}, but not LPT. 

Building upon the formalism presented in \citep{Wang:2013hwa} and having already laid the foundation in Section \ref{BiasMG}, we proceed to expand the GSM (\ref{xiGSM}) to predict the anisotropic redshift-space correlation function for biased tracers in MG, modeling its ingredients with CLPT. To do so, we need to express the two velocity moments in CLPT, as we have already done with $\xi(r)$, which is performed below. 

\subsection{Velocity Moments in Modified Gravity}

With regards to the calculation of the velocity moments in MG, the main point of divergence from the corresponding approach in GR employed in \citep{Wang:2013hwa,Vlah:2016bcl}, lies in the fact that the LPT growth factors are also scale-dependent in this case, as we saw in section \ref{LPTMG}. Keeping this in mind, below we present the main results and summarize how they differ from their GR counterparts, with the details shown in the Appendix \ref{AppendixGSM}. The relative peculiar velocity between two tracers at Eulerian positions $\bold{x}_2$ and $\bold{x}_1$, is
\begin{equation}\label{veldif}
\frac{\bold{v}_n(\bold{x}_2) - \bold{v}_n(\bold{x}_1)}{a H} = \frac{\bold{\dot{\Psi}}_{2n}- \bold{\dot{\Psi}}_{1n}}{H} = \frac{\bold{\dot{\Delta}}_n}{H},
\end{equation}
where we made use of the fact that $\bold{v} = a \bold{\dot{\Psi}}$. In GR, one typically uses, as we also saw in section \ref{DirectLPT}, the fact that $\bold{\Psi}^{(n)}\propto D^n(a)$ in the EDS approximation, which gives $\bold{\dot{\Psi}}^{(n)}= n f_0 H \bold{\Psi}^{(n)}$, so as to simplify (\ref{veldif}), which is $\it{not}$ the case in MG; here $\bold{\dot{\Delta}}_n$ needs to be evaluated numerically. Following standard practice, one may then define the velocity generating function \citep{PhysRevD.70.083007,Wang:2013hwa}
\begin{equation}\label{velgen}
Z(\bold{r},\bold{J}) = \langle [1+ \delta_X(\bold{x})] [1+\delta_X(\bold{x+r})] e^{\bold{J}\cdot \frac{\bold{\dot{\Delta}}}{H}}\rangle,
\end{equation}
with $\xi_X(r) = Z(r,0) -1 $. In the case of a local Lagrangian bias (\ref{delJacX}) and after the usual Fourier transforms (as in \citep{Aviles:2018qot,Valogiannis:2019xed}), it can be expressed as:
\begin{equation}\label{velgen2}
Z(\bold{r},\bold{J})= \int d^3q\int \frac{d^3k}{\left(2 \pi\right)^3}e^{i\bold{k}\cdot(\bold{q}-\bold{r})} \int \frac{d^2\Lambda_1}{\left(2 \pi\right)^2}\frac{d^2\Lambda_2}{\left(2 \pi\right)^2}\tilde{F}_1 \tilde{F}_2\langle e^{i\left[\lambda_1 \delta_1+\lambda_2 \delta_2+\eta_1 \nabla^2\delta_1+\eta_2 \nabla^2\delta_2+\bold{k}\cdot \bold{\Delta}+\bold{J}\cdot \frac{\bold{\dot{\Delta}}}{H})\right]}\rangle,
\end{equation}
where we defined $\Lambda_1=(\lambda_1,\eta_1)$, $\tilde{F}_1=\tilde{F}(\Lambda_1)$ and $\Lambda_2=(\lambda_2,\eta_2)$, $\tilde{F}_2=\tilde{F}(\Lambda_2)$, as in \citep{Aviles:2018saf}. Given a generating functional, the velocity moments of order $p$ can then be straightforwardly evaluated as \citep{Wang:2013hwa}:
\begin{equation}\label{velmom}
\begin{aligned}
& \langle [1+ \delta_X(\bold{x})] [1+\delta_X(\bold{x+r})] \biggl( \prod_{k=1}^{p}[\bold{v}_{i_k}(\bold{x+r})-\bold{v}_{i_k}(\bold{x})]\biggr) \rangle = \\
& = \prod_{k=1}^{p}\biggl(-i \frac{\partial}{\partial \bold{J}_{i_k}} \biggr)Z(\bold{r},\bold{J})\Bigr\rvert_{\bold{J}=0 }  \\
& =   \int d^3q\int \frac{d^3k}{\left(2 \pi\right)^3}e^{i\bold{k}\cdot(\bold{q}-\bold{r})} \int \frac{d^2\Lambda_1}{\left(2 \pi\right)^2}\frac{d^2\Lambda_2}{\left(2 \pi\right)^2}\tilde{F}_1 \tilde{F}_2\times \langle \prod_{k=1}^{p}\biggl(\frac{\bold{\dot{\Delta}}_{i_k}}{H})\biggr) e^{i\left[\lambda_1 \delta_1+\lambda_2 \delta_2+\eta_1 \nabla^2\delta_1+\eta_2 \nabla^2\delta_2+\bold{k}\cdot \bold{\Delta}\right]}\rangle\\ 
& = \int d^3q M_{p,(i_1,.,i_p)} (\bold{r},\bold{q}), \\
\end{aligned}
\end{equation}
where in the last line we defined $M_{p,(i_1,.,i_p)}$ as the integrand quantity.  

The real-space mean pairwise velocity along the pair separation vector, $\hat{r}$, is defined as :
\begin{equation}\label{meanv12}
v_{12}(r)  = v_{12,n} \hat{r}_n= \frac{\langle [1+ \delta_X(\bold{x})] [1+\delta_X(\bold{x+r})] [\bold{v}_{n}(\bold{x+r})-\bold{v}_{n}(\bold{x})] \rangle }{\langle [1+ \delta_X(\bold{x})] [1+\delta_X(\bold{x+r})]\rangle }\hat{r}_n.
\end{equation}
The denominator of (\ref{meanv12}) is simply equal to $1+\xi_X(r)$ from (\ref{xiXfinal}), whereas the numerator represents the galaxy-number weighted average pairwise velocity. Given the definition (\ref{velmom}), and using the CLPT scheme for MG discussed in Section \ref{BiasMG}, we have \citep{Wang:2013hwa}:
\begin{equation}\label{v12n}
v_{12,n} = \frac{ \int d^3q M_{1,n} (\bold{r},\bold{q})}{1+\xi_X(r)},
\end{equation}
with
\begin{eqnarray}
\label{M1def}
M_{1,n} (\bold{r},\bold{q}) &=& f_0 \frac{e^{-\frac{1}{2} (q_i - r_i) (A^{-1}_{L})_{ij}(q_j - r_j)}} {\left(2\pi\right)^{3/2} |A_{L}|^{1/2}} \times \nonumber\\
 && \Biggl(  2 b_1\dot{U}^{(1)}_n - g_i \dot{A}_{in}  + b_2 \dot{U}^{20} + b_1^2 \dot{U}^{11} - \frac{1}{2} G_{ij} \dot{W}_{ijk} -2 b_1g_i \dot{A}^{10}_{in} \nonumber\\
&&+ 2b_1 b_2 \xi_L \dot{U}^{(1)}_n - 2[b_2 + b_1^2]g_i U^{(1)}_i \dot{U}^{(1)}_n - b_1^2\xi_Lg_i \dot{A}^{(1)}_{in} - 2 b_1 G_{ij} U^{(1)}_i\dot{A}^{(1)}_{in}\nonumber\\
&& -2b_{\nabla^2 \delta}\mathcal{B}_{2,n} \Biggr),
\end{eqnarray}
where we defined
\begin{align}\label{dotfuncs}
 \dot{U}_i(\vq) \equiv \frac{1}{f_0 H}\langle \delta(\vq_1) \dot{\Delta}\rangle, \qquad  & 
 \dot{A}_{ij}(\vq) \equiv \frac{1}{f_0 H}\langle \Delta_i \dot{\Delta}_j\rangle \nonumber\\
 \dot{U}^{20}_i  \equiv \frac{\langle \delta^2(\vq_1) \dot{\Delta}_i\rangle}{H f_0}, \qquad &
 \dot{U}^{11}_i  \equiv \frac{\langle \delta(\vq_1)\delta(\vq_2) \dot{\Delta}_i\rangle}{H f_0}, \nonumber\\
 \dot{A}_{ij}^{10} = \frac{\langle \delta(\vq_1) \Delta_i \dot{\Delta}_j\rangle}{f_0 H}, \qquad &
\dot{W}_{ijk} = \frac{\langle \Delta_i \Delta_j \dot{\Delta}_k\rangle}{f_0 H}, \nonumber\\
 \mathcal{B}_{2,n} = -\nabla_n \xi_L(q). \qquad & \nonumber\\    
\end{align}
The expressions for the new correlators (\ref{dotfuncs}) are presented in the Appendix \ref{AppendixGSM}. Here we briefly stress that, even though these definitions are the same as in the GR case \citep{Wang:2013hwa,Vlah:2016bcl}, the functions (\ref{dotfuncs}) take different values in MG, because of the different evolution of the LPT displacement field (manifesting itself in the $\Delta$ and $\dot{\Delta}$ functions). Similarly, this is also the case for the $M_{1,n}$ function in (\ref{M1def}), that depends on these functions. The quantity entering (\ref{xiGSM}) is actually the pairwise velocity along the line-of-sight, rather than the separation vector, which simply accounts to multiplying $v_{12}$ from (\ref{meanv12}) by $\mu$. 

The pairwise velocity dispersion along the line-of-sight is defined as 
\begin{equation}\label{sig12}
\hat{\sigma}^2_{12}(r,\mu)= \frac{\langle [1+ \delta_X(\bold{x})] [1+\delta_X(\bold{x+r})] [v_{z}(\bold{x+r})-v_{z}(\bold{x})]^2 \rangle }{\langle [1+ \delta_X(\bold{x})] [1+\delta_X(\bold{x+r})]\rangle },
\end{equation}
which is commonly decomposed into components parallel  $\hat{\sigma}^2_{\parallel}$ and perpendicular $\hat{\sigma}^2_{\perp}$ to the pair separation vector, r, in which case 
\begin{equation}\label{sig12dec}
\hat{\sigma}^2_{12}(r,\mu)= \mu^2\hat{\sigma}^2_{\parallel} + (1-\mu^2)\hat{\sigma}^2_{\perp}.
\end{equation}
The two components can be calculated after taking projections of the second velocity moment, the pairwise velocity dispersion tensor, given by \citep{Wang:2013hwa}:
\begin{equation}\label{sig12mn}
\hat{\sigma}^2_{12,nm} = \frac{ \int d^3q M_{2,nm} (\bold{r},\bold{q})}{1+\xi_X(r)},
\end{equation}
with
\begin{eqnarray}
\label{M2def}
M_{2,nm} (\bold{r},\bold{q}) &=& f_0^2 \frac{e^{-\frac{1}{2} (q_i - r_i) (A^{-1}_{L})_{ij}(q_j - r_j)}} {\left(2\pi\right)^{3/2} |A_{L}|^{1/2}} \times \nonumber\\
 && \Biggl(  2\left[b_1^2+b_2\right]\left(\dot{U}^{(1)}_n\dot{U}^{(1)}_m\right)  -2 b_1\left(\dot{A}^{(1)}_{in} g_i \dot{U}^{(1)}_m + \dot{A}^{(1)}_{im} g_i \dot{U}^{(1)}_n\right) -\dot{A}^{(1)}_{im}\dot{A}^{(1)}_{jn}G_{ij}\nonumber\\
&&+ \ddot{A}_{nm} + b_1^2 \xi_L \ddot{A}^{(1)}_{nm} -2 b_1 U^{(1)}_i g_i\ddot{A}^{(1)}_{nm} + 2 b_1 \ddot{A}^{10}_{nm} -\ddot{W}_{inm}g_i\Biggr),
\end{eqnarray}
and where we additionally defined
\begin{equation}\label{ddotfuncs}
\ddot{A}_{ij} = \frac{\langle  \dot{\Delta}_i \dot{\Delta}_j\rangle}{f_0^2 H^2}, \qquad \ddot{A}_{ij}^{10} =  \frac{\langle \delta(\vq_1) \dot{\Delta}_i \dot{\Delta}_j\rangle}{f_0^2 H^2}, \qquad \ddot{W}_{ijk} = \frac{\langle \Delta_i \dot{\Delta}_j \dot{\Delta}_k\rangle}{f_0^2 H^2}. 
\end{equation}
The expressions for the functions (\ref{ddotfuncs}) are presented in the Appendix \ref{AppendixGSM}. It is worth adding here, that all $b_{\nabla^2 \delta}$ terms identified in $\sigma^2_{12,nm}$, are multiplications of U functions and terms that although are order $P_L$, have a size similar to 1-loop terms \citep{Vlah:2016bcl}, which is why they are dropped and thus absent in (\ref{M2def}). More importantly, even though we do not generally include EFT corrections in our biasing scheme, we emphasize on the fact that one of the leading EFT counterterms in (\ref{M2def}) is of the form $\alpha_\sigma \delta_{nm}$ \citep{Vlah:2016bcl}, which corresdonds to the correction:
\begin{equation}\label{asEFT}
\hat{\sigma}^2_{12,nm} \rightarrow \hat{\sigma}^2_{12,nm} + \alpha_\sigma \frac{1+\xi_{matter}^{0-loop}}{1+\xi_{X}^{1-loop}} \delta_{nm},
\end{equation}
that leads to a constant shift, $\alpha_{\sigma}$, at large scales. This naturally accommodates the need to add a constant shift to the PT prediction for $\sigma^2_{12,nm}$ so as to match the values extracted from N-body simulations, as was found in \citep{doi:10.1111/j.1365-2966.2011.19379.x,Wang:2013hwa,Vlah:2016bcl} for GR and as we will also show to be the case for MG, in the next section.
\\ Having obtained $\hat{\sigma}^2_{12}$, we get the cumulant version $\sigma^2_{12,nm} = \hat{\sigma}^2_{12,nm} - v_{12,n}v_{12,m}$  \citep{Vlah:2016bcl}, which is then projected as:
\begin{equation}\label{sigparperp}
\sigma_\parallel^2(r) =  \hat{r}_n\hat{r}_m\sigma^2_{12,nm}(\ve r), \qquad \sigma_\perp^2(r) =  \frac{1}{2}(\delta_{nm}+\hat{r}_n\hat{r}_m)\sigma^2_{12,nm}(\ve r).
\end{equation}
The combination of (\ref{sig12dec})-(\ref{sigparperp}) gives us $\sigma^2_{12}(r,\mu)$ that is the final necessary ingredient to enter eqn (\ref{xiGSM}). Finally, as was the case for $\xi_X(r)$ in (\ref{xiXfinal}), we eventually consider the velocity moments $v_{12}$ and $\sigma^2_{12}$ with a local in matter density Lagrangian bias (meaning $b_{\nabla^2\delta}=0)$, the bias factors $b_1$ and $b_2$ of which are evaluated using the analytical model discussed in section \ref{BiasMG}. 

\section{Results}\label{results}

The objective in this work is to assess and compare configuration-space (redshift-space) predictions coming from our analytical model, against those extracted from the Group I and Group II N-body simulations, that we  introduced in section \ref{Nbody}.  

In section \ref{sec:pwvel}, we first confirm that the velocity information entering the GSM (\ref{xiGSM}), the real-space pairwise velocity and the pairwise velocity dispersion for halos, in (\ref{v12n}) and (\ref{sig12mn}), are accurately captured by our CLPT implementation in MG. 

In section \ref{sec:twoptcf}, we assess the predictions for the redshift-space 2-point correlation function for halos using both the Direct Langrangian and GSM approaches. Using the PBS formalism for the bias values, the 3 ingredients used as input in the GSM expression (\ref{xiGSM}), are calculated through (\ref{xiXfinal}), (\ref{v12n}) and (\ref{sig12mn}).  The real-space 2-point correlation function for halos in MG has already been cross-checked and confirmed in \citep{Valogiannis:2019xed}, using the same CLPT and bias schemes against the same set of simulations. 

To perform the various integrations, we use a suitably modified version of the public C++ code released by \citep{Vlah:2016bcl}, to incorporate the modifications to allow deviations from GR. This extends our previous work in \citep{Valogiannis:2019xed}, made publicly available in \url{https://github.com/CornellCosmology/bias_MG_LPT_products}. The code accepts as input the linear power spectrum and the LPT growth factors up to $3^{rd}$ order, evaluated for each MG model and cosmology. The linear power spectra are calculated using the publicly available code \verb|CAMB| \citep{Lewis:1999bs}, while the growth factors are extracted from the \verb|MATHEMATICA| notebooks released in the above GitHub repository.

\subsection{Halo Pairwise Velocity Statistics}\label{sec:pwvel}

We begin this section by comparing the CLPT predictions for the real-space pairwise velocity of halos in MG, obtained through (\ref{meanv12})-(\ref{M1def}), against the results from the N-body simulations. 

\begin{figure}[tbp]
\centering
\includegraphics[width=1.0\textwidth]{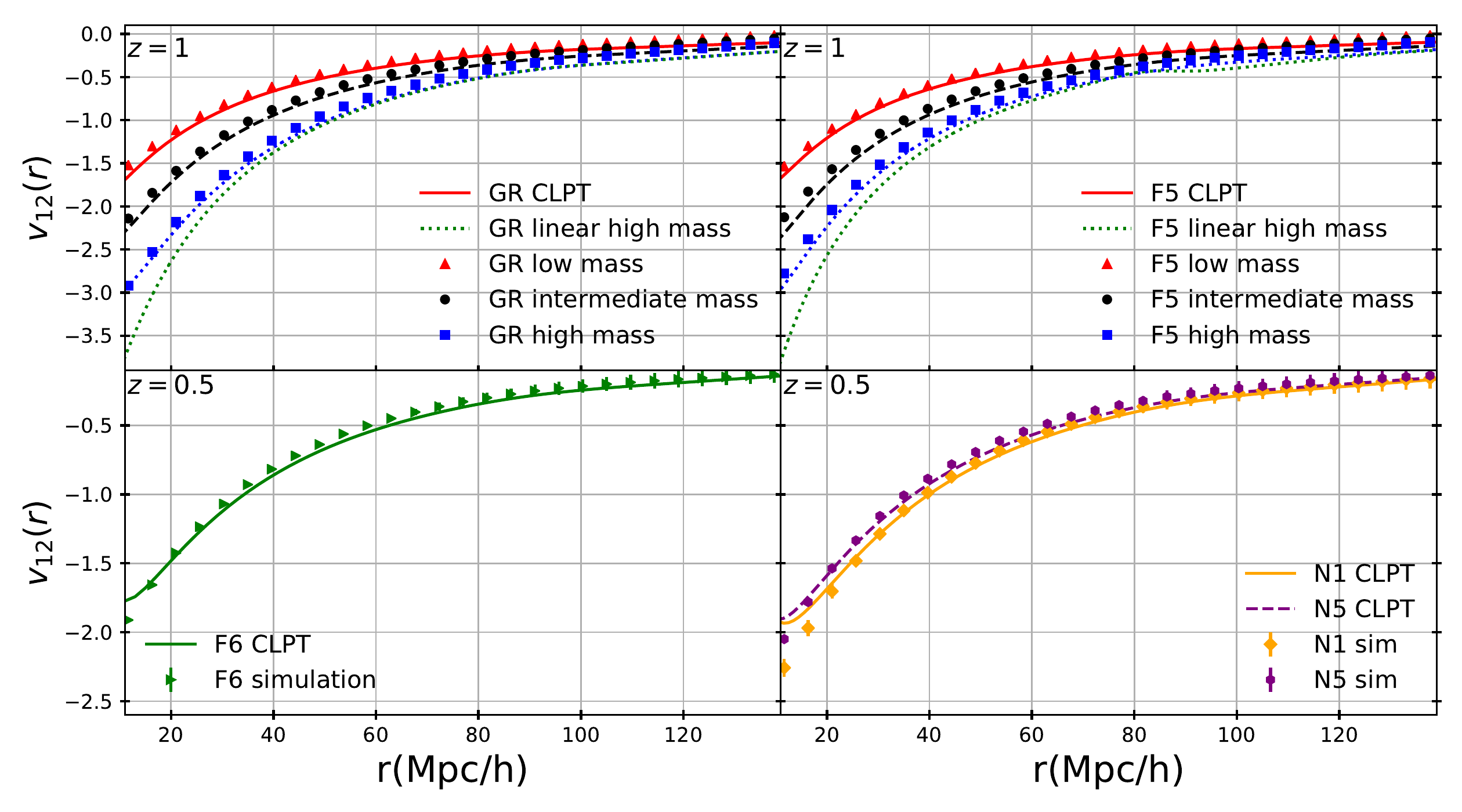}
\caption{\label{fig:1} Real-space pairwise velocity for GR [top left] and the F5 model [top right] at $z=1$, obtained by the Group II simulations in the low mass [blue square], intermediate mass [black circle] and high mass [red triangle] bins, and for F6 [bottom left] and the N1 \& N5 models [bottom right, orange diamond and purple hexagon respectively] at $z=0.5$ from the Group I simulations. Lines denote the theoretical predictions from 1-loop CLPT, through (\ref{meanv12})-(\ref{M1def}), for each corresponding model. The green dotted lines in the upper panels represent the linear prediction for the high mass bin.}
\end{figure}

In Fig. \ref{fig:1} we compare the analytical CLPT  predictions for the real-space pairwise velocity of halos in MG, obtained through (\ref{meanv12})-(\ref{M1def}), against the results from the N-body simulations. In the top panels we show comparisons  for GR and the F5 MG model  for the 3 mass bins in the $z=1$ snapshot of the Group II simulations. The F5 CLPT curves are found to trace the shape of the pairwise velocity well, across a wide range of halo masses, achieving the same level of agreement as in the known GR case, down to scales of $r\sim10$ Mpc/h. 
The 1-loop CLPT result significantly improves upon the accuracy of the linear-theory prediction.
These results are consistent with what was observed in the corresponding GR case in \citep{Wang:2013hwa}. 

Performing the same comparison against the F6 and N1 $\&$ N5 models of the Group I snapshots at $z=0.5$, shown in the bottom panel of Fig. \ref{fig:1}, we again find that CLPT has the same level of agreement as previously, but down to scales of $r\sim17$ Mpc/h for the F6 $\&$ N5 models and for $r\sim20$ Mpc/h for the N1 case. Given that this comparison is now performed at a lower redshift than in the Group II case, where the expected scale where nonlinearities become significant (and PT fails) is larger, this result is expected. 

The bias parameters we use here, derived by fitting the ST parameters to the simulated halo mass function, are a key factor in achieving accurate predictions for the pairwise velocity, which is a necessary requirement for accurate predictions using the GSM approach. An alternative option is to treat the biases as free parameters, that give the best-fit to the simulations, as for example done in \citep{2013MNRAS.429.1674C,Wang:2013hwa,Vlah:2016bcl}.

We next investigate how well our CLPT framework performs in capturing the halo pairwise velocity dispersion from the simulations, for all MG models we consider. 
When performing this comparison, in terms of the dispersion components parallel ($\hat{\sigma}^2_{\parallel}$) and perpendicular ($\hat{\sigma}^2_{\perp}$) to the pair separation vector, we uncover the existence of a persistent offset between the theoretical curves and the simulated values, manifesting itself in all cases we study. This phenomenon has been observed in the context of GR, both when using Eulerian \citep{doi:10.1111/j.1365-2966.2011.19379.x} and Lagrangian \citep{Wang:2013hwa,Vlah:2016bcl} PT to model $\sigma^2$ and in MG \citep{Bose:2017dtl}, which we also find to be the case in our CLPT implementation for MG models. This mismatch is attributed to small-scale contributions to the halo velocity dispersion, that are impossible to capture analytically  \citep{doi:10.1111/j.1365-2966.2011.19379.x,Wang:2013hwa}. It was observed that simply adding a constant offset so as to match the two predictions at large scales suffices to get an accurate perturbative representation of $\sigma^2$. We should also note, at this point, that this feature was also discussed by \citep{PhysRevD.70.083007}, who noted that nonlinear contributions to the velocity dispersion contribute a constant in the large-scale limit, that cannot be captured by linear theory. In \citep{Vlah:2016bcl}, that considered EFT corrections in the context of the GSM, it was shown that one of the EFT counter-terms contributing to the velocity dispersion is of the form (\ref{asEFT}), which reduces to a constant at large scales and naturally accommodates for the need to correct this offset with a constant free parameter. In our work, and in agreement with the above results, we find the CLPT prediction from (\ref{sig12mn}) $\&$ (\ref{M2def}), combined with a constant offset to match the largest simulation bin at $r=137.8$ Mpc/h, to be able to capture the shape of $\hat{\sigma}^2_{\parallel}$ and $\hat{\sigma}^2_{\perp}$ very well for all the modified gravity models, as well as GR. The values of the constants added for each model are reported in Table~\ref{table1}.  

When looking at Table~\ref{table1}, one notices that the Zel'dovich result for $\sigma^2$ always falls a little short of the simulated value at the largest spatial bin and needs to be corrected by a small positive constant. Trying to correct for this deficit by including the 1-loop contributions, however, always results at a relatively larger overestimate of the large-scale value of the dispersion, which then needs to corrected by adding a large negative constant, contrary to the Zel'dovich results that need only a small positive offset. This behavior has been observed before for $\Lambda$CDM in \citep{doi:10.1111/j.1365-2966.2011.19379.x,Wang:2013hwa,Uhlemann:2015hqa,Vlah:2016bcl} (see in particular fig.4 and the discussion after eq.~44 in \citep{Uhlemann:2015hqa}), with the authors of \citep{Uhlemann:2015hqa} arguing that the better performance of linear theory, relatively over CLPT, should be considered an accident because one expects large corrections to $\sigma^2$ due to the presence of zero-lag correlators. Furthermore, \citep{Vlah:2016bcl} also reported large values for the offset (see table 1 of that paper), while the same method was applied to fit real data in \citep{2012MNRAS.426.2719R,Icaza-Lizaola:2019zgk}. We further find parallels with what was observed in \citep{Vlah:2014nta,Valogiannis:2019xed} for predictions of the dispersion of the LPT displacement field. There, it was shown that 1-loop LPT tends to over-predict the LPT displacement-field dispersion, compared to the simulations, because this quantity depends on zero-lag correlators, that are hard to model perturbatively. Linear theory, on the other hand, was found to give a closer estimate, a little short of the simulated value, as we also see to be the case when modeling the pairwise velocity dispersion. If we take the large-scale limit of the CLPT expression for the pairwise velocity dispersion (\ref{sig12mn}) $\&$ (\ref{M2def}), we find that
\begin{equation} \label{sigmaLSlimit}
 \hat{\sigma}_{\parallel}^2 , \hat{\sigma}_{\perp}^2 \longrightarrow f_0^2 (\ddot{X}^L_{\infty} + \ddot{X}^\text{loop}_{\infty} + 2 b_1 \ddot{X}^{10}_{\infty}),
\end{equation}
with $\ddot{X}_{\infty}$ and $\ddot{X}^{10}_{\infty}$ the constant limits of functions $\ddot{X}(q)$ and $\ddot{X}^{10}(q)$, defined in (\ref{ddotXdefs},\ref{ddotX10defs}), as $q \rightarrow \infty$.
As a result of (\ref{sigmaLSlimit}), we see that the two pairwise velocity dispersion components $\hat{\sigma}^2_{\parallel}$ and $\hat{\sigma}^2_{\perp}$ also involve zero-lag quantities in their large-scale limit, which explains the connection that can be drawn between these two cases. As in \citep{Valogiannis:2019xed}, we find this overestimation to be more pronounced in MG, compared to the GR case, getting progressively more pronounced with higher deviations from GR, which explains why the degree of analytical over-estimation in the LPT predictions becomes more severe, as one considers models that are less screened. We also notice that the degree of analytical over-estimation increases with halo mass in the Group II snapshots. This is also explained from (\ref{sigmaLSlimit}) where we see that the 1-loop correction of the large-scale limit of the velocity dispersion depends on the the linear halo bias, through the third term. Higher halo masses correspond to a larger value for the linear bias $b_1$, which in turn makes the third term larger, relative to the lower mass cases, resulting in an overall even larger overestimation in this case. The fact that the bias dependence appears beyond the linear level, explains why the shifts required for the Zel'dovich predictions are very similar across the different mass bins for both the GR and the F5 snapshots of the Group II simulations. Also, this bias dependence implies that, for a fixed halo mass range, and given that the bias $b_1$ increases with redshift, this overestimation can become relatively more pronounced at higher z. \footnote{At this point, however, we should be careful \textit{not} to directly compare the results reported for the two redshifts of Table~\ref{table1}, as they refer to two snapshots with different mass ranges, which also correspond to different cosmologies.} Finally, in Table \ref{table1} we notice that 1-loop LPT predicts increasing values for the large-scale limit of the pairwise velocity dispersion, as we move towards higher halo masses, whereas the opposite trend is reflected in the Group II simulations. Even though this is interesting, we cannot certainly say whether this trend is statistically significant, as there is only one available realization and thus we defer this investigation to future work, when more simulations become available. 

\begin{table}[t!]
\begin{center}
\begin{tabular}{ | l | C{4em}| C{4em} | C{4em} | C{4em} | C{4em} |}
\hline
\multirow{2}{*}{Model} & \multicolumn{3}{c|}{$\hat{\sigma}^2_{\parallel}$($r=$100Mpc/h)[(Mpc/h)$^2$] }&  \multicolumn{2}{c|}{Shift ($r$=137.8Mpc/h) }
\\
\cline{2-6}
& LPT & Zel.& Sim. & LPT & Zel.
\\
\hline\hline
Group I: GR & 36.2 & $23.8$ & $27.9$ & $-7.7$ & 4.0
\\ \hline
Group I: F6 & $43.3$& 25.0 & $27.8$ & $-15.5$ & 2.5
\\ 
Group I: N1 & $44.6$ & $29.5$ & $36.2$ & $-8.6$ & 6.5
\\
Group I: N5 & $37.6$ & $25.1$ & $29.7$ & $-7.5$ & 4.3
\\ \hline\hline
Group II: GR low-mass & $28.6$ & $19.9$ & $20.9$  & $-8.3$ & $1.4$
\\
Group II: GR mid-mass & $32.4$ & $19.8$ & $20.7$ & $-11.8$ & $1.0 $
\\
Group II: GR hi-mass & $37.6$ & $19.8$ & $20.5$  & $-17.3$ & $0.5$
\\ \hline
Group II: F5 low-mass & $33.2$ & $21.8$ & $22.5$ &$ -10.5$ & $0.5$ 
\\
Group II: F5 mid-mass & $38.5$ & $21.8$ & $21.8$  & $-16.5$ & $0.0$
\\
Group II: F5 hi-mass & $43.4$ & $21.8$ & $20.9$  & $-22.5$& $-1.0$
\\
\hline
\end{tabular}
\end{center}
\caption{\label{table1} The left-hand columns  compare the values for $\hat{\sigma}_{\parallel}^2$,  the pairwise velocity dispersion parallel to the separation vector, predicted by  1-loop LPT [$1^{st}$ column] 
and the Zel'dovich approximations [$2^{nd}$ col.] of  (\ref{M2def}) with those obtained from the N-body simulations [$3^{rd}$ col.], at $r=100 Mpc/h$ for all the gravity models. 
The right-hand columns present the values of the constant shifts, added to theoretical predictions to give large scale agreement with simulations (at the largest bin center $r=$137.8 Mpc/h), 
for  the 1-loop $\&$ Zel'dovich LPT results from (\ref{sig12mn}) $\&$ (\ref{M2def}).}
\end{table}

\begin{figure}[tbp]
\centering 
{\includegraphics[width=0.49\textwidth]{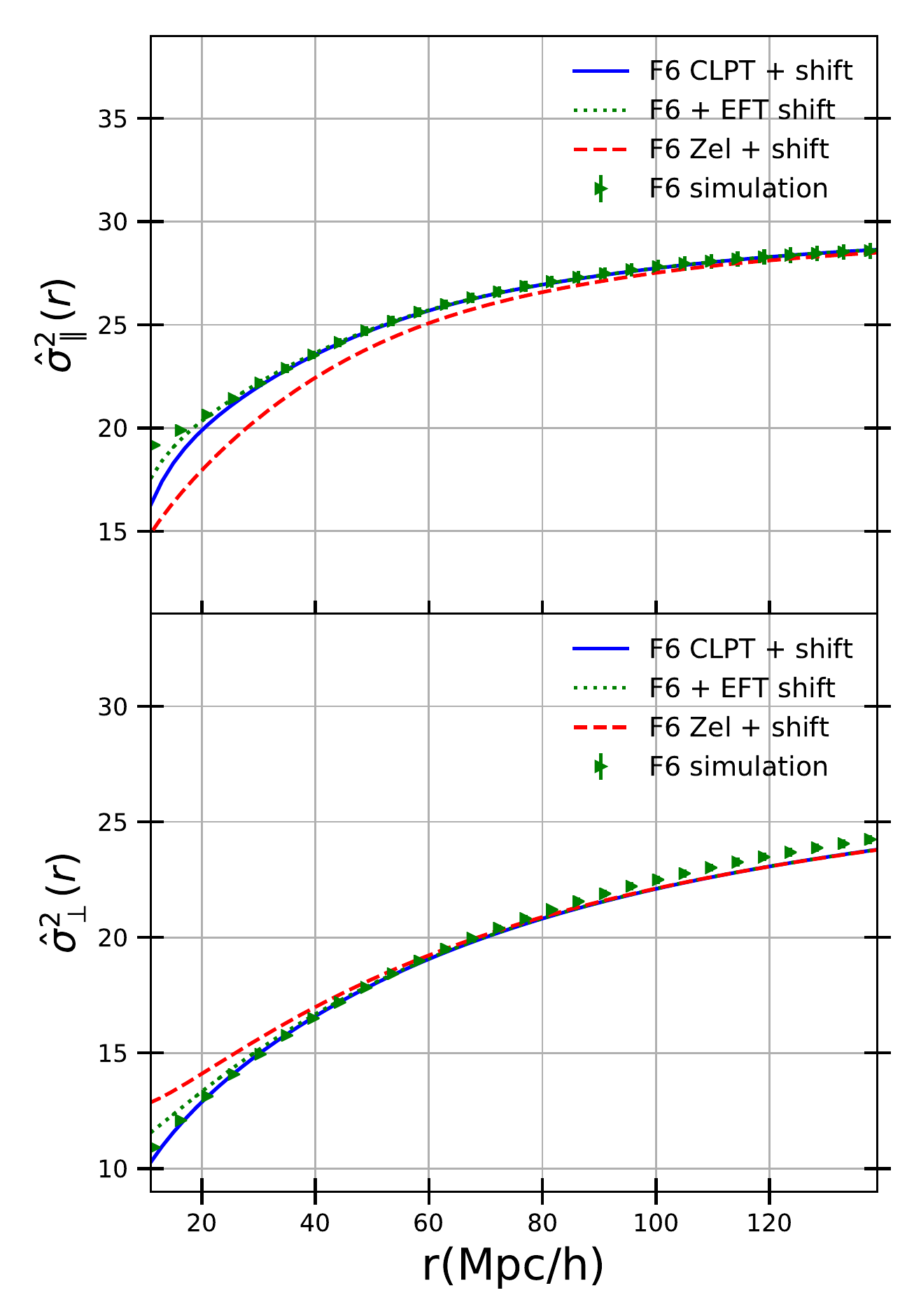}
\includegraphics[width=0.49\textwidth]{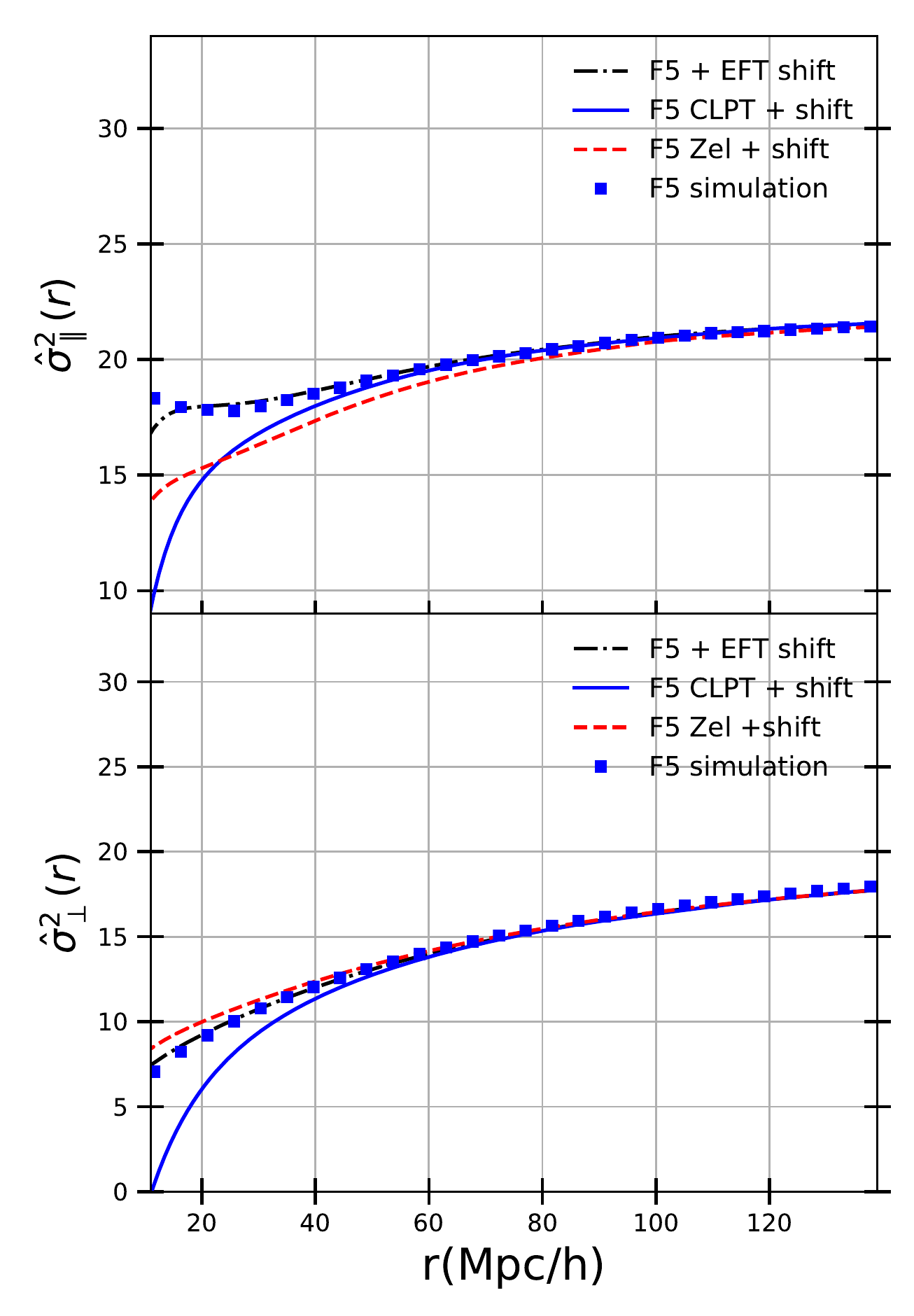}}
\caption{\label{fig:2} The pairwise velocity dispersion parallel [top] and perpendicular [bottom] to the pair separation vector for the F6 model at $z=0.5$ from 
the Group I simulations [left panels, green right triangles]] and the F5 model at $z=1$ from the high mass bin of the Group II simulations [right panels, blue squares]. In all panels the comparison to the theory predictions is shown for the 1-loop [solid blue line] and the Zel'dovich [red dashed line] CLPT predictions from (\ref{sig12mn}) $\&$ (\ref{M2def}), shifted by a constant, as well as from the 1-loop CLPT prediction shifted by a correction term given by EFT [green dotted line], as in (\ref{asEFT}). The values of the applied constant shifts are reported in Table~\ref{table1}.}
\end{figure}

In the left panel of Fig. \ref{fig:2}, the $\hat{\sigma}^2_{\parallel}$ and $\hat{\sigma}^2_{\perp}$ CLPT predictions from (\ref{sig12mn}) $\&$ (\ref{M2def})  are compared against the simulations for the F6 model at $z=0.5$. It is found that the 1-loop result, shifted by a constant, significantly improves upon the (also shifted) Zel'dovich approximation and remains consistent (within 1-$\sigma$ errorbars) with the simulations down to $r\sim30$ Mpc/h for $\hat{\sigma}^2_{\parallel}$ and $r\sim20$ Mpc/h for $\hat{\sigma}^2_{\perp}$. When we shift the 1-loop CLPT result by the EFT counter-term (\ref{asEFT}), the accuracy is further improved and the results remain consistent down to smaller $r$. If we perform the same comparison for the F5 Group II snapshot at $z=1$, however, as done for the high mass bin in the right panel of Fig. \ref{fig:2}, we find that the constant that needs to be added to the 1-loop curve is very large and negative, comparable to the large-scale amplitude of $\sigma^2$. This leads to negative and unphysical values for the dispersion components at low $r$ . In particular, the perpendicular component of the pairwise velocity dispersion, $\sigma^2_{\perp}$, becomes negative at $r=11.2, 6.1, 2.5$ Mpc/h for the high mass, intermediate mass and low mass bins of the F5 snapshot, respectively. The behavior is qualitatively similar for $\sigma^2_{\parallel}$, but the values remain physical (positive) down to slightly smaller scales. The reason this happened at $z=1$ is that, since it is an earlier cosmological time, the velocity dispersion is smaller compared to the $z=0.5$ case and comparable to the negative constant that needs to be added in order to adjust the CLPT prediction. To overcome this issue, which manifests itself for all three $z=1$ halo mass bins and for both MG and GR scenarios, we can model the dispersion components using the other two approximations, the (shifted by a positive constant) Zel'dovich curve and/or the EFT-shifted 1-loop result, which are both better behaved at all scales of interest, as can be seen in Fig. \ref{fig:2}.  In \citep{Vlah:2016bcl} another approach to avoid the issues associated with $\sigma^2<0$  was proposed, namely to keep the linear part of the dispersion in the exponent and determinant of (\ref{xiGSM}) and expand out the higher orders.  As we discuss  below, we did not find it to be necessary to adopt this approach in order to get accurate quadrupole predictions for the models we considered. Thus we do not consider this approach in our work. 

\begin{figure}[tbp]
\centering
\includegraphics[width=1.0\textwidth]{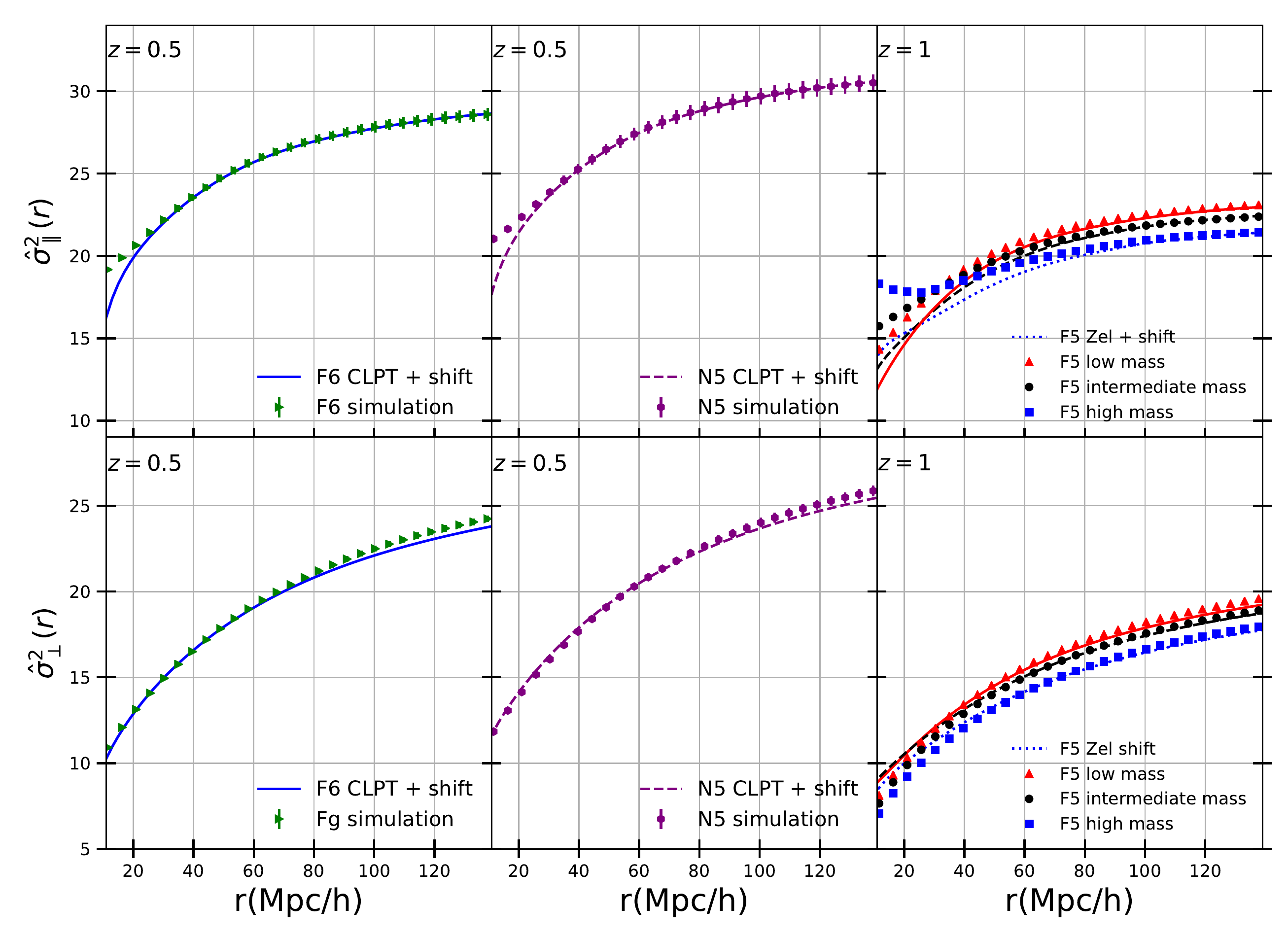}
\caption{\label{fig:3} The pairwise velocity dispersion parallel [top] and perpendicular [bottom] to the pair separation vector for the F6 model at $z=0.5$ [green right triagnles] in the left panel, for the N5 model at $z=0.5$ [purple hexagons] in the middle panel and for the F5 model at $z=1$ in the right panel. In the right panel, the results are shown in the low mass [blue squares], intermediate mass [black circles] and the high mass [red triangles] bins identified in the Group II simulations. For the $z=0.5$ case, the lines represent the 1-loop CLPT prediction from (\ref{sig12mn}) $\&$ (\ref{M2def}) for each model, shifted by a constant, whereas in the $z=1$ case, the lines show the prediction given by the Zel'dovich approximation in each bin, shifted by a constant. The values of the constant shifts are reported in Table~\ref{table1}.}
\end{figure}

In Fig. \ref{fig:3}, we perform the same comparison between theory and simulations, with respect to $\hat{\sigma}^2_{\parallel}$ and $\hat{\sigma}^2_{\perp}$, for the N5 Group I snapshot at $z=0.5$ and also for all 3 mass bins of the F5 Group II case at $z=1$, finding very similar results as in Fig. \ref{fig:2}. We note that the reason that we only show the shifted Zel'dovich results in the F5 Group II case of the right panel, is because it is this choice that will give the best match with the simulations, with respect to the values of the quadrupole of the redshift-space correlation function, as we will see below. In this work we choose a shift constant value to match $\hat{\sigma}^2_{\parallel}$ at large scales. This gives very good agreement for both components, with only a small mismatch between the large-scale trends of the theory and simulation-derived values of $\hat{\sigma}^2_{\perp}$ in Figs \ref{fig:2} $\&$ \ref{fig:3} as was also noted in \citep{doi:10.1111/j.1365-2966.2011.19379.x,Wang:2013hwa,Vlah:2016bcl}. 

In the following analysis of the correlations functions, we use 1-loop CLPT to model the 3 ingredients entering the GSM expression (\ref{xiGSM}), with the pairwise velocity dispersion shifted by a constant to match the simulations at the largest $r$ bin. For the Group I simulations the LPT predicted value is shifted down by a constant. To compare to the Group II simulations, at higher redshift, since the constant shift to the predicted velocity dispersion leads to negative dispersion measurements at small separations, we use the (shifted) Zel'dovich result for the velocity dispersion, together with the 1-loop expressions for $\xi(r)$ and $v_{12}(r)$, as inputs into the GSM expressions. 

\subsection{Halo Redshift-Space 2-point Correlation Function}\label{sec:twoptcf}

\begin{figure}[tbp]
\centering 
\includegraphics[width=.6\textwidth]{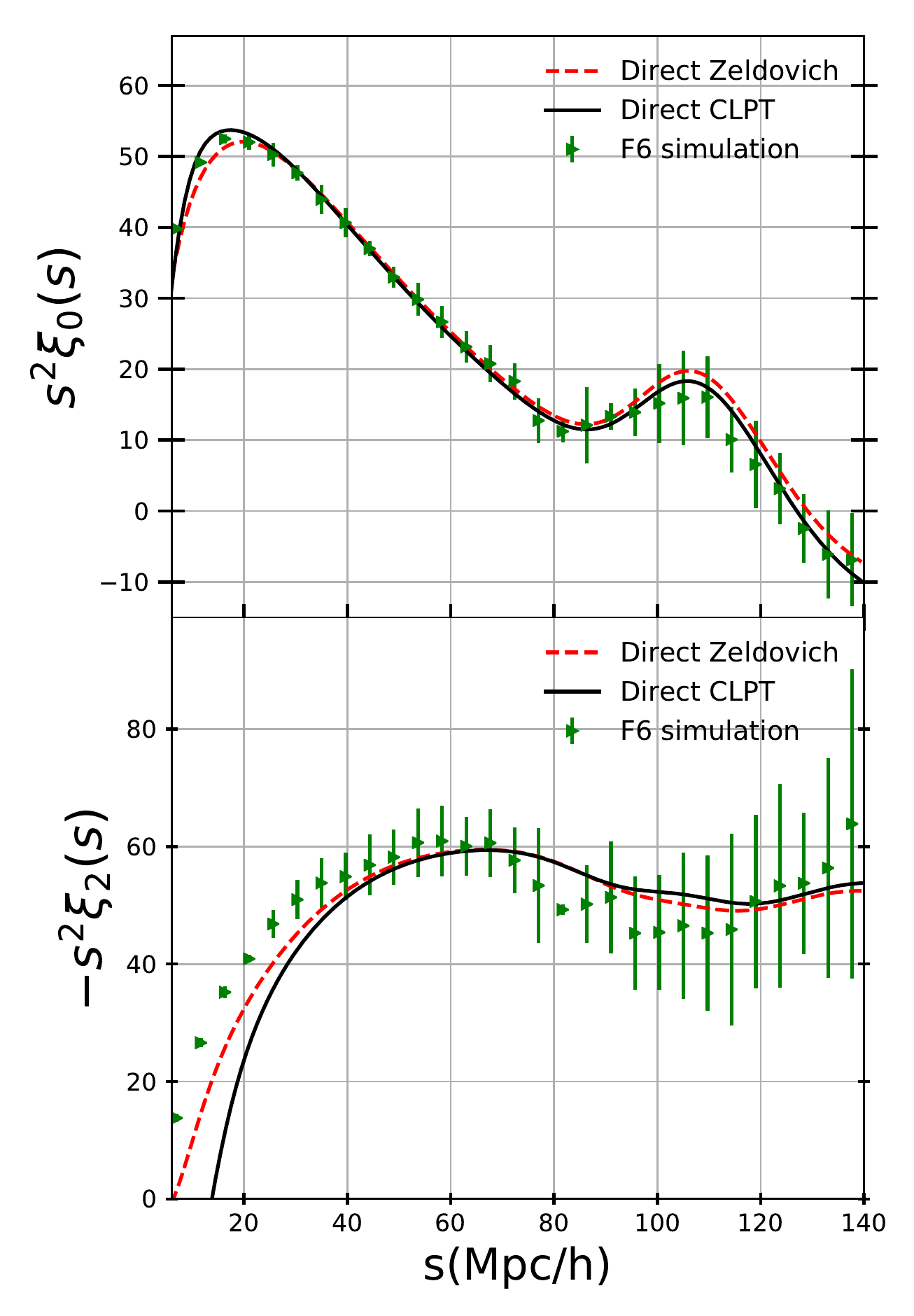}
\caption{\label{fig:4} The monopole [Top] and quadrupole [Bottom] of the redshift-space correlation function for the F6 model at $z=0.5$, as obtained by the Group I simulations [green right triangles] and by the direct Lagrangian approach using 1-loop CLPT [solid black line] and the Zel'dovich approximation [red dashed line]. }
\end{figure}

In this section, we present our predictions for the anisotropic redshift-space 2-point correlation function for halos, as obtained from the various analytical approaches considered and compare how well they capture the results from the MG N-body simulations. Given the directional dependence induced by RSD, the correlation function will now not only depend on the redshift-space separation $s=\sqrt{s_{\perp}^2+s_{\parallel}^2}$ , but also on the cosine $\mu_s = \hat{z}\cdot\hat{s} = \frac{s_{\parallel}}{s}$ (not to be confused with $\mu = \frac{y}{r}$). Following common practice, the 2D anisotropic correlation function can then be expanded in a basis of Legendre polynomials, $P_l(\mu_s)$, as
\begin{equation}
\xi(s,\mu_s)=\sum_l \xi_l(s) P_l(\mu_s),
\end{equation}
where the multiples of order $l$ can then be obtained from
\begin{equation}\label{mult}
\xi_l(s)  = \frac{2 l +1}{2} \int_{-1}^{1} d\mu_s \xi(s,\mu_s) P_l(\mu_s).
\end{equation}
Our comparison will focus on the first three non-vanishing multipoles, the monopole, the quadrupole and the hexadecapole, for which $l=\{0,2,4\}$ and 
$P_l(\mu_s)=\{1,(3\mu_s^2-1)/2,(35 \mu_s^4-30\mu_s^2+3)/8 \}$, respectively.

Having obtained $\xi(s,\mu_s)$, either from an analytical model or the simulations, we integrate (\ref{mult}) to get the multipoles. In the case of the Direct LPT and the GSM approaches, we perform a Gauss-Legendre integration scheme for the angular part, and trapezoidal quadrature for the radial part, whereas to extract this information from the simulations we use Simpson's rule. 

In  section \ref{DirectLPT},  we described the Direct Lagrangian approach to the redshift-space correlation function for halos in MG. 
As is shown in the upper panel of Fig. \ref{fig:4}, we see that this approach, both at the Zel'dovich level and including 1-loop corrections, can capture the monopole reasonably well, 
down to at least scales of $r\sim20$ Mpc/h for the Group I F6 sample at $z=0.5$. When performing the same comparison with respect to the quadrupole, however, 
and as shown in the bottom panel of Fig. \ref{fig:4}, we find that the Direct approach performs very poorly, failing to follow the simulation trend at scales lower than $r<50$ Mpc/h. 
Moreover, we notice that adding loop contributions to the linear, Zel'dovich approximation, does not improve the quadrupole analytical result, instead it performs even worse. 
This counterintuitive outcome is not new in the literature, but has been observed in \cite{White:2014gfa} for GR (see Figs.~2 and 3 of that work), and here shown
to also be the case in MG cosmologies, which motivates pursuing another avenue towards a precise modeling of redshift-space anisotropies, by means of the scale-dependent GSM approach.

\begin{figure}[tbp]
\centering 
\includegraphics[width=1.0\textwidth]{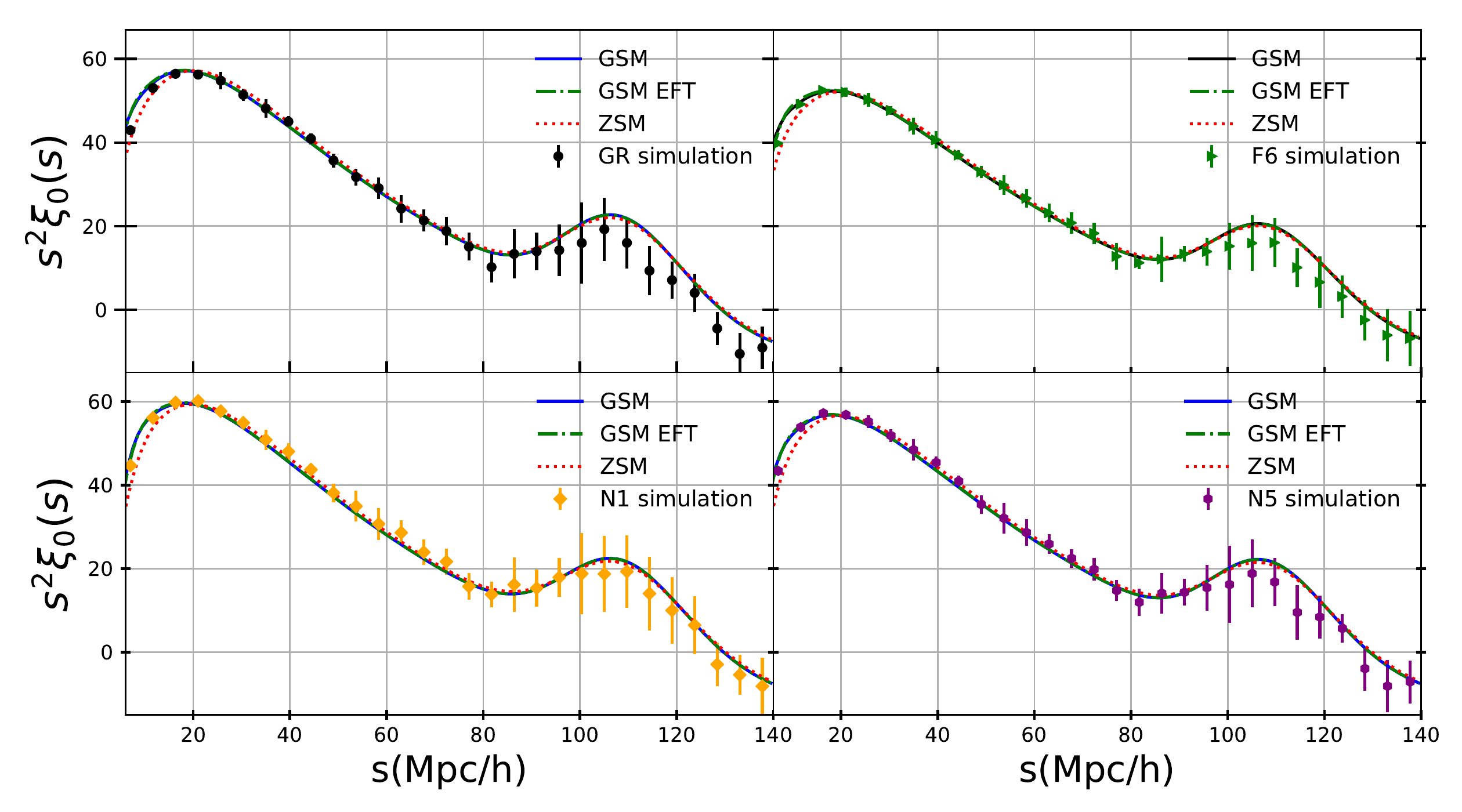}
\caption{\label{fig:5} The monopole of the redshift-space two-point correlation function for GR [black circles] in the top left panel, for the F6 model [green right triangles] in the top right panel, 
for the N1 model [orange diamonds] in bottom left panel and for N5 model [purple hexagons] in the bottom right panel, as obtained from the Group I simulations at $z=0.5$. Furthermore, for each model we plot the theoretical predictions given by the Gaussian Streaming Model (GSM) up to 1-loop order [black solid lines], by the Zel'dovich Streaming Model (ZSM) [magenta dotted lines] and by the GSM with the 1-loop velocity dispersion shifted by the EFT counter-term (\ref{asEFT}) [green dot-dash line].}
\end{figure}

\begin{figure}[tbp]
\centering 
\includegraphics[width=.6\textwidth]{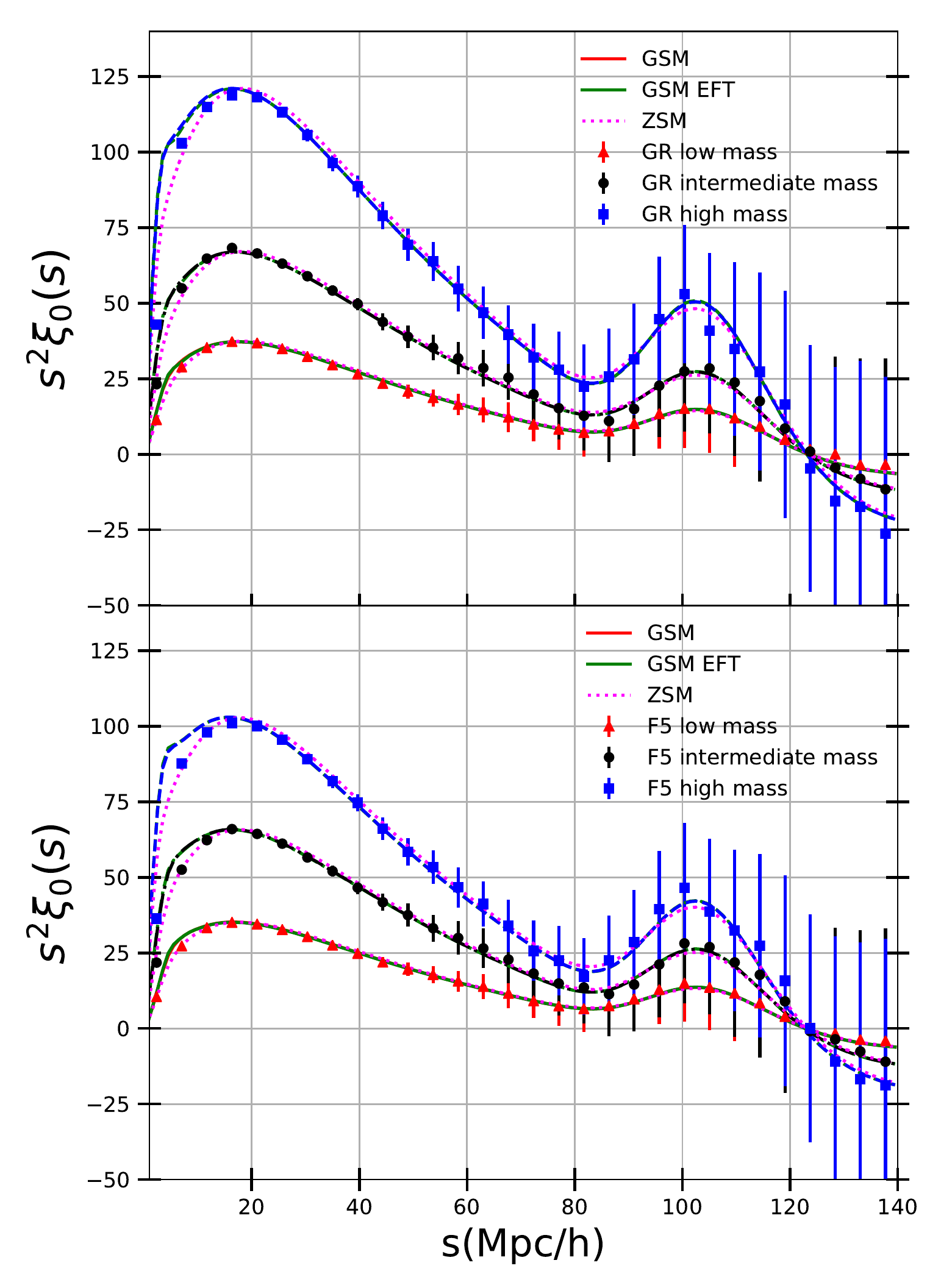}
\caption{\label{fig:6}  The monopole of the redshift-space two-point correlation function for GR [top] and for the F5 model [bottom], as obtained in the low mass [red triangles], the intermediate mass [black circles] and the high mass [blue squares] halo bins identified in the Group II simulations at $z=1$. Furthermore, for each model and mass bin we plot the theoretical predictions given by the Gaussian Streaming Model (GSM) up to 1-loop order with the shifted Zel'dovich dispersion [solid red (low mass), black dot-dash (intermediate mass) and blue dash (high mass) lines], by the Zel'dovich Streaming Model (ZSM) [magenta dotted lines] and by the GSM with the 1-loop velocity dispersion shifted by the EFT counter-term (\ref{asEFT}) [green solid, dot-dash and dash lines]. }
\end{figure}

We proceed to evaluate the performance of the GSM approach in MG, given by (\ref{xiGSM}). For the comparison, the ``GSM'' curves are obtained following a constant shift to the LPT-predicted velocity dispersion for the lower redshift, Group I, simulations, and a shift to the Zel'dovich-predicted velocity dispersion for the higher redshift, Group II sims. When the 1-loop result for the velocity dispersion is shifted by the EFT term (\ref{asEFT}), the theoretical curve is labeled as ``GSM EFT'' in all figures. Finally, we consider the GSM predictions when simply using the Zel'dovich linear (LPT) theory to approximate all 3 ingredients of the model, the Zel'dovich Streaming Model (ZSM) \citep{White:2014gfa}.

In Fig. \ref{fig:5},  the GSM prediction is shown to perform very well, across the spectrum of MG models probed in the Group I simulations, in capturing the monopole of the correlation function, down to scales of $r\sim15$ Mpc/h. This is consistent with findings for the real-space monopole in \citep{Valogiannis:2019xed}. The differences between the results using the EFT shift and the Zel'dovich approximation are rather small, and for larger scales well within the $1-\sigma$ error bars, demonstrating consistency between the monopole predictions from the different approaches. 

The same level of consistency is observed when comparing against the $z=1$ snapshot of the Group II simulations, as apparent in Fig. \ref{fig:6}, this time across all 3 mass halo bins identified in the sample. The GSM result with the shifted Zel'dovich dispersion remains consistent with the simulated monopole for a wide range of scales, including both the BAO region and also the power-law regime, down to  $r\sim20$ Mpc/h. Adding the EFT shift to the 1-loop velocity dispersion causes an almost indistinguishable change to the theoretical prediction, but the Zel'dovich approximation performs considerably better at scales $r<20$ Mpc/h. The latter has also been observed when studying the real-space counterpart in \citep{Valogiannis:2019xed}. 

\begin{figure}[tbp]
\centering 
\includegraphics[width=1.0\textwidth]{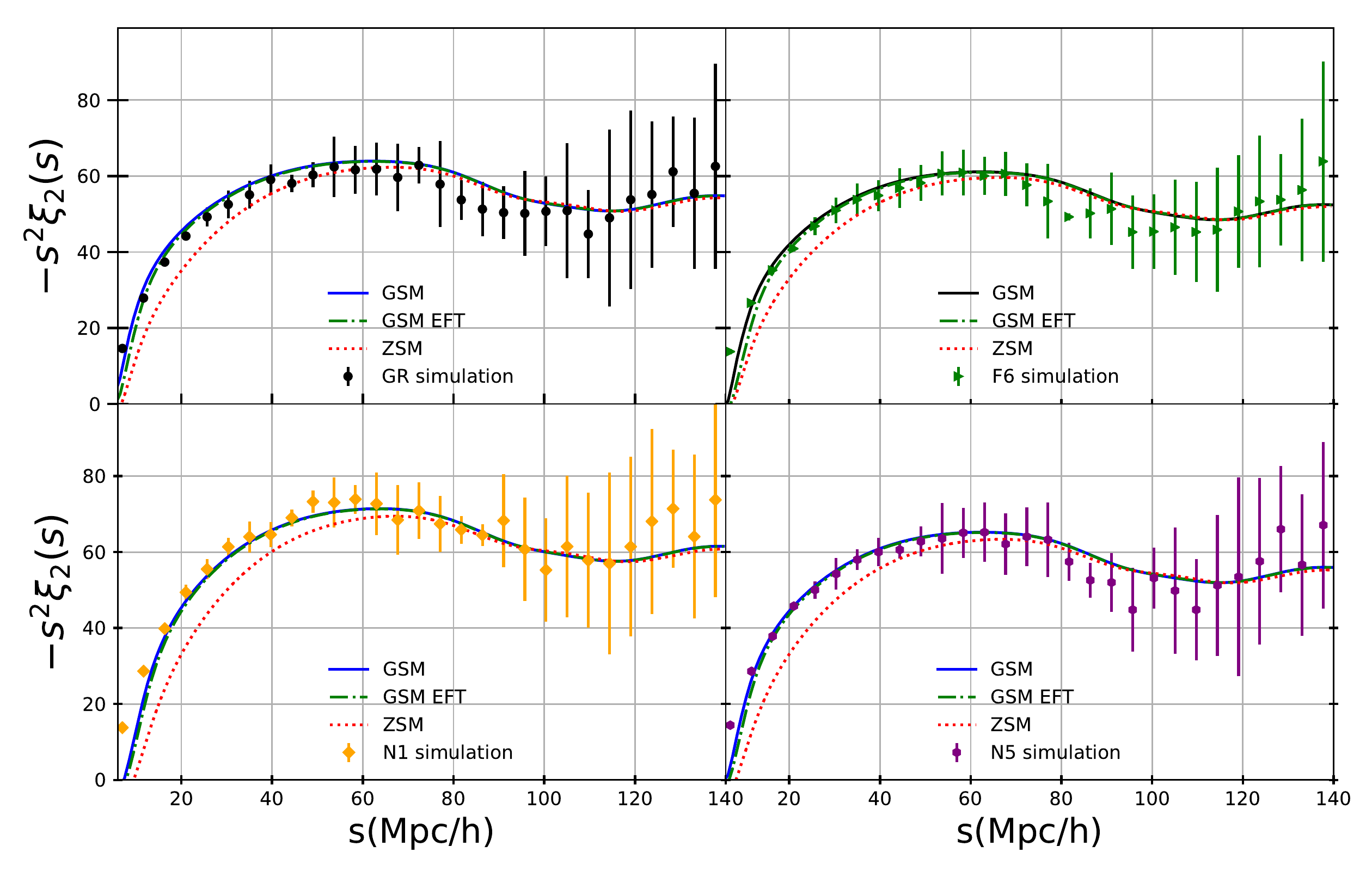}
\caption{\label{fig:7} The quadrupole of the redshift-space two-point correlation function for GR [black circles] in the top left panel, for the F6 model [green right triangles] in the top right panel, for the N1 model [orange diamonds] in bottom left panel and for N5 model [purple hexagons] in the bottom right panel, as obtained from the Group I simulations at $z=0.5$. Furthermore, for each model we plot the theoretical predictions given by the Gaussian Streaming Model (GSM) up to 1-loop order [black solid lines], by the Zel'dovich Streaming Model (ZSM) [magenta dotted lines] and by the GSM with the 1-loop velocity dispersion shifted by the EFT counter-term (\ref{asEFT}) [green dot-dash line]. }
\end{figure}

\begin{figure}[tbp]
\centering 
\includegraphics[width=.6\textwidth]{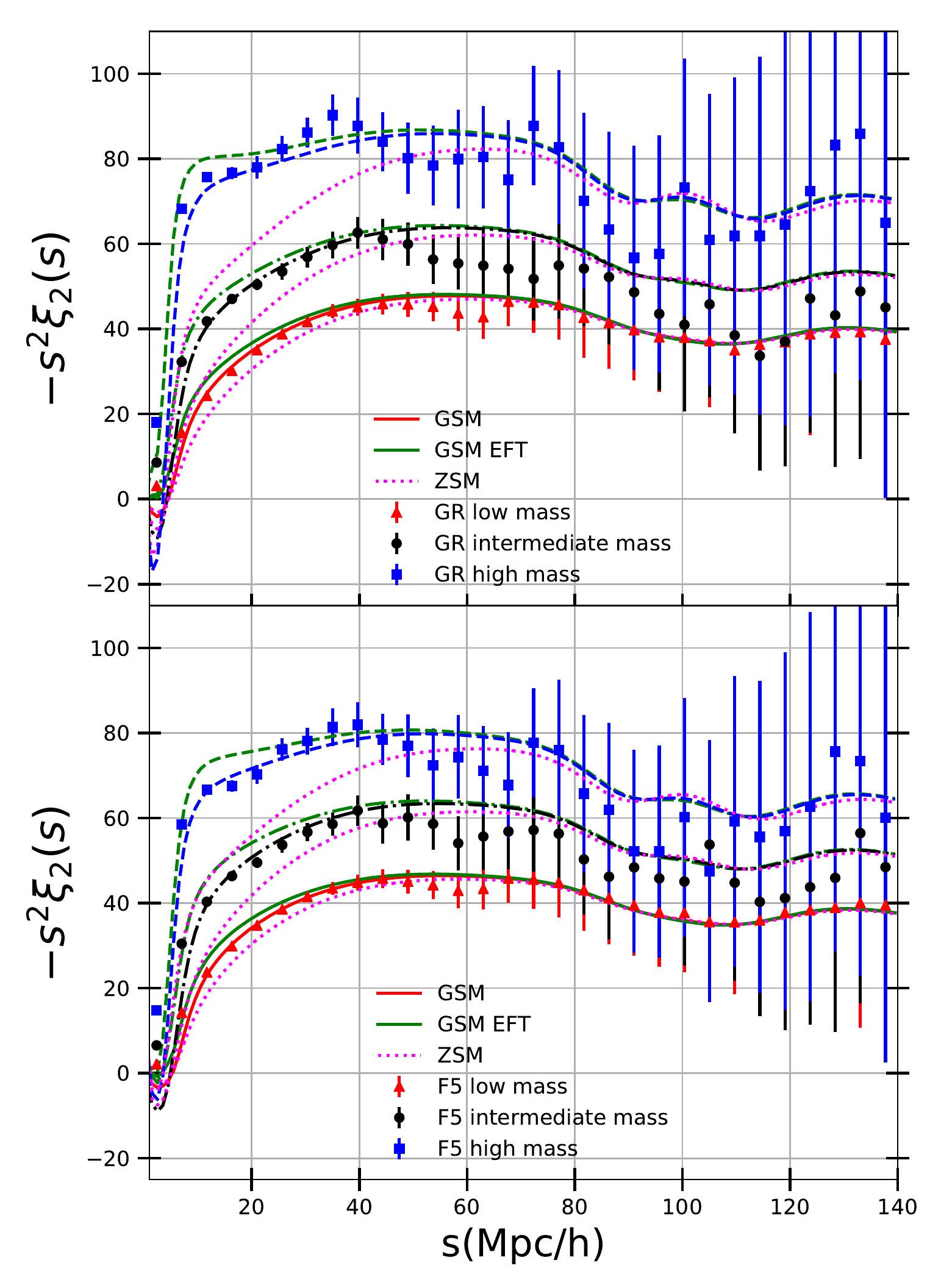}
\caption{\label{fig:8} The quadrupole of the redshift-space two-point correlation function for GR [top] and for the F5 model [bottom], as obtained in the low mass [red triangles], the intermediate mass [black circles] and the high mass [blue squares] bins identified in the Group II simulations at $z=1$. Furthermore, for each model and mass bin we plot the theoretical predictions given by the Gaussian Streaming Model (GSM) up to 1-loop order with the shifted Zel'dovich dispersion [solid red (low mass), black dot-dash (intermediate mass) and blue dash (high mass) lines], by the Zel'dovich Streaming Model (ZSM) [magenta dotted lines] and by the GSM with the 1-loop velocity dispersion shifted by the EFT counter-term (\ref{asEFT}) [green solid, dot-dash and dash lines].}
\end{figure}

\begin{figure}[tbp]
\centering 
\includegraphics[width=.9\textwidth]{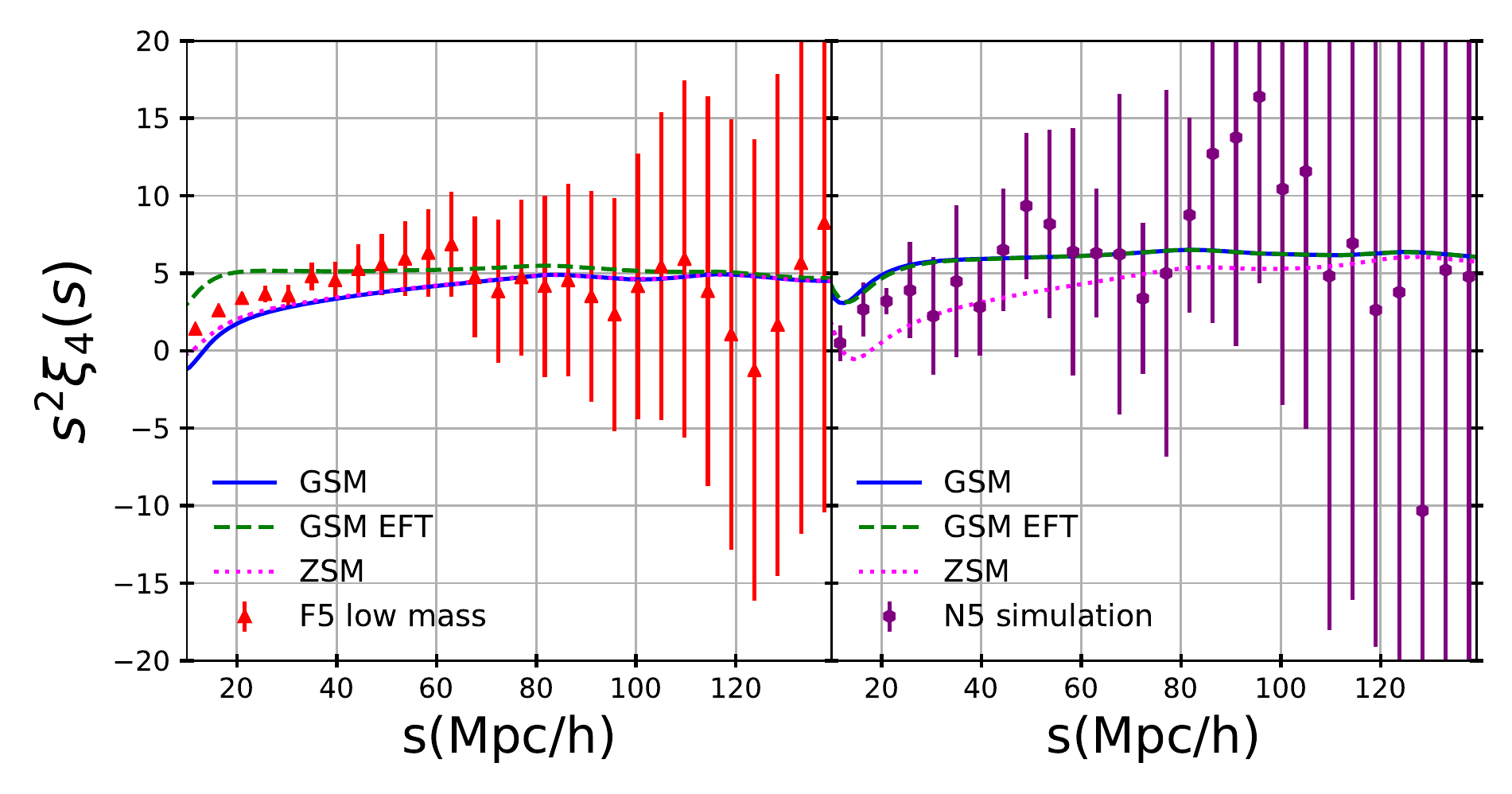}
\caption{\label{fig:9} The hexadecapole of the redshift-space two-point correlation function as obtained in the F5 low mass bin identified in the Group II simulations [left], 
as well as in the N5 model of the Group I snapshot [right]. Furthermore, for each case we plot the theoretical predictions given by the Gaussian Streaming Model (GSM) up to 1-loop order [solid blue line], 
by the Zel'dovich Streaming Model (ZSM) [magenta dotted lines] and by the GSM with the 1-loop velocity dispersion shifted by the EFT counter-term (\ref{asEFT}) [green dashed]. }
\end{figure}

Moving on to the redshift-space quadrupole and starting with the Group I simulations, as shown in Fig. \ref{fig:7}, we determine that the GSM achieves a significant improvement compared to the Direct Lagrangian approach of the previous section, with the theoretical prediction remaining consistent with the N-body simulations, down to scales of at least $r\sim17$ Mpc/h, for all cases. Adding the EFT shift to the velocity dispersion further improves the accuracy at small scales, with the difference being practically indistinguishable at scales $r>40$ Mpc/h. The ZSM result, however, performs much more poorly in this case,  for all models, demonstrating the need to include the 1-loop corrections for an accurate prediction of the quadrupole.

 As in the monopole case, the same level of consistency is observed when comparing against the Group II simulations, which is done in Fig. \ref{fig:8}, where we see that the 1-loop GSM result with the shifted Zel'dovich dispersion accurately captures the simulated quadrupole, for all mass bins and both in GR and the F5 MG model. Even for the high mass bin, that contains much less halos, and is inevitably noisier, the trend of the simulation data points is clearly traced by our GSM prediction. The GSM prediction obtained using the EFT shift to the velocity dispersion seems to perform noticeably worse, in this snapshot, for all mass bins. Just like in the Group I case of Fig. \ref{fig:7}, the ZSM seems to be inadequate at accurately capturing the quadrupole at quasi-linear scales. 
 
Finally, we compare our analytical predictions with respect to the hexadecapole of the anisotropic correlation function for two cases: the F5 low mass bin in the Group II simulations, as well as the N5 model in the Group I snapshot, both of which are shown in Fig. \ref{fig:9}. While the noise increases significantly between the quadrupole and the hexadecapole, so we should take these results only as indicative, we find that the GSM prediction traces the simulated hexadecapole well  down to scales of at least $r\sim17$ Mpc/h, but with an offset of a few percent. This offset is for theoretical predictions with bias values obtained with the PBS formalism; an alternative might be to allow biases to vary and fit to the data. 

These results overall demonstrate that the GSM can serve as an invaluable tool to model the anisotropic correlation function of halos in MG models, extending the success of this approach beyond the previously explored domain of GR cosmologies. It is worth emphasizing, also, that since the Lagrangian bias factors are calculated through our analytical model, the only free parameter needed to get accurate quadrupole predictions has been the constant offset added to the components of the pairwise velocity dispersion of halos, a value that can be easily determined through a single large-scale measurement of $\sigma^2$ (either from N-body simulations or observations).

\section{Conclusions}\label{Conclusions}

In this work, we expanded the Gaussian Streaming Model to predict the redshift-space anisotropic correlation function for biased tracers in Modified Gravity models. This is the first time, to our knowledge, that the effects of both redshift-space distortions and halo bias have been jointly studied analytically for scenarios that go beyond GR. 

We build upon our previous work on the study of biased tracers in MG using LPT \citep{Aviles:2018saf,Valogiannis:2019xed}, and employ the Convolution LPT resummation scheme, with a local Lagrangian bias, to analytically describe the necessary missing ingredients that enter the GSM for such models: the real-space halo pairwise velocity and its scale-dependent dispersion. The approach includes analytic determination of the bias parameters using the PBS formalism with fitted Sheth-Tormen parameters.  Through cross-checking our analytical predictions across a large suite of state-of-the-art N-body simulations for the f(R) Hu-Sawicki and the nDGP MG models, we find that the 1-loop CLPT prediction for the real-space halo pairwise velocity remains consistent with the simulated data for the scales of interest, a fact that is crucial for the accuracy of the GSM prediction. 


When performing the same comparison with respect to the halo pairwise velocity dispersion, however,  we find that the CLPT approach is able to match the simulated curve only if an offset is 
added to the theoretical dispersion result. In doing this, we have followed two different strategies; namely, we added a phenomenological constant shift to it, as proposed in \cite{doi:10.1111/j.1365-2966.2011.19379.x}, and 
included a leading order EFT effect that contributes as a scale dependent shift, tending to a constant large scales \cite{Vlah:2016bcl}. We have seen that although theoretically well-motivated, the EFT prescription does not
necessarily work better than the constant shift, which is evident in the quadrupole of the Group II of simulations (Fig.~\ref{fig:8}). The pairwise dispersion is nonlinear in nature, as has been settled down since the seminal work of Scoccimarro \cite{PhysRevD.70.083007}, 
such that it would be not surprising that higher order contributions were important, which also could be the reason of why their parallel and perpendicular to the line of sight 
components require slightly different offsets to match simulations, as it is done in \cite{doi:10.1111/j.1365-2966.2011.19379.x,Wang:2013hwa}.

In establishing that the CLPT approach can accurately predict the ingredients that enter the GSM expression, we proceed to evaluate the performance of the model against the N-body simulations, with respect to the redshift-space monopole, the quadrupole and the hexadecapole of the anisotropic correlation function of halos.  Unlike direct Lagrangian approaches, which prove to be significantly inaccurate, we find that the 1-loop GSM approach can successfully capture the redshift-space quadrupole for all MG models, remaining consistent with the corresponding results from the simulations down to scales of at least $r\sim17$ Mpc/h and including the BAO peak. In the hexadecapole case, the GSM prediction traces the shape of simulated result down to equally small scales, but with a few percent offset. 

Our analysis overall shows that our theoretical analytical predictions using the GSM implementation achieve strong agreement with the non-linear simulated data for a variety of MG models, across different levels of screening and different screening mechanisms, and across a wide range of halo masses. We emphasize the fact that this agreement occurs across all scales down into the mildly non-linear regime, through simply matching a single shift parameter in the  halo pairwise velocity dispersion, that can be determined through comparing the theory with a large-scale (linear regime) observed measurement. The predictions come with bias parameters determined by the PBS formalism but an alternative is to consider the bias factors as free parameters, that could be fit with the simulations. The approach is one that has great potential for making accurate clustering predictions for upcoming spectroscopic large scale structure surveys. 

In this work, we have followed the commonly used approach in GSM-PT that fits a constant or EFT shift to the pairwise velocity dispersion obtained from the simulations. 
On the other hand, in applications of the GSM to surveys, a constant shift is added directly into eq. (\ref{xiGSM}) by means of a substitution 
$\sigma^2_{12}(r) \longrightarrow \sigma^2_{12}(r) + \sigma^2_\text{FoG}$, and with $\sigma^2_\text{FoG}$ considered a free parameter to be fitted 
with observations; see e.g., \cite{2012MNRAS.426.2719R,Icaza-Lizaola:2019zgk}. In order to utilize the GSM approach for parameter inferences on real data, these works treated the offset as a nuisance parameter with a broad prior of $0-40$ applied. We also find, in agreement with these results in the context in GR, that adding large offsets is an inherent part of ensuring a good fit with the data. Understanding the connection between the offset and the underlying nonlinear clustering in MG, as well as the impact of marginalizing over offsets in parameter estimation are areas of definite interest in future work.

The framework  developed  is  flexible and could be easily applied to any scalar-tensor theory of interest. Future lines of improvement include exploring how more general bias schemes, like for example including tidal bias terms, in conjunction with corrections from EFT \citep{Vlah:2016bcl} can serve to further improve the accuracy of the analytical model. It would be also very interesting to see how including higher-order cumulants to the GSM expansion, as in \citep{Uhlemann:2015hqa}, could yield additional predictive power. Last but not least, our perturbative treatment can be used to disentangle modifications to gravity with estimators beyond the standard 2-point statistics, such as the marked correlation function \citep{White:2016yhs,Valogiannis:2017yxm,Hernandez-Aguayo:2018yrp,Armijo:2018urs,Satpathy:2019nvo} and higher order statistics. We plan to address these natural extensions in future work. 

In this era of precision cosmology, the next-stage cosmological surveys, such as DESI, EUCLID and the LSST, will thoroughly explore the LSS of the universe providing an opportunity to shed light on the dark sector. This highlights the need to compliment such observational endeavors with efficient analytical approaches to characterize the distinctive signatures of theoretical models of the dark sector that the observations can constrain. In this work we have shown that the GSM, previously only explored for GR-based cosmologies, can also serve as a valuable predictive tool to probe cosmic modifications to gravity in redshift-space as well as real-space, to explore modified gravity cosmologies with biased tracers observed through spectroscopic and photometric surveys. 

\acknowledgments
We wish to thank Baojiu Li for kindly providing the \verb|ELEPHANT| simulations, on behalf of \citep{Cautun:2017tkc} and for numerous discussions on available simulations. We would also like to thank Wojciech Hellwing for kindly making available the $n$DGP simulations, on behalf of \citep{Hellwing:2017pmj}, as well as Christian Arnold, together with the other authors of \citep{Arnold:2018nmv}, for kindly providing their Lightcone simulations. We are also grateful to Martin White for providing valuable insights on perturbation theory and available simulations and also to Uros Seljak, Emanuele Castorina and Steven Chen for useful discussions. Last but not least, we want to thank David Alonso for useful discussions on calculating correlation functions with \verb|CUTE|. The work of Georgios Valogiannis and Rachel Bean is supported by DoE grant DE-SC0011838, NASA ATP grants NNX14AH53G and 80NSSC18K0695,  NASA ROSES grant 12-EUCLID12-0004 and funding related to the WFIRST Science Investigation Team. Alejandro Aviles acknowledges partial support from Conacyt Fronteras Grant No. 281 and Conacyt Grant No. 283151.

\appendix
\section*{Appendix}
\section{GSM functions in MG}\label{AppendixGSM}

In this appendix, we give expressions for the necessary ingredients to construct the Gaussian streaming model in LPT. That is, we find 1D integral expressions for (\ref{dotfuncs}) and (\ref{ddotfuncs})
in cosmologies beyond $\Lambda$CDM. Before displaying all these equations, as an example we consider the ``$q$-function'' $\dot{U}_i(\vq)$:
\begin{align}
\dot{U}_i(\vq) = \frac{1}{f_0 H} \langle \delta_L(\vq_1) \dot{\Delta}_i\rangle = \int \frac{d^3k_1d^3k_2}{ (2\pi)^6}  e^{i\vk_1\cdot \vq_1} \big(e^{i\vk_2 \cdot\vq_2}-e^{i\vk_2 \cdot \vq_1}\big) \langle \delta_L(\vk_1) \dot{\Psi}_i(\vk_2)\rangle_c . 
\end{align}
Up to 1-loop, $ \langle \delta_L(\vk_1) \dot{\Psi}_i(\vk_2)\rangle_c \simeq H f(k_2) \langle \delta_L(\vk_1) \Psi^{(1)}_i(\vk_2)\rangle + 3 H f(k_2) \langle \delta_L(\vk_1) \Psi^{(3)}_i(\vk_2)\rangle$, where for illustrative
purposes we approximate $\dot{\Psi}^{(n)}(\vk) = n H f(\vk) \Psi^{(n)}(\vk)$. After some straightforward manipulations one obtains (see, e.g., \cite{2013MNRAS.429.1674C})
\begin{equation}
 \dot{U}_i(\vq) =-  \hat{q}_i \int \frac{dk}{2\pi^2} k \frac{f(k)}{f_0} \left[ P_L(k)+ \frac{7}{5} R_1(k) \right] j_1(kq)
\end{equation}
where the ``$k$-function'' $R_1$  is \cite{Matsubara:2007wj}
\begin{equation}
R_1(k) =  \frac{21}{10}P_L(k) \int \Dk{p} \vk \cdot \ve L^{(3)}(\vk,-\vp,\vp) P_L(p)
\end{equation}
with $\ve L^{(3)}$ the third order LPT kernel. 
In the rest of this appendix we will find all the functions given by  (\ref{dotfuncs}) and (\ref{ddotfuncs}) $\it{without}$ the use of the above approximation, which is not as accurate in MG models, as it is in $\Lambda$CDM.

\subsection{LPT kernels} \label{app:GSM_LPTK}

The Lagrangian displacement field is formally expanded as $ \Psi_i(\vq,t) = \sum_{n=0}^{\infty} \Psi^{(n)}_i(\vq).$ In Fourier space, the $n$th order Lagrangian displacement is the weighted convolution of $n$ linear density fields,
\begin{equation}
 \Psi^{(n)}_i(\vk,t) = \frac{i}{n!} \int\frac{d^3k_1\cdots d^3k_n}{(2\pi)^{3(n-1)}} 
 \dD(\vk-\vk_{1\cdots n})L^{(n)}_{i}(\vk_1,\dots,\vk_n;t) \delta(\vk_1,t)\cdots\delta(\vk_n,t).
\end{equation}
The first order kernel is $\mathbf{L}^{(1)}(\vk)=\vk/k^2$, as can be seen from (\ref{eq:ZeldispsolMG}). Higher perturbative orders are obtained
by solving iteratively (\ref{eq:modpoisson}) \cite{Aviles:2017aor}. For  $\mathbf{L}^{(2)}$ we have 
\begin{equation} \label{L2LPTK}
 L^{(2)}_i(\vk_1,\vk_2,t) = \frac{3}{7} \frac{(\vk_1+\vk_2)_i}{|\vk_1+\vk_2|^2}
 \left[ \mA(\vk_1,\vk_2,t) - \mB(\vk_1,\vk_2,t) \frac{(\vk_1\cdot\vk_2)^2}{k_1^2k_2^2} \right]
\end{equation}
where the $\mA$ and $\mB$ functions are obtained by solving second order differential equations, and are given in (2.19) of \cite{Aviles:2018saf}; see also \cite{Aviles:2017aor}.
In $\Lambda$CDM we have that functions $\mA$ and $\mB$ are equal and only time-dependent, while for EdS, or more generally for $\Lambda$CDM models
with $f^2=\Omega_m$, these functions are exactly 1, and hence the kernels become time independent. Since $\mA|_{\Lambda\text{CDM}}(z=0) \simeq 1.01$, this
is usually approximated as 1, in what is called the static kernels approximation, yielding subpercent errors at quasilinear scales. In MG, the departure from $\mA,\mB \simeq 1$ is larger than the percent level; moreover, the $\mA$ function carries the non-linear terms responsible for screening mechanisms, such that the use of the static kernels approximation in MG misses this important property of MG. By isotropy we can write $\mA,\mB(k_1,k_2,x=\hat{\vk}_1\cdot\hat{\vk}_2)$ and perform a Legendre expansion on the angle $x$. Such procedure shows that the monopole of these functions is the dominant term, other multipoles give smaller contributions, and hence the decomposition in (\ref{L2LPTK}) is not as arbitrary as it may look at first sight. The case of DGP, is even simpler, because $\mB$ is only time dependent, and $\mA$ has only a monopole and a small quadrupole, the latter given exclusively by the screening non-linear terms in the Klein-Gordon equation \cite{Aviles:2018qot}. 

The expression for the third order kernel $L^{(3)}_i$ is quite large and we do not reproduce it here, but we refer the reader to ref.~\cite{Aviles:2017aor}. All these functions are time-dependent, and for compactness we will omit to write it explicitly in the following. 

In LPT-RSD models one finds the time derivative of the Lagrangian displacement field. 
As was discussed in the main text, in $\Lambda$CDM, a good approximation
is given by $\dot{\Psi}^{(n)}(\vq) \simeq n H f \Psi(\vq)$, which is no longer possible in MG since the linear growth function $D^{(1)}$ is scale-dependent, as was discussed in the main text. A general expression is
\begin{equation} \label{dotPsi}
 \dot{\Psi}^{(n)}_i(\vk) = \frac{i}{n!} n f_0 H\int \frac{d^3k_1\cdots d^3k_n}{(2 \pi)^{3(n-1)}} \delta_\text{D}(\vk-\vk_{1\cdots n})
 L^{f(n)}_i(\vk_1,\cdots,\vk_n) \delta(\vk_1) \cdots \delta(\vk_n),
\end{equation}
with $\vk_{1\cdots n}=\vk_1+\cdots + \vk_n$, and kernels
\begin{align} \label{defGf}
  L^{f(n)}_i(\vk_1,\cdots,\vk_n) &= \frac{f(k_1)+\cdots + f(k_n)}{n f_0} L^{(n)}_i(\vk_1,\cdots,\vk_n) 
              + \frac{1}{n f_0} L^{'(n)}_i(\vk_1,\cdots,\vk_n) \nonumber\\
     &\equiv  \frac{\vk_{1\cdots n}^i}{k_{1\cdots n}^2}C_n \Gamma_n^f(\vk_1,\cdots,\vk_n), 
\end{align}
where we remind that $f(k,t)=d \ln D^{(1)}(k,t)/d \ln a$ is 
the growth factor at a scale $k$, and $f_0\equiv f(k=0,t)$ is the large-scale growth factor, usually coinciding with that of $\Lambda$CDM. The last line in the above equation serves to define the scalar kernels $\Gamma_n^f$ and a set of numbers $C_n$; for convenience we choose $C_1=1$, $C_2=3/7$. We can use a weaker version of the static approximation and neglect the second term in the second equality of (\ref{defGf}); however, we find that $\ve L'$ is about the same order as the corrections introduced by MG to the EdS kernels $\ve L|_\text{EdS}$. Hence, to be consistent we have to keep both terms in that equation. 

Analogously as we introduced $\Gamma^f_n$ functions, for the Lagrangian displacements we define 
\begin{equation}\label{defGk}
C_n\Gamma_n(\vk_1,\dots,\vk_n) = k^i_{1\cdots n}L_i^{(n)}(\vk_1,\dots,\vk_n),
\end{equation}
and hence these $\Gamma_n$ functions are the kernels of the longitudinal component of the Lagrangian displacement,
 $\vk\cdot\Psi^{(n)}(\vk)$. Since for 1-loop, 2-point statistics, the transverse components project out, one can use the scalar $\Gamma$ instead of vector $\ve L$ kernels without loss of generality. 

The first order scalar kernels are $\Gamma_1(\vk)=1$ and $\Gamma_1^f(\vk)=f(k)/f_0$. To second order 
\begin{align} \label{Gamma2}
\Gamma_2[\vp_1,\vp_2] &= \left[\mA(\vp_1,\vp_2) - \mB(\vp_1,\vp_2) \frac{(\vp_1 \cdot \vp_2)^2}{p_1^2 p_2^2}\right]
                       =  \frac{7}{3}\frac{ D^{(2)}(\vp_1,\vp_2) }{ D_+(p_1)D_+(p_2) }, \\     
\Gamma^f_2[\vp_1,\vp_2] 
     &= \Gamma_2[\vp_1,\vp_2] \frac{f(p_1) + f(p_2)}{2 f_0} 
     +  \frac{1}{2f_0 H_0}\left[\dot{\mA}(\vp_1,\vp_2) - \dot{\mB}(\vp_1,\vp_2) \frac{(\vp_1 \cdot \vp_2)^2}{p_1^2 p_2^2}\right], \nonumber\\
      &=  \frac{1}{2f_0}\frac{7}{3}\frac{   \frac{d\,}{d\ln a} D^{(2)}(\vp_1,\vp_2) }{ D_+(p_1)D_+(p_2) },
\end{align}
where $\mA,\mB=\mA,\mB(\vp_1,\vp_2)$, and $D^{(2)}$ is the second order growth function defined in \cite{Aviles:2017aor}. The third order kernels are
\begin{align}
C_3 \Gamma_3[\vp_1,\vp_2,\vp_3] &= \frac{D_+^{(3)s}(\vp_1,\vp_2,\vp_3)}{D_+(\vp_1)D_+(\vp_2)D_+(\vp_3)},  \\
C_3 \Gamma^f_3[\vp_1,\vp_2,\vp_3] &= \frac{1}{3f_0} \frac{\frac{d\,}{d \ln a} D_+^{(3)s}(\vp_1,\vp_2,\vp_3)}{D_+(\vp_1)D_+(\vp_2)D_+(\vp_3)}.
\end{align}
with the (symmetric) third order growth function $D_+^{(3)s}$ as given in \cite{Aviles:2017aor}. Actually, we will not use the value of $C_3$ at all, so 
we can let it free. But by defining $C_2=3/7$ we make the notation simpler in the following sections.

We notice that the approximation of static kernels, usually taken in  $\Lambda$CDM and exact for EdS, corresponds to
\begin{equation}
 \Gamma_n^f \simeq  \Gamma_n,  \qquad \text{($\Lambda$CDM)},
\end{equation}
and therefore, the functions presented in the
rest of this appendix can be recasted in their well-known, $\Lambda$CDM counterparts by omitting the ``$f$'' labels.

\subsection{k functions} \label{app:GSM_q}

The $Q_n(k)$ and $R_n(k)$ scalar functions, introduced first in \cite{Matsubara:2007wj,Matsubara:2008wx}, are the building blocks of LPT statistics. These are constructed out of
$N$-point functions of linear density fields, Lagrangian displacements and their derivatives, as for example $\langle \delta_L(\vk_1) \Psi^{(n)}(\vk_2) \dot{\Psi}^{(m)}(\vk_3)\rangle$, with $(n)$ and $(m)$ denoting perturbative orders; see \cite{Matsubara:2007wj,Matsubara:2008wx}. We do not write here the definitions of these polyspectra, but we refer the reader
to the above referenced works and to \cite{Valogiannis:2019xed,Aviles:2018saf} for MG. Here we extend those definitions to include time derivatives of Lagrangian displacements.

The only $k$-functions involving third order Lagrangian displacements are
\begin{align}
R_1(k)   &\equiv \int \Dk{p}  \frac{21}{10} C_3\Gamma_3(\vk,-\vp,\vp)  P_L(p)P_L(k), \label{R1} \\
R_1^f(k) &\equiv \int \Dk{p}  \frac{21}{10} C_3\Gamma_3^f(\vk,-\vp,\vp)   P_L(p)P_L(k). \label{R1f}
\end{align}
Expressions involving just one time derivative, and that it is operating on one Lagrangian displacement to second order are denoted with a label ``$f$'':
\begin{align}
Q_1^f(k) &\equiv \int \Dk{p} \Gamma_2[\vp,\vk-\vp]\Gamma_2^f[\vp,\vk-\vp] P_L(|\vk-\vp|)P_L(p) \label{Q1f}\\
Q_{2}^f(k) &= \int \Dk{p} \frac{(\vk\cdot \vp)\vk \cdot (\vk-\vp)}{p^2 |\vk-\vp|^2} \Gamma^f_2[\vp,\vk - \vp] P_L(|\vk-\vp|) P_L(p),\\
Q_5^f(k) &= \int \Dk{p} \frac{\vk \cdot \vp}{p^2}\Gamma^f_2[\vp,\vk - \vp] P_L(|\vk-\vp|) P_L(p),\\
Q_8^f(k) &= \int \Dk{p} \Gamma^f_2[\vp,\vk - \vp] P_L(|\vk-\vp|) P_L(p),\\
Q_{I}^f(k) &= \int \Dk{p} \frac{((\vk\cdot \vp)\vk - k^2 \vp) \cdot (\vk-\vp)}{p^2 |\vk-\vp|^2} \Gamma^f_2[\vp,\vk - \vp] P_L(|\vk-\vp|) P_L(p),\\
R_{2}^f(k) &= \int \Dk{p} \frac{\vk \cdot \vp \, \vk \cdot(\vk-\vp)}{p^2|\vk-\vp|^2} \Gamma^f_2[\vk,-\vp]  P_L(k) P_L(p), \\
R_{1+2}^f(k) &= \int \Dk{p} \frac{\vk \cdot(\vk-\vp)}{|\vk-\vp|^2} \Gamma^f_2[\vp,\vk]  P_L(k) P_L(p),\\
R_{I}^f(k) &= \int \Dk{p} \frac{((\vk\cdot \vp)\vk - k^2 \vp) \cdot (\vk-\vp)}{p^2 |\vk-\vp|^2} \Gamma^f_2[\vk,-\vp]  P_L(k) P_L(p),
\end{align}
The usual, ``undotted'', $Q$ and $R$ functions are obtained by replacing $\Gamma_2^f$ by $\Gamma_2$ in the above equations.
Functions $R_I$, $Q_I$ and $R_{1+2}$ are equal to $R_1$, $Q_1$ and $R_1+R_2$ respectively for EdS kernels. In ref.\cite{Valogiannis:2019xed} these are named
as $[R_1]_\text{MG}$, $[Q_1]_\text{MG}$ and $[R_1+ R_2]_\text{MG}$.

Now, the ``$f*$'' $k$-functions contain one derivative in a linear field, and no derivatives in the other fields. These are given by 
\begin{align}
Q_{I}^{f*}(k) &= \int \Dk{p} \frac{((\vk\cdot \vp)\vk - k^2 \vp) \cdot (\vk-\vp)}{p^2 |\vk-\vp|^2} \Gamma_2[\vp,\vk - \vp] \frac{f(p)}{f_0}P_L(|\vk-\vp|) P_L(p),\\
Q_{2}^{f*}(k) &= \int \Dk{p} \frac{(\vk\cdot \vp)\vk \cdot (\vk-\vp)}{p^2 |\vk-\vp|^2} \Gamma_2[\vp,\vk - \vp]  \frac{f(p)}{f_0} P_L(|\vk-\vp|) P_L(p),\\
Q_5^{f*}(k) &= \int \Dk{p} \frac{\vk \cdot \vp}{p^2}\Gamma_2[\vp,\vk - \vp] \frac{f(p)}{f_0} P_L(|\vk-\vp|) P_L(p),\\
R_{2}^{f*}(k) &= \int \Dk{p} \frac{\vk \cdot \vp \, \vk \cdot(\vk-\vp)}{p^2|\vk-\vp|^2} \Gamma_2[\vk,-\vp]  \frac{f(p)}{f_0} P_L(k) P_L(p), \\
R_{1+2}^{f*}(k) &= \int \Dk{p} \frac{\vk \cdot(\vk-\vp)}{|\vk-\vp|^2} \Gamma_2[\vk,-\vp]  \frac{f(p)}{f_0} P_L(k) P_L(p),\\
R_{I}^{f*}(k) &= \int \Dk{p} \frac{((\vk\cdot \vp)\vk - k^2 \vp) \cdot (\vk-\vp)}{p^2 |\vk-\vp|^2} \Gamma_2[\vk,-\vp] \frac{f(p)}{f_0}  P_L(k) P_L(p).
\end{align}
While the ``$ff$'' $k$-functions ---that contain two time derivatives, one in a linear field and the other in a  second order field--- are obtained by replacing
$\Gamma_2 \rightarrow \Gamma_2^f$ in the above equations. 
The exception to this rule is $Q_1$, where a label ``$ff$'' denotes that the two second order Lagrangian displacements are differentiated,
\begin{equation}
Q_1^{ff}(k) = \int \Dk{p} \Gamma^f[\vp,\vk - \vp] \Gamma^f[\vp,\vk - \vp] P_L(|\vk-\vp|) P_L(p).
\end{equation}
Also, there are functions that contain two derivatives, and both in linear displacement fields, denoted with an ``$f*f*$'' label, these are
\begin{align}
Q_{I}^{f*f*}(k) &= \int \Dk{p} \frac{((\vk\cdot \vp)\vk - k^2 \vp) \cdot (\vk-\vp)}{p^2 |\vk-\vp|^2} \Gamma_2[\vp,\vk - \vp] 
\frac{f(|\vk-\vp|)}{f_0}\frac{f(p)}{f_0} P_L(|\vk-\vp|)  P_L(p),\\
Q_{2}^{f*f*}(k) &= \int \Dk{p} \frac{(\vk\cdot \vp)\vk \cdot (\vk-\vp)}{p^2 |\vk-\vp|^2} \Gamma_2[\vp,\vk - \vp] 
\frac{f(|\vk-\vp|)}{f_0} \frac{f(p)}{f_0} P_L(|\vk-\vp|)  P_L(p).
\end{align}

The $q$-functions of (\ref{dotfuncs}) and (\ref{ddotfuncs}) can be recasted as 1D integrations of the above $Q$ and $R$ functions, which we will do in the upcoming section. The main differences with these functions and the corresponding in $\Lambda$CDM, are the use of the kernels $\Gamma$, differing in the percent level in MG and standard cosmologies, and more importantly, the appereance of factors $f(k)/f_0$ that can be as large as $\sim 1.1$.

\subsection{q functions} \label{app:GSM_q}

Several $q$ dependent functions should be calculated before computing the correlation function, 
the pairwise velocity and the pairwise velocity dispersion. These are
the ``$U$'', ``$A$'' and ``$W$'' functions defined in (\ref{dotfuncs}) and (\ref{ddotfuncs})

The $U_i(q)$ vector functions can be written as 
\begin{equation}
 U_i(\vq) = U(q)\hat{q}_i,
\end{equation}
with
\begin{equation}
\begin{aligned}\label{UdotMG}
\dot{U}(q) &=- \frac{1}{2\pi^2}\int dk \, k \left[ \frac{f(k)}{f_0} P_L(k)+ \frac{5}{7} R^f_1(k) \right] j_1(kq), \\
\dot{U}^{20}(q)  &=  -\frac{3}{7 \pi^2} \int dk \, k  Q_8^f(k) j_1(kq),  \\
\dot{U}^{11}(q)  &=  -\frac{6}{7 \pi^2} \int dk \, k  R_{1+2}^f(k) j_1(kq).
\end{aligned}
\end{equation}
In $\Lambda$CDM $R^f,Q^f \simeq R,Q$ and $f(k) =f_0$, hence we obtain the standard results $\dot{U} = U^{(1)} + 3 U^{(3)}$, $\dot{U}^{20}=2 U^{20}$ and $\dot{U}^{11}=2 U^{11}$; see \cite{Wang:2013hwa}.

The $A$ functions are decomposed as
\begin{equation}
 \dot{A}_{ij}(\vq) =   \dot{X}(q)\delta_{ij}   +  \dot{Y}(q)\hat{q}_i\hat{q}_j,\qquad 
 \ddot{A}_{ij}(\vq) =   \ddot{X}(q)\delta_{ij}   +  \ddot{Y}(q)\hat{q}_i\hat{q}_j,
\end{equation}
with
\begin{align}\label{ddotXdefs}
&\dot{X}(q) = \frac{1}{\pi^2}\int dk 
  \left[ \frac{f(k)}{f_0} P_L(k) + \frac{9}{49} Q^f_1(k) + \frac{5}{21} \frac{f(k)}{f_0} R_1(k) + \frac{5}{7} R^f_1(k) \right]
  \left[\frac{1}{3} -  \frac{j_1(k q)}{k q} \right], \nonumber\\
& \dot{Y}(q) = \frac{1}{\pi^2}\int dk 
  \left[ \frac{f(k)}{f_0} P_L(k) +\frac{9}{49} Q^f_1(k) + \frac{5}{21} \frac{f(k)}{f_0} R_1(k) + \frac{5}{7} R^f_1(k) \right]
  j_2(kq),\nonumber\\   
&\ddot{X}(q) = \frac{1}{\pi^2}\int dk 
  \left[ \left(\frac{f(k)}{f_0}\right)^2 P_L(k) + \frac{18}{49} Q^{ff}_1(k)  + \frac{10}{7} \frac{f(k)}{f_0} R_1^f(k) \right]
  \left[\frac{1}{3} -  \frac{j_1(k q)}{k q} \right], \nonumber\\
&\ddot{Y}(q) = \frac{1}{\pi^2}\int dk 
  \left[ \left(\frac{f(k)}{f_0}\right)^2 P_L(k) + \frac{18}{49} Q^{ff}_1(k)  + \frac{10}{7} \frac{f(k)}{f_0} R_1^f(k)\right]
  j_2(kq), 
\end{align}
and the $A^{10}$ functions as
\begin{equation}
 \dot{A}_{ij}^{10}(\vq) =   \dot{X}^{10}(q)\delta_{ij}   +  \dot{Y}^{10}(q)\hat{q}_i\hat{q}_j,\qquad 
 \ddot{A}^{10}_{ij}(\vq) =   \ddot{X}^{10}(q)\delta_{ij}   +  \ddot{Y}^{10}(q)\hat{q}_i\hat{q}_j,
\end{equation}
with
\begin{align} \label{ddotX10defs}
&\dot{X}_{10}(q) =\frac{1}{\pi^2}\int_0^\infty dk \frac{1}{28}\Bigg[ 2  (2R^f_I-2R^f_2+R^{f*}_I-R^{f*}_2) + 3 (2 R^f_I + R^{f*}_I ) j_0(kq) \nonumber\\
   &\qquad - 3( 2R^f_I + 4 R^f_2 + 4 R^f_{1+2} + 4 Q^f_5 + R^{f*}_I + 2 R^{f*}_2 + 2 R^{f*}_{1+2} + 2 Q^{f*}_5) \frac{j_1(kq)}{kq} \Bigg], \nonumber\\
&\dot{Y}_{10}(q) = \frac{1}{\pi^2}\int_0^\infty dk \frac{3}{28 } \Big[ 2 R^f_I +4 R^f_2 + 4 R^f_{1+2} + 4 Q^f_5 \nonumber\\
  &\qquad \qquad \qquad + R^{f*}_I +2 R^{f*}_2 + 2 R^{f*}_{1+2} + 2 Q^{f*}_5\Big] j_2(kq), \nonumber\\
&\ddot{X}_{10}(q) =\frac{1}{\pi^2}\int_0^\infty dk \frac{4}{28}\Bigg[   2(R^{ff}_I-R^{ff}_2) + 3  R^{ff}_I  j_0(kq) \nonumber\\
&\qquad \qquad \qquad- 3( R^{ff}_I + 2 R^{ff}_2 + 2 R^{ff}_{1+2} + 2 Q^{ff}_5) \frac{j_1(kq)}{kq} \Bigg], \nonumber\\
&\ddot{Y}_{10}(q) = \frac{1}{\pi^2}\int_0^\infty dk \frac{4\times 3}{28 } 
\Big[  R^{ff}_I +2 R^{ff}_2 + 2 R^{ff}_{1+2} + 2 Q^{ff}_5\Big] j_2(kq).
\end{align}

Now, the $W$ functions have the form
\begin{equation}
 \dot{W}_{ijk} = W^{(11\dot{2})}_{ijk} + W^{(12\dot{1})}_{ijk} + W^{(21\dot{1})}_{ijk}, \quad  \ddot{W}_{ijk} = W^{(1\dot{1}\dot{2})}_{ijk} + W^{(1\dot{2}\dot{1})}_{ijk} + W^{(2\dot{1}\dot{1})}_{ijk},
\end{equation}
where the dot over a number indicates that the Lagrangian displacement of that order should be differentiated.
Except for the undotted case, these cannot be decomposed as $W_{ijk} = V\hat{q}_{\{i}\delta_{jk\}} + T \hat{q}_i \hat{q}_j \hat{q}_k$, because in general
$\dot{W}_{ijk} \neq \dot{W}_{(jki)}$. However, we do have $ \dot{W}_{ijk}= \dot{W}_{(ij)k}$ and $ \ddot{W}_{ijk}= \ddot{W}_{i(jk)}$. These symmetries allow us to
decompose
\begin{align}
W_{ijk} &= V(q) \hat{q}_{\{i}\delta_{jk\}} + T(q) \hat{q}_i \hat{q}_j \hat{q}_k, \\
 \dot{W}_{ijk}(\vq) &  =   \dot{V}_1(q)\big(\hat{q}_{i}\delta_{jk} + \hat{q}_{j}\delta_{ki}\big) + 
           \dot{V}_3(q)\hat{q}_{k}\delta_{ij} +  \dot{T}(q) \hat{q}_i \hat{q}_j \hat{q}_k, \\
\ddot{W}_{ijk}(\vq) &=   \ddot{V}_1(q)\hat{q}_{i}\delta_{jk} + 
           \dot{V}_3(q)\big( \hat{q}_{j}\delta_{ki} + \hat{q}_{k}\delta_{ij}\big) +  \ddot{T}(q) \hat{q}_i \hat{q}_j \hat{q}_k,
\end{align}
with  $V(q)$ and $T(q)$ given in refs.~\cite{Valogiannis:2019xed,Aviles:2018saf} for MG, and
\begin{align}
\dot{T}(q)&= -\int \frac{dk}{\pi^2}\frac{3}{7 k} \Bigg[ 2 Q_2^f +2 Q_2^{f*} + Q_I^{f} + Q_I^{f*} +2\frac{f(k)}{f_0} R_2 +4 R_2^f \nonumber\\
&\qquad  +2 R_2^{f*} + \frac{f(k)}{f_0} R_I + 2 R_I^{f} + R_I^{f*} \Bigg] j_3(k q), \nonumber\\
\dot{V}_1(q) &=\int \frac{dk}{\pi^2}\frac{3}{70 k} \Bigg[ 4 Q_2^f +4 Q_2^{f*} +2 Q_I^{f} - 3 Q_I^{f*} +4\frac{f(k)}{f_0} R_2 +8 R_2^f  \nonumber\\
&\qquad  +4 R_2^{f*} + 2 \frac{f(k)}{f_0} R_I - 6 R_I^{f} - 3 R_I^{f*} \Bigg] j_1(k q) - \frac{1}{5} \dot{T}(q), \nonumber\\
\dot{V}_3(q)&= \dot{V}_1(q)  -\int \frac{dk}{\pi^2}\frac{3}{14 k} \Bigg[ 2Q_1^f - Q_1^{f*} + \frac{f(k)}{f_0}R_I - 2 R_I^f \Bigg]j_1(k q),
\end{align}
and
\begin{align}
\ddot{T}(q)&= -\int \frac{dk}{\pi^2}\frac{3}{14 k} \Bigg[ 8 Q_2^{ff} +2 Q_2^{f*f*} +4 Q_I^{ff} + Q_I^{f*f*} +8\frac{f(k)}{f_0} R_2^f +8 R_2^{ff} \nonumber\\
&\qquad  +4 \frac{f(k)}{f_0}  R_2^{f*} + 4 \frac{f(k)}{f_0} R_I^f + 4 R_I^{ff} +2 \frac{f(k)}{f_0}  R_I^{f*} \Bigg] j_3(k q), \nonumber\\
\ddot{V}_1(q) &=\int \frac{dk}{\pi^2}\frac{3}{35 k} \Bigg[ 4 Q_2^{ff} + Q_2^{f*f*} +2 Q_I^{ff} - 2 Q_I^{f*f*} +4\frac{f(k)}{f_0} R_2^f +4 R_2^{ff}  \nonumber\\
&\qquad  +2 \frac{f(k)}{f_0} R_2^{f*} + 2 \frac{f(k)}{f_0} R_I^f - 8 R_I^{ff} + \frac{f(k)}{f_0} R_I^{f*} \Bigg] j_1(k q) - \frac{1}{5} \ddot{T}(q), \nonumber\\
\dot{V}_3(q)  &= \dot{V}_1(q)  -\int \frac{dk}{\pi^2}\frac{3}{14 k} \Bigg[ 2Q_1^{ff} - Q_1^{f*f*} -4 R_I^{ff} + \frac{f(k)}{f_0}(2 R_I^f + R_I^{f*}) \Bigg]j_1(k q),
\end{align}
which complete our search for 1D integral expressions for the functions defined in (\ref{dotfuncs}) and (\ref{ddotfuncs}).

We end this section by emphasizing that the $\Lambda$CDM standard results are recovered by making the substitutions $Q,R^{f,f*,f*f*,ff}=Q,R$. 

\subsection{Tidal bias} \label{app:GSM_tidal}

In this subsection we introduce tidal bias following \cite{Vlah:2016bcl}. This is achieved by adding $s^2=s_{ij} s_{ij}$ as an argument to the Lagrangian biasing function $F$ [see (\ref{biasF})], with the shear tensor 
\begin{equation}
s_{ij}(\vq) = \left( \frac{\partial_i \partial_j}{\partial^2} -\frac{1}{3}\delta_{ij} \right) \delta(\vq).
\end{equation}

Almost all expressions related to tidal bias contain only linear fields, such that in the integrals of Appendix D of \cite{Vlah:2016bcl}, they only need the substitution $P_L \rightarrow (f(k)/f) P_L$ for dotted functions and 
$P_L \rightarrow (f(k)/f)^2 P_L$ for double-dotted functions . 
The only new, substantially different function, is 
\begin{equation}
V_i^{10} =\langle s^2(\vq_1) \Psi_i^{(2)}(\vq_2) \rangle_c = -\frac{3}{7} \int \frac{dk}{2\pi^2} Q_{s^2}(k) j_1(kq), 
\end{equation}
with
\begin{equation}
Q_{s^2}(k) = \int \Dk{p} \Gamma_2(\vk-\vp, \vp) \left[ \frac{\big((\vk-\vp)\cdot \vp \big)^2}{p^2|\vk-\vp|^2} - \frac{1}{3} \right] P_L(|\vk-\vp|) P_L(p).  
\end{equation}
This does not reduce to the result of \cite{Vlah:2016bcl} for $\Lambda$CDM. Instead, it differs by a $1/2$ factor: $V_i^{10} = \frac{1}{2} V_i^{10 \,[\text{That work}]}$.

\section{Direct Lagrangian approach in MG}\label{AppendixDirect}
In this Appendix section, we explain how the Direct Lagrangian approach to RSD, laid out in section \ref{DirectLPT}, will be implemented in MG theories and point out the differences with respect to the GR case. In particular, starting with the simpler GR case, in section \ref{DirectLPT} we explained how the LPT field evolves as $\bold{\Psi}^{(n)}\propto D^n(a)$, giving $\bold{\dot{\Psi}}^{(n)}= n f_0 H \bold{\Psi}^{(n)}$, which allows us to map the LPT field to redshift space, order by order, through 
\begin{equation}\label{PsiRSD2app}
\Psi_i^{s(n)}=\left(\delta_{ij}+n f_0\hat{z}_i\hat{z}_j\right)\Psi_j^{(n)},
\end{equation}
where we made use of (\ref{Psimap}). Equation (\ref{PsiRSD2app}) then allows us to easily ``Directly'' map each of the Lagrangian correlators (\ref{eq:correl}) to redshift space. Focusing on the function $ U_i(\vq) = U(q)\hat{q}_i$, as an example, and expanding order by order as $U(q) = U^{(1)}(q) + U^{(3)}(q)+..$, we get 
\begin{equation}\label{CorrelRSDapp}
U_i^{s(n)}=\left(\delta_{ij}+n f_0\hat{z}_i\hat{z}_j\right)U_j^{(n)},
\end{equation}
where $U_i^{s(n)}$ denotes the redshift space version of $U^{(n)}(q)$.

The above derivation does not hold in MG theories however, because of the scale-dependent growth factors that are introduced, which means that $\bold{\dot{\Psi}}^{(n)} \neq n f H \bold{\Psi}^{(n)}$. In this case, and as explained in detail in the previous appendix section \ref{AppendixGSM}, $\bold{\dot{\Psi}}^{(n)}$ is instead given by (\ref{dotPsi}), combined with (\ref{defGf}), which leads to the mapping
\begin{equation}\label{PsiRSDMG}
\Psi_i^{s(n)}=\Psi_i^{(n)}+\hat{z}_i\hat{z}_j\frac{d\Psi_j^{(n)}}{d\ln a} = \Psi_i^{(n)}+\hat{z}_i\hat{z}_j\frac{\dot{\Psi}_j^{(n)}}{H(a)}.
\end{equation}
From the definitions (\ref{eq:correl}), we will now get
\begin{equation}\label{CorrelRSDMG}
U_i^{s(n)}=U_i^{(n)}+f_0\hat{z}_i\hat{z}_j\dot{U}_j^{(n)},
\end{equation}
where $\dot{U}_j^{(n)}$ is given in (\ref{UdotMG}). In the GR limit, $\dot{U}(q) = U^{(1)}(q) + 3 U^{(3)}(q)$ and (\ref{CorrelRSDMG}) reduces back to the GR expression (\ref{CorrelRSDapp}). The rest of the correlators (\ref{eq:correl}) can be similarly mapped to their redshift space expressions, following the same procedure, combined with the ``dot'' functions presented in appendix \ref{AppendixGSM}. 



\bibliographystyle{JHEP} 

\providecommand{\href}[2]{#2}\begingroup\raggedright\endgroup



\begin{thebibliography}{100}

\bibitem{Eisenstein:2005su}
{\scshape SDSS Collaboration} collaboration, \emph{{Detection of the baryon
  acoustic peak in the large-scale correlation function of SDSS luminous red
  galaxies}}, \href{https://doi.org/10.1086/466512}{\emph{Astrophys.J.}
  {\bfseries 633} (2005) 560}
  [\href{https://arxiv.org/abs/astro-ph/0501171}{{\ttfamily
  astro-ph/0501171}}].

\bibitem{Percival:2007yw}
W.~J. Percival et~al., \emph{{Measuring the Baryon Acoustic Oscillation scale
  using the SDSS and 2dFGRS}},
  \href{https://doi.org/10.1111/j.1365-2966.2007.12268.x}{\emph{Mon. Not. Roy.
  Astron. Soc.} {\bfseries 381} (2007) 1053}
  [\href{https://arxiv.org/abs/0705.3323}{{\ttfamily 0705.3323}}].

\bibitem{Percival:2009xn}
W.~J. Percival et~al., \emph{{Baryon Acoustic Oscillations in the Sloan Digital
  Sky Survey Data Release 7 Galaxy Sample}},
  \href{https://arxiv.org/abs/0907.1660}{{\ttfamily 0907.1660}}.

\bibitem{Kazin:2014qga}
E.~A. Kazin, J.~Koda, C.~Blake and N.~Padmanabhan, \emph{{Improved WiggleZ Dark
  Energy Survey Distance Measurements to z = 1 with Reconstruction of the
  Baryonic Acoustic Feature}},
  \href{https://arxiv.org/abs/1401.0358}{{\ttfamily 1401.0358}}.

\bibitem{Spergel:2013tha}
D.~Spergel, N.~Gehrels, J.~Breckinridge, M.~Donahue, A.~Dressler et~al.,
  \emph{{Wide-Field InfraRed Survey Telescope-Astrophysics Focused Telescope
  Assets WFIRST-AFTA Final Report}},
  \href{https://arxiv.org/abs/1305.5422}{{\ttfamily 1305.5422}}.

\bibitem{Ade:2013zuv}
{\scshape Planck Collaboration} collaboration, \emph{{Planck 2013 results. XVI.
  Cosmological parameters}},  \href{https://arxiv.org/abs/1303.5076}{{\ttfamily
  1303.5076}}.

\bibitem{Ade:2015xua}
{\scshape Planck} collaboration, \emph{{Planck 2015 results. XIII. Cosmological
  parameters}},
  \href{https://doi.org/10.1051/0004-6361/201525830}{\emph{Astron. Astrophys.}
  {\bfseries 594} (2016) A13}
  [\href{https://arxiv.org/abs/1502.01589}{{\ttfamily 1502.01589}}].

\bibitem{Perlmutter:1998np}
{\scshape Supernova Cosmology Project} collaboration, \emph{{Measurements of
  Omega and Lambda from 42 high redshift supernovae}},
  \href{https://doi.org/10.1086/307221}{\emph{Astrophys. J.} {\bfseries 517}
  (1999) 565} [\href{https://arxiv.org/abs/astro-ph/9812133}{{\ttfamily
  astro-ph/9812133}}].

\bibitem{Riess:2004nr}
{\scshape Supernova Search Team} collaboration, \emph{{Type Ia supernova
  discoveries at z > 1 from the Hubble Space Telescope: Evidence for past
  deceleration and constraints on dark energy evolution}},
  \href{https://doi.org/10.1086/383612}{\emph{Astrophys. J.} {\bfseries 607}
  (2004) 665} [\href{https://arxiv.org/abs/astro-ph/0402512}{{\ttfamily
  astro-ph/0402512}}].

\bibitem{Weinberg:1988cp}
S.~Weinberg, \emph{{The Cosmological Constant Problem}},
  \href{https://doi.org/10.1103/RevModPhys.61.1}{\emph{Rev. Mod. Phys.}
  {\bfseries 61} (1989) 1}.

\bibitem{Koyama:2015vza}
K.~Koyama, \emph{{Cosmological Tests of Modified Gravity}},
  \href{https://doi.org/10.1088/0034-4885/79/4/046902}{\emph{Rept. Prog. Phys.}
  {\bfseries 79} (2016) 046902}
  [\href{https://arxiv.org/abs/1504.04623}{{\ttfamily 1504.04623}}].

\bibitem{Ishak:2018his}
M.~Ishak, \emph{{Testing General Relativity in Cosmology}},
  \href{https://arxiv.org/abs/1806.10122}{{\ttfamily 1806.10122}}.

\bibitem{Ferreira:2019xrr}
P.~G. Ferreira, \emph{{Cosmological Tests of Gravity}},
  \href{https://arxiv.org/abs/1902.10503}{{\ttfamily 1902.10503}}.

\bibitem{Will:2005va}
C.~M. Will, \emph{{The Confrontation between general relativity and
  experiment}}, \href{https://doi.org/10.12942/lrr-2006-3}{\emph{Living Rev.
  Rel.} {\bfseries 9} (2006) 3}
  [\href{https://arxiv.org/abs/gr-qc/0510072}{{\ttfamily gr-qc/0510072}}].

\bibitem{TheLIGOScientific:2017qsa}
{\scshape Virgo, LIGO Scientific} collaboration, \emph{{GW170817: Observation
  of Gravitational Waves from a Binary Neutron Star Inspiral}},
  \href{https://doi.org/10.1103/PhysRevLett.119.161101}{\emph{Phys. Rev. Lett.}
  {\bfseries 119} (2017) 161101}
  [\href{https://arxiv.org/abs/1710.05832}{{\ttfamily 1710.05832}}].

\bibitem{Goldstein:2017mmi}
A.~Goldstein et~al., \emph{{An Ordinary Short Gamma-Ray Burst with
  Extraordinary Implications: Fermi-GBM Detection of GRB 170817A}},
  \href{https://doi.org/10.3847/2041-8213/aa8f41}{\emph{Astrophys. J.}
  {\bfseries 848} (2017) L14}
  [\href{https://arxiv.org/abs/1710.05446}{{\ttfamily 1710.05446}}].

\bibitem{Savchenko:2017ffs}
V.~Savchenko et~al., \emph{{INTEGRAL Detection of the First Prompt Gamma-Ray
  Signal Coincident with the Gravitational-wave Event GW170817}},
  \href{https://doi.org/10.3847/2041-8213/aa8f94}{\emph{Astrophys. J.}
  {\bfseries 848} (2017) L15}
  [\href{https://arxiv.org/abs/1710.05449}{{\ttfamily 1710.05449}}].

\bibitem{Monitor:2017mdv}
{\scshape Virgo, Fermi-GBM, INTEGRAL, LIGO Scientific} collaboration,
  \emph{{Gravitational Waves and Gamma-rays from a Binary Neutron Star Merger:
  GW170817 and GRB 170817A}},
  \href{https://doi.org/10.3847/2041-8213/aa920c}{\emph{Astrophys. J.}
  {\bfseries 848} (2017) L13}
  [\href{https://arxiv.org/abs/1710.05834}{{\ttfamily 1710.05834}}].

\bibitem{GBM:2017lvd}
{\scshape GROND, SALT Group, OzGrav, DFN, INTEGRAL, Virgo, Insight-Hxmt, MAXI
  Team, Fermi-LAT, J-GEM, RATIR, IceCube, CAASTRO, LWA, ePESSTO, GRAWITA,
  RIMAS, SKA South Africa/MeerKAT, H.E.S.S., 1M2H Team, IKI-GW Follow-up, Fermi
  GBM, Pi of Sky, DWF (Deeper Wider Faster Program), Dark Energy Survey,
  MASTER, AstroSat Cadmium Zinc Telluride Imager Team, Swift, Pierre Auger,
  ASKAP, VINROUGE, JAGWAR, Chandra Team at McGill University, TTU-NRAO, GROWTH,
  AGILE Team, MWA, ATCA, AST3, TOROS, Pan-STARRS, NuSTAR, ATLAS Telescopes,
  BOOTES, CaltechNRAO, LIGO Scientific, High Time Resolution Universe Survey,
  Nordic Optical Telescope, Las Cumbres Observatory Group, TZAC Consortium,
  LOFAR, IPN, DLT40, Texas Tech University, HAWC, ANTARES, KU, Dark Energy
  Camera GW-EM, CALET, Euro VLBI Team, ALMA} collaboration,
  \emph{{Multi-messenger Observations of a Binary Neutron Star Merger}},
  \href{https://doi.org/10.3847/2041-8213/aa91c9}{\emph{Astrophys. J.}
  {\bfseries 848} (2017) L12}
  [\href{https://arxiv.org/abs/1710.05833}{{\ttfamily 1710.05833}}].

\bibitem{Lombriser:2015sxa}
L.~Lombriser and A.~Taylor, \emph{{Breaking a Dark Degeneracy with
  Gravitational Waves}},
  \href{https://doi.org/10.1088/1475-7516/2016/03/031}{\emph{JCAP} {\bfseries
  1603} (2016) 031} [\href{https://arxiv.org/abs/1509.08458}{{\ttfamily
  1509.08458}}].

\bibitem{Lombriser:2016yzn}
L.~Lombriser and N.~A. Lima, \emph{{Challenges to Self-Acceleration in Modified
  Gravity from Gravitational Waves and Large-Scale Structure}},
  \href{https://doi.org/10.1016/j.physletb.2016.12.048}{\emph{Phys. Lett.}
  {\bfseries B765} (2017) 382}
  [\href{https://arxiv.org/abs/1602.07670}{{\ttfamily 1602.07670}}].

\bibitem{Sakstein:2017xjx}
J.~Sakstein and B.~Jain, \emph{{Implications of the Neutron Star Merger
  GW170817 for Cosmological Scalar-Tensor Theories}},
  \href{https://doi.org/10.1103/PhysRevLett.119.251303}{\emph{Phys. Rev. Lett.}
  {\bfseries 119} (2017) 251303}
  [\href{https://arxiv.org/abs/1710.05893}{{\ttfamily 1710.05893}}].

\bibitem{Ezquiaga:2017ekz}
J.~M. Ezquiaga and M.~ZumalacÃ¡rregui, \emph{{Dark Energy After GW170817:
  Dead Ends and the Road Ahead}},
  \href{https://doi.org/10.1103/PhysRevLett.119.251304}{\emph{Phys. Rev. Lett.}
  {\bfseries 119} (2017) 251304}
  [\href{https://arxiv.org/abs/1710.05901}{{\ttfamily 1710.05901}}].

\bibitem{Creminelli:2017sry}
P.~Creminelli and F.~Vernizzi, \emph{{Dark Energy after GW170817 and
  GRB170817A}},
  \href{https://doi.org/10.1103/PhysRevLett.119.251302}{\emph{Phys. Rev. Lett.}
  {\bfseries 119} (2017) 251302}
  [\href{https://arxiv.org/abs/1710.05877}{{\ttfamily 1710.05877}}].

\bibitem{Baker:2017hug}
T.~Baker, E.~Bellini, P.~G. Ferreira, M.~Lagos, J.~Noller and I.~Sawicki,
  \emph{{Strong constraints on cosmological gravity from GW170817 and GRB
  170817A}}, \href{https://doi.org/10.1103/PhysRevLett.119.251301}{\emph{Phys.
  Rev. Lett.} {\bfseries 119} (2017) 251301}
  [\href{https://arxiv.org/abs/1710.06394}{{\ttfamily 1710.06394}}].

\bibitem{Horndeski1974}
G.~W. Horndeski, \emph{Second-order scalar-tensor field equations in a
  four-dimensional space},
  \href{https://doi.org/10.1007/BF01807638}{\emph{International Journal of
  Theoretical Physics} {\bfseries 10} (1974) 363}.

\bibitem{PhysRevD.84.064039}
C.~Deffayet, X.~Gao, D.~A. Steer and G.~Zahariade, \emph{From $k$-essence to
  generalized galileons},
  \href{https://doi.org/10.1103/PhysRevD.84.064039}{\emph{Phys. Rev. D}
  {\bfseries 84} (2011) 064039}.

\bibitem{Khoury:2010xi}
J.~Khoury, \emph{{Theories of Dark Energy with Screening Mechanisms}},
  \href{https://arxiv.org/abs/1011.5909}{{\ttfamily 1011.5909}}.

\bibitem{Khoury:2013tda}
J.~Khoury, \emph{{Les Houches Lectures on Physics Beyond the Standard Model of
  Cosmology}},  \href{https://arxiv.org/abs/1312.2006}{{\ttfamily 1312.2006}}.

\bibitem{VAINSHTEIN1972393}
A.~Vainshtein, \emph{To the problem of nonvanishing gravitation mass},
  \href{https://doi.org/http://dx.doi.org/10.1016/0370-2693(72)90147-5}{\emph{Physics
  Letters B} {\bfseries 39} (1972) 393 }.

\bibitem{Babichev:2013usa}
E.~Babichev and C.~Deffayet, \emph{{An introduction to the Vainshtein
  mechanism}},
  \href{https://doi.org/10.1088/0264-9381/30/18/184001}{\emph{Class. Quant.
  Grav.} {\bfseries 30} (2013) 184001}
  [\href{https://arxiv.org/abs/1304.7240}{{\ttfamily 1304.7240}}].

\bibitem{PhysRevD.69.044026}
J.~Khoury and A.~Weltman, \emph{Chameleon cosmology},
  \href{https://doi.org/10.1103/PhysRevD.69.044026}{\emph{Phys. Rev. D}
  {\bfseries 69} (2004) 044026}.

\bibitem{PhysRevLett.93.171104}
J.~Khoury and A.~Weltman, \emph{Chameleon fields: Awaiting surprises for tests
  of gravity in space},
  \href{https://doi.org/10.1103/PhysRevLett.93.171104}{\emph{Phys. Rev. Lett.}
  {\bfseries 93} (2004) 171104}.

\bibitem{Wang:2012kj}
J.~Wang, L.~Hui and J.~Khoury, \emph{{No-Go Theorems for Generalized Chameleon
  Field Theories}},
  \href{https://doi.org/10.1103/PhysRevLett.109.241301}{\emph{Phys. Rev. Lett.}
  {\bfseries 109} (2012) 241301}
  [\href{https://arxiv.org/abs/1208.4612}{{\ttfamily 1208.4612}}].

\bibitem{Burrage:2017qrf}
C.~Burrage and J.~Sakstein, \emph{{Tests of Chameleon Gravity}},
  \href{https://doi.org/10.1007/s41114-018-0011-x}{\emph{Living Rev. Rel.}
  {\bfseries 21} (2018) 1} [\href{https://arxiv.org/abs/1709.09071}{{\ttfamily
  1709.09071}}].

\bibitem{PhysRevLett.104.231301}
K.~Hinterbichler and J.~Khoury, \emph{Screening long-range forces through local
  symmetry restoration},
  \href{https://doi.org/10.1103/PhysRevLett.104.231301}{\emph{Phys. Rev. Lett.}
  {\bfseries 104} (2010) 231301}.

\bibitem{Olive:2007aj}
K.~A. Olive and M.~Pospelov, \emph{{Environmental dependence of masses and
  coupling constants}},
  \href{https://doi.org/10.1103/PhysRevD.77.043524}{\emph{Phys. Rev.}
  {\bfseries D77} (2008) 043524}
  [\href{https://arxiv.org/abs/0709.3825}{{\ttfamily 0709.3825}}].

\bibitem{Dvali:2010jz}
G.~Dvali, G.~F. Giudice, C.~Gomez and A.~Kehagias, \emph{{UV-Completion by
  Classicalization}},
  \href{https://doi.org/10.1007/JHEP08(2011)108}{\emph{JHEP} {\bfseries 08}
  (2011) 108} [\href{https://arxiv.org/abs/1010.1415}{{\ttfamily 1010.1415}}].

\bibitem{Levi:2013gra}
{\scshape DESI collaboration} collaboration, \emph{{The DESI Experiment, a
  whitepaper for Snowmass 2013}},
  \href{https://arxiv.org/abs/1308.0847}{{\ttfamily 1308.0847}}.

\bibitem{Abell:2009aa}
{\scshape LSST Science Collaborations, LSST Project} collaboration, \emph{{LSST
  Science Book, Version 2.0}},
  \href{https://arxiv.org/abs/0912.0201}{{\ttfamily 0912.0201}}.

\bibitem{Laureijs:2011gra}
{\scshape EUCLID Collaboration} collaboration, \emph{{Euclid Definition Study
  Report}},  \href{https://arxiv.org/abs/1110.3193}{{\ttfamily 1110.3193}}.

\bibitem{Ishak:2019aay}
M.~Ishak et~al., \emph{{Modified Gravity and Dark Energy models Beyond
  $w(z)$CDM Testable by LSST}},
  \href{https://arxiv.org/abs/1905.09687}{{\ttfamily 1905.09687}}.

\bibitem{1984ApJ...284L...9K}
N.~{Kaiser}, \emph{{On the spatial correlations of Abell clusters}},
  \href{https://doi.org/10.1086/184341}{\emph{The Astrophysical Journal}
  {\bfseries 284} (1984) L9}.

\bibitem{10.1093/mnras/227.1.1}
N.~Kaiser, \emph{{Clustering in real space and in redshift space}},
  \href{https://doi.org/10.1093/mnras/227.1.1}{\emph{Monthly Notices of the
  Royal Astronomical Society} {\bfseries 227} (1987) 1}
  [\href{https://arxiv.org/abs/http://oup.prod.sis.lan/mnras/article-pdf/227/1/1/18522208/mnras227-0001.pdf}{{\ttfamily
  http://oup.prod.sis.lan/mnras/article-pdf/227/1/1/18522208/mnras227-0001.pdf}}].

\bibitem{1988MNRAS.235..715E}
G.~{Efstathiou}, C.~S. {Frenk}, S.~D.~M. {White} and M.~{Davis},
  \emph{{Gravitational clustering from scale-free initial conditions}},
  \href{https://doi.org/10.1093/mnras/235.3.715}{\emph{Monthly Notices of the
  Royal Astronomical Society} {\bfseries 235} (1988) 715}.

\bibitem{Bernardeau:2001qr}
F.~Bernardeau, S.~Colombi, E.~Gaztanaga and R.~Scoccimarro, \emph{{Large scale
  structure of the universe and cosmological perturbation theory}},
  \href{https://doi.org/10.1016/S0370-1573(02)00135-7}{\emph{Phys. Rept.}
  {\bfseries 367} (2002) 1}
  [\href{https://arxiv.org/abs/astro-ph/0112551}{{\ttfamily
  astro-ph/0112551}}].

\bibitem{2009PhRvD..80d3531C}
J.~{Carlson}, M.~{White} and N.~{Padmanabhan}, \emph{{Critical look at
  cosmological perturbation theory techniques}},
  \href{https://doi.org/10.1103/PhysRevD.80.043531}{\emph{Physical Review D}
  {\bfseries 80} (2009) 043531}
  [\href{https://arxiv.org/abs/0905.0479}{{\ttfamily 0905.0479}}].

\bibitem{Tassev:2013pn}
S.~Tassev, M.~Zaldarriaga and D.~Eisenstein, \emph{{Solving Large Scale
  Structure in Ten Easy Steps with COLA}},
  \href{https://doi.org/10.1088/1475-7516/2013/06/036}{\emph{JCAP} {\bfseries
  1306} (2013) 036} [\href{https://arxiv.org/abs/1301.0322}{{\ttfamily
  1301.0322}}].

\bibitem{Valogiannis:2016ane}
G.~Valogiannis and R.~Bean, \emph{{Efficient simulations of large scale
  structure in modified gravity cosmologies with comoving Lagrangian
  acceleration}}, \href{https://doi.org/10.1103/PhysRevD.95.103515}{\emph{Phys.
  Rev.} {\bfseries D95} (2017) 103515}
  [\href{https://arxiv.org/abs/1612.06469}{{\ttfamily 1612.06469}}].

\bibitem{1983ApJ...267..465D}
M.~{Davis} and P.~J.~E. {Peebles}, \emph{{A survey of galaxy redshifts. V - The
  two-point position and velocity correlations}},
  \href{https://doi.org/10.1086/160884}{\emph{The Astrophysical Journal}
  {\bfseries 267} (1983) 465}.

\bibitem{10.1093/mnras/258.3.581}
J.~A. Peacock, \emph{{Errors on the measurement of ? via cosmological
  dipoles}}, \href{https://doi.org/10.1093/mnras/258.3.581}{\emph{Monthly
  Notices of the Royal Astronomical Society} {\bfseries 258} (1992) 581}
  [\href{https://arxiv.org/abs/http://oup.prod.sis.lan/mnras/article-pdf/258/3/581/3777991/mnras258-0581.pdf}{{\ttfamily
  http://oup.prod.sis.lan/mnras/article-pdf/258/3/581/3777991/mnras258-0581.pdf}}].

\bibitem{doi:10.1111/j.1365-2966.2011.19379.x}
B.~A. Reid and M.~White, \emph{Towards an accurate model of the redshift-space
  clustering of haloes in the quasi-linear regime},
  \href{https://doi.org/10.1111/j.1365-2966.2011.19379.x}{\emph{Monthly Notices
  of the Royal Astronomical Society} {\bfseries 417} (2011) 1913}.

\bibitem{1995ApJ...448..494F}
K.~B. {Fisher}, \emph{{On the Validity of the Streaming Model for the
  Redshift-Space Correlation Function in the Linear Regime}}, {\emph{The
  Astrophysical Journal, eprint = {astro-ph/9412081}, keywords = {COSMOLOGY:
  LARGE-SCALE STRUCTURE OF UNIVERSE, COSMOLOGY: THEORY, GALAXIES: DISTANCES AND
  REDSHIFTS}, year = 1995, month = aug, volume = 448, pages = {494}, doi =
  {10.1086/175980}, adsurl =
  {https://ui.adsabs.harvard.edu/abs/1995ApJ...448..494F}, adsnote = {Provided
  by the SAO/NASA Astrophysics Data System}} }.

\bibitem{PhysRevD.70.083007}
R.~Scoccimarro, \emph{Redshift-space distortions, pairwise velocities, and
  nonlinearities},
  \href{https://doi.org/10.1103/PhysRevD.70.083007}{\emph{Phys. Rev. D}
  {\bfseries 70} (2004) 083007}.

\bibitem{Wang:2013hwa}
L.~Wang, B.~Reid and M.~White, \emph{{An analytic model for redshift-space
  distortions}}, \href{https://doi.org/10.1093/mnras/stt1916}{\emph{Mon. Not.
  Roy. Astron. Soc.} {\bfseries 437} (2014) 588}
  [\href{https://arxiv.org/abs/1306.1804}{{\ttfamily 1306.1804}}].

\bibitem{Zeldovich:1969sb}
{\relax Ya}.~B. Zeldovich, \emph{{Gravitational instability: An Approximate
  theory for large density perturbations}}, {\emph{Astron. Astrophys.}
  {\bfseries 5} (1970) 84}.

\bibitem{1989A&A...223....9B}
T.~{Buchert}, \emph{{A class of solutions in Newtonian cosmology and the
  pancake theory}}, {\emph{Astron. Astrophys.} {\bfseries 223} (1989) 9}.

\bibitem{Bouchet:1994xp}
F.~R. Bouchet, S.~Colombi, E.~Hivon and R.~Juszkiewicz, \emph{{Perturbative
  Lagrangian approach to gravitational instability}}, {\emph{Astron.
  Astrophys.} {\bfseries 296} (1995) 575}
  [\href{https://arxiv.org/abs/astro-ph/9406013}{{\ttfamily
  astro-ph/9406013}}].

\bibitem{Hivon:1994qb}
E.~Hivon, F.~R. Bouchet, S.~Colombi and R.~Juszkiewicz, \emph{{Redshift
  distortions of clustering: A Lagrangian approach}}, {\emph{Astron.
  Astrophys.} {\bfseries 298} (1995) 643}
  [\href{https://arxiv.org/abs/astro-ph/9407049}{{\ttfamily
  astro-ph/9407049}}].

\bibitem{Taylor:1996ne}
A.~N. Taylor and A.~J.~S. Hamilton, \emph{{Nonlinear cosmological power spectra
  in real and redshift space}},
  \href{https://doi.org/10.1093/mnras/282.3.767}{\emph{Mon. Not. Roy. Astron.
  Soc.} {\bfseries 282} (1996) 767}
  [\href{https://arxiv.org/abs/astro-ph/9604020}{{\ttfamily
  astro-ph/9604020}}].

\bibitem{Matsubara:2007wj}
T.~Matsubara, \emph{{Resumming Cosmological Perturbations via the Lagrangian
  Picture: One-loop Results in Real Space and in Redshift Space}},
  \href{https://doi.org/10.1103/PhysRevD.77.063530}{\emph{Phys. Rev.}
  {\bfseries D77} (2008) 063530}
  [\href{https://arxiv.org/abs/0711.2521}{{\ttfamily 0711.2521}}].

\bibitem{Matsubara:2008wx}
T.~Matsubara, \emph{{Nonlinear perturbation theory with halo bias and
  redshift-space distortions via the Lagrangian picture}},
  \href{https://doi.org/10.1103/PhysRevD.78.109901,
  10.1103/PhysRevD.78.083519}{\emph{Phys. Rev.} {\bfseries D78} (2008) 083519}
  [\href{https://arxiv.org/abs/0807.1733}{{\ttfamily 0807.1733}}].

\bibitem{2013MNRAS.429.1674C}
J.~{Carlson}, B.~{Reid} and M.~{White}, \emph{{Convolution Lagrangian
  perturbation theory for biased tracers}},
  \href{https://doi.org/10.1093/mnras/sts457}{\emph{Monthly Notices of the
  Royal Astronomical Society} {\bfseries 429} (2013) 1674}
  [\href{https://arxiv.org/abs/1209.0780}{{\ttfamily 1209.0780}}].

\bibitem{Matsubara:2015ipa}
T.~Matsubara, \emph{{Recursive Solutions of Lagrangian Perturbation Theory}},
  \href{https://doi.org/10.1103/PhysRevD.92.023534}{\emph{Phys. Rev.}
  {\bfseries D92} (2015) 023534}
  [\href{https://arxiv.org/abs/1505.01481}{{\ttfamily 1505.01481}}].

\bibitem{Uhlemann:2015hqa}
C.~Uhlemann, M.~Kopp and T.~Haugg, \emph{{Edgeworth streaming model for
  redshift space distortions}},
  \href{https://doi.org/10.1103/PhysRevD.92.063004}{\emph{Phys. Rev.}
  {\bfseries D92} (2015) 063004}
  [\href{https://arxiv.org/abs/1503.08837}{{\ttfamily 1503.08837}}].

\bibitem{Bianchi:2016qen}
D.~Bianchi, W.~Percival and J.~Bel, \emph{{Improving the modelling of
  redshift-space distortions? II. A pairwise velocity model covering large and
  small scales}}, \href{https://doi.org/10.1093/mnras/stw2243}{\emph{Mon. Not.
  Roy. Astron. Soc.} {\bfseries 463} (2016) 3783}
  [\href{https://arxiv.org/abs/1602.02780}{{\ttfamily 1602.02780}}].

\bibitem{Vlah:2015sea}
Z.~Vlah, M.~White and A.~Aviles, \emph{{A Lagrangian effective field theory}},
  \href{https://doi.org/10.1088/1475-7516/2015/09/014}{\emph{JCAP} {\bfseries
  1509} (2015) 014} [\href{https://arxiv.org/abs/1506.05264}{{\ttfamily
  1506.05264}}].

\bibitem{Vlah:2016bcl}
Z.~Vlah, E.~Castorina and M.~White, \emph{{The Gaussian streaming model and
  convolution Lagrangian effective field theory}},
  \href{https://doi.org/10.1088/1475-7516/2016/12/007}{\emph{JCAP} {\bfseries
  1612} (2016) 007} [\href{https://arxiv.org/abs/1609.02908}{{\ttfamily
  1609.02908}}].

\bibitem{10.1093/mnras/stx196}
P.~Arnalte-Mur, W.~A. Hellwing and P.~Norberg, \emph{{Real- and redshift-space
  halo clustering in f(R) cosmologies}},
  \href{https://doi.org/10.1093/mnras/stx196}{\emph{Monthly Notices of the
  Royal Astronomical Society} {\bfseries 467} (2017) 1569}
  [\href{https://arxiv.org/abs/http://oup.prod.sis.lan/mnras/article-pdf/467/2/1569/10874417/stx196.pdf}{{\ttfamily
  http://oup.prod.sis.lan/mnras/article-pdf/467/2/1569/10874417/stx196.pdf}}].

\bibitem{PhysRevD.94.084022}
A.~Barreira, A.~G. S\'anchez and F.~Schmidt, \emph{Validating estimates of the
  growth rate of structure with modified gravity simulations},
  \href{https://doi.org/10.1103/PhysRevD.94.084022}{\emph{Phys. Rev. D}
  {\bfseries 94} (2016) 084022}.

\bibitem{Hernandez-Aguayo:2018oxg}
C.~Hern~$\'{a}$ndez Aguayo, J.~Hou, B.~Li, C.~M. Baugh and A.~G.
  S~$\'{a}$nchez, \emph{{Large-scale redshift space distortions in modified
  gravity theories}}, \href{https://doi.org/10.1093/mnras/stz516}{\emph{Mon.
  Not. Roy. Astron. Soc.} {\bfseries 485} (2019) 2194}
  [\href{https://arxiv.org/abs/1811.09197}{{\ttfamily 1811.09197}}].

\bibitem{PhysRevD.79.123512}
K.~Koyama, A.~Taruya and T.~Hiramatsu, \emph{Nonlinear evolution of the matter
  power spectrum in modified theories of gravity},
  \href{https://doi.org/10.1103/PhysRevD.79.123512}{\emph{Phys. Rev. D}
  {\bfseries 79} (2009) 123512}.

\bibitem{Taruya:2013quf}
A.~Taruya, K.~Koyama, T.~Hiramatsu and A.~Oka, \emph{{Beyond consistency test
  of gravity with redshift-space distortions at quasilinear scales}},
  \href{https://doi.org/10.1103/PhysRevD.89.043509}{\emph{Phys. Rev.}
  {\bfseries D89} (2014) 043509}
  [\href{https://arxiv.org/abs/1309.6783}{{\ttfamily 1309.6783}}].

\bibitem{PhysRevD.88.023527}
P.~Brax and P.~Valageas, \emph{Impact on the power spectrum of screening in
  modified gravity scenarios},
  \href{https://doi.org/10.1103/PhysRevD.88.023527}{\emph{Phys. Rev. D}
  {\bfseries 88} (2013) 023527}.

\bibitem{PhysRevD.90.123515}
A.~Taruya, T.~Nishimichi, F.~Bernardeau, T.~Hiramatsu and K.~Koyama,
  \emph{Regularized cosmological power spectrum and correlation function in
  modified gravity models},
  \href{https://doi.org/10.1103/PhysRevD.90.123515}{\emph{Phys. Rev. D}
  {\bfseries 90} (2014) 123515}.

\bibitem{PhysRevD.92.063522}
E.~Bellini and M.~Zumalac\'arregui, \emph{Nonlinear evolution of the baryon
  acoustic oscillation scale in alternative theories of gravity},
  \href{https://doi.org/10.1103/PhysRevD.92.063522}{\emph{Phys. Rev. D}
  {\bfseries 92} (2015) 063522}.

\bibitem{Fasiello:2017bot}
M.~Fasiello and Z.~Vlah, \emph{{Screening in perturbative approaches to LSS}},
  \href{https://doi.org/10.1016/j.physletb.2017.08.032}{\emph{Phys. Lett.}
  {\bfseries B773} (2017) 236}
  [\href{https://arxiv.org/abs/1704.07552}{{\ttfamily 1704.07552}}].

\bibitem{Bose:2017dtl}
B.~Bose and K.~Koyama, \emph{{A Perturbative Approach to the Redshift Space
  Correlation Function: Beyond the Standard Model}},
  \href{https://doi.org/10.1088/1475-7516/2017/08/029}{\emph{JCAP} {\bfseries
  1708} (2017) 029} [\href{https://arxiv.org/abs/1705.09181}{{\ttfamily
  1705.09181}}].

\bibitem{Bose:2016qun}
B.~Bose and K.~Koyama, \emph{{A Perturbative Approach to the Redshift Space
  Power Spectrum: Beyond the Standard Model}},
  \href{https://doi.org/10.1088/1475-7516/2016/08/032}{\emph{JCAP} {\bfseries
  1608} (2016) 032} [\href{https://arxiv.org/abs/1606.02520}{{\ttfamily
  1606.02520}}].

\bibitem{Bose:2018zpk}
B.~Bose and A.~Taruya, \emph{{The one-loop matter bispectrum as a probe of
  gravity and dark energy}},
  \href{https://doi.org/10.1088/1475-7516/2018/10/019}{\emph{JCAP} {\bfseries
  1810} (2018) 019} [\href{https://arxiv.org/abs/1808.01120}{{\ttfamily
  1808.01120}}].

\bibitem{Aviles:2018saf}
A.~Aviles, M.~A. Rodriguez-Meza, J.~De-Santiago and J.~L. Cervantes-Cota,
  \emph{{Nonlinear evolution of initially biased tracers in modified gravity}},
  \href{https://doi.org/10.1088/1475-7516/2018/11/013}{\emph{JCAP} {\bfseries
  1811} (2018) 013} [\href{https://arxiv.org/abs/1809.07713}{{\ttfamily
  1809.07713}}].

\bibitem{Aviles:2018qot}
A.~Aviles, J.~L. Cervantes-Cota and D.~F. Mota, \emph{{Screenings in Modified
  Gravity: a perturbative approach}},
  \href{https://doi.org/10.1051/0004-6361/201834383}{\emph{Astron. Astrophys.}
  {\bfseries 622} (2019) A62}
  [\href{https://arxiv.org/abs/1810.02652}{{\ttfamily 1810.02652}}].

\bibitem{Valogiannis:2019xed}
G.~Valogiannis and R.~Bean, \emph{{Convolution Lagrangian perturbation theory
  for biased tracers beyond general relativity}},
  \href{https://doi.org/10.1103/PhysRevD.99.063526}{\emph{Phys. Rev.}
  {\bfseries D99} (2019) 063526}
  [\href{https://arxiv.org/abs/1901.03763}{{\ttfamily 1901.03763}}].

\bibitem{Aviles:2017aor}
A.~Aviles and J.~L. Cervantes-Cota, \emph{{Lagrangian perturbation theory for
  modified gravity}},
  \href{https://doi.org/10.1103/PhysRevD.96.123526}{\emph{Phys. Rev.}
  {\bfseries D96} (2017) 123526}
  [\href{https://arxiv.org/abs/1705.10719}{{\ttfamily 1705.10719}}].

\bibitem{1986ApJ...304...15B}
J.~M. {Bardeen}, J.~R. {Bond}, N.~{Kaiser} and A.~S. {Szalay}, \emph{{The
  statistics of peaks of Gaussian random fields}},
  \href{https://doi.org/10.1086/164143}{\emph{The Astrophysical Journal}
  {\bfseries 304} (1986) 15}.

\bibitem{PhysRevD.88.023515}
F.~Schmidt, D.~Jeong and V.~Desjacques, \emph{Peak-background split,
  renormalization, and galaxy clustering},
  \href{https://doi.org/10.1103/PhysRevD.88.023515}{\emph{Phys. Rev. D}
  {\bfseries 88} (2013) 023515}.

\bibitem{Mo:1996cn}
H.~J. Mo, Y.~P. Jing and S.~D.~M. White, \emph{{High-order correlations of
  peaks and halos: A Step toward understanding galaxy biasing}},
  \href{https://doi.org/10.1093/mnras/284.1.189}{\emph{Mon. Not. Roy. Astron.
  Soc.} {\bfseries 284} (1997) 189}
  [\href{https://arxiv.org/abs/astro-ph/9603039}{{\ttfamily
  astro-ph/9603039}}].

\bibitem{Hu:2007nk}
W.~Hu and I.~Sawicki, \emph{{Models of f(R) Cosmic Acceleration that Evade
  Solar-System Tests}},
  \href{https://doi.org/10.1103/PhysRevD.76.064004}{\emph{Phys. Rev.}
  {\bfseries D76} (2007) 064004}
  [\href{https://arxiv.org/abs/0705.1158}{{\ttfamily 0705.1158}}].

\bibitem{Dvali:2000hr}
G.~R. Dvali, G.~Gabadadze and M.~Porrati, \emph{{4-D gravity on a brane in 5-D
  Minkowski space}},
  \href{https://doi.org/10.1016/S0370-2693(00)00669-9}{\emph{Phys. Lett.}
  {\bfseries B485} (2000) 208}
  [\href{https://arxiv.org/abs/hep-th/0005016}{{\ttfamily hep-th/0005016}}].

\bibitem{DeFelice:2010aj}
A.~De~Felice and S.~Tsujikawa, \emph{{f(R) theories}},
  \href{https://doi.org/10.12942/lrr-2010-3}{\emph{Living Rev. Rel.} {\bfseries
  13} (2010) 3} [\href{https://arxiv.org/abs/1002.4928}{{\ttfamily
  1002.4928}}].

\bibitem{Carroll:2003wy}
S.~M. Carroll, V.~Duvvuri, M.~Trodden and M.~S. Turner, \emph{{Is cosmic speed
  - up due to new gravitational physics?}},
  \href{https://doi.org/10.1103/PhysRevD.70.043528}{\emph{Phys. Rev.}
  {\bfseries D70} (2004) 043528}
  [\href{https://arxiv.org/abs/astro-ph/0306438}{{\ttfamily
  astro-ph/0306438}}].

\bibitem{Brax:2008hh}
P.~Brax, C.~van~de Bruck, A.-C. Davis and D.~J. Shaw, \emph{{f(R) Gravity and
  Chameleon Theories}},
  \href{https://doi.org/10.1103/PhysRevD.78.104021}{\emph{Phys. Rev.}
  {\bfseries D78} (2008) 104021}
  [\href{https://arxiv.org/abs/0806.3415}{{\ttfamily 0806.3415}}].

\bibitem{Koyama:2007zz}
K.~Koyama, \emph{{Ghosts in the self-accelerating universe}},
  \href{https://doi.org/10.1088/0264-9381/24/24/R01}{\emph{Class. Quant. Grav.}
  {\bfseries 24} (2007) R231}
  [\href{https://arxiv.org/abs/0709.2399}{{\ttfamily 0709.2399}}].

\bibitem{Cautun:2017tkc}
M.~Cautun, E.~Paillas, Y.-C. Cai, S.~Bose, J.~Armijo, B.~Li et~al., \emph{{The
  Santiago Harvard Edinburgh Durham void comparison I. SHEDding
  light on chameleon gravity tests}},
  \href{https://doi.org/10.1093/mnras/sty463}{\emph{Mon. Not. Roy. Astron.
  Soc.} {\bfseries 476} (2018) 3195}
  [\href{https://arxiv.org/abs/1710.01730}{{\ttfamily 1710.01730}}].

\bibitem{1475-7516-2012-01-051}
B.~Li, G.-B. Zhao, R.~Teyssier and K.~Koyama, \emph{Ecosmog : an efficient code
  for simulating modified gravity}, {\emph{Journal of Cosmology and
  Astroparticle Physics} {\bfseries 2012} (2012) 051}.

\bibitem{Bose:2016wms}
S.~Bose, B.~Li, A.~Barreira, J.-h. He, W.~A. Hellwing, K.~Koyama et~al.,
  \emph{{Speeding up $N$-body simulations of modified gravity: Chameleon
  screening models}},
  \href{https://doi.org/10.1088/1475-7516/2017/02/050}{\emph{JCAP} {\bfseries
  1702} (2017) 050} [\href{https://arxiv.org/abs/1611.09375}{{\ttfamily
  1611.09375}}].

\bibitem{Li:2013nua}
B.~Li, G.-B. Zhao and K.~Koyama, \emph{{Exploring Vainshtein mechanism on
  adaptively refined meshes}},
  \href{https://doi.org/10.1088/1475-7516/2013/05/023}{\emph{JCAP} {\bfseries
  1305} (2013) 023} [\href{https://arxiv.org/abs/1303.0008}{{\ttfamily
  1303.0008}}].

\bibitem{Barreira:2015xvp}
A.~Barreira, S.~Bose and B.~Li, \emph{{Speeding up N-body simulations of
  modified gravity: Vainshtein screening models}},
  \href{https://doi.org/10.1088/1475-7516/2015/12/059}{\emph{JCAP} {\bfseries
  1512} (2015) 059} [\href{https://arxiv.org/abs/1511.08200}{{\ttfamily
  1511.08200}}].

\bibitem{Hellwing:2017pmj}
W.~A. Hellwing, K.~Koyama, B.~Bose and G.-B. Zhao, \emph{{Revealing modified
  gravity signals in matter and halo hierarchical clustering}},
  \href{https://doi.org/10.1103/PhysRevD.96.023515}{\emph{Phys. Rev.}
  {\bfseries D96} (2017) 023515}
  [\href{https://arxiv.org/abs/1703.03395}{{\ttfamily 1703.03395}}].

\bibitem{2013ApJ...762..109B}
P.~S. {Behroozi}, R.~H. {Wechsler} and H.-Y. {Wu}, \emph{{The ROCKSTAR
  Phase-space Temporal Halo Finder and the Velocity Offsets of Cluster Cores}},
  \href{https://doi.org/10.1088/0004-637X/762/2/109}{\emph{The Astrophysical
  Journal} {\bfseries 762} (2013) 109}
  [\href{https://arxiv.org/abs/1110.4372}{{\ttfamily 1110.4372}}].

\bibitem{Arnold:2018nmv}
C.~Arnold, P.~Fosalba, V.~Springel, E.~Puchwein and L.~Blot, \emph{{The
  modified gravity lightcone simulation project I: Statistics of matter and
  halo distributions}},  \href{https://arxiv.org/abs/1805.09824}{{\ttfamily
  1805.09824}}.

\bibitem{doi:10.1093/mnras/stt1575}
E.~Puchwein, M.~Baldi and V.~Springel, \emph{Modified-gravity-gadget: a new
  code for cosmological hydrodynamical simulations of modified gravity models},
  \href{https://doi.org/10.1093/mnras/stt1575}{\emph{Monthly Notices of the
  Royal Astronomical Society} {\bfseries 436} (2013) 348}.

\bibitem{2001MNRAS.328..726S}
V.~{Springel}, S.~D.~M. {White}, G.~{Tormen} and G.~{Kauffmann},
  \emph{{Populating a cluster of galaxies - I. Results at [formmu2]z=0}},
  \href{https://doi.org/10.1046/j.1365-8711.2001.04912.x}{\emph{Monthly Notices
  of the Royal Astronomical Society} {\bfseries 328} (2001) 726}
  [\href{https://arxiv.org/abs/astro-ph/0012055}{{\ttfamily
  astro-ph/0012055}}].

\bibitem{2012arXiv1210.1833A}
D.~{Alonso}, \emph{{CUTE solutions for two-point correlation functions from
  large cosmological datasets}}, {\emph{arXiv e-prints} (2012) arXiv:1210.1833}
  [\href{https://arxiv.org/abs/1210.1833}{{\ttfamily 1210.1833}}].

\bibitem{Desjacques:2016bnm}
V.~Desjacques, D.~Jeong and F.~Schmidt, \emph{{Large-Scale Galaxy Bias}},
  \href{https://doi.org/10.1016/j.physrep.2017.12.002}{\emph{Phys. Rept.}
  {\bfseries 733} (2018) 1} [\href{https://arxiv.org/abs/1611.09787}{{\ttfamily
  1611.09787}}].

\bibitem{PhysRevD.83.083518}
T.~Matsubara, \emph{Nonlinear perturbation theory integrated with nonlocal
  bias, redshift-space distortions, and primordial non-gaussianity},
  \href{https://doi.org/10.1103/PhysRevD.83.083518}{\emph{Phys. Rev. D}
  {\bfseries 83} (2011) 083518}.

\bibitem{Aviles:2018thp}
A.~Aviles, \emph{{Renormalization of Lagrangian bias via spectral parameters}},
  \href{https://doi.org/10.1103/PhysRevD.98.083541}{\emph{Phys. Rev.}
  {\bfseries D98} (2018) 083541}
  [\href{https://arxiv.org/abs/1805.05304}{{\ttfamily 1805.05304}}].

\bibitem{1991ApJ...379..440B}
J.~R. {Bond}, S.~{Cole}, G.~{Efstathiou} and N.~{Kaiser}, \emph{{Excursion set
  mass functions for hierarchical Gaussian fluctuations}},
  \href{https://doi.org/10.1086/170520}{\emph{The Astrophysical Journal}
  {\bfseries 379} (1991) 440}.

\bibitem{Mo:1995cs}
H.~J. Mo and S.~D.~M. White, \emph{{An Analytic model for the spatial
  clustering of dark matter halos}},
  \href{https://doi.org/10.1093/mnras/282.2.347}{\emph{Mon. Not. Roy. Astron.
  Soc.} {\bfseries 282} (1996) 347}
  [\href{https://arxiv.org/abs/astro-ph/9512127}{{\ttfamily
  astro-ph/9512127}}].

\bibitem{Sheth:1998xe}
R.~K. Sheth and G.~Lemson, \emph{{Biasing and the distribution of dark matter
  haloes}}, \href{https://doi.org/10.1046/j.1365-8711.1999.02378.x}{\emph{Mon.
  Not. Roy. Astron. Soc.} {\bfseries 304} (1999) 767}
  [\href{https://arxiv.org/abs/astro-ph/9808138}{{\ttfamily
  astro-ph/9808138}}].

\bibitem{Sheth:1999mn}
R.~K. Sheth and G.~Tormen, \emph{{Large scale bias and the peak background
  split}}, \href{https://doi.org/10.1046/j.1365-8711.1999.02692.x}{\emph{Mon.
  Not. Roy. Astron. Soc.} {\bfseries 308} (1999) 119}
  [\href{https://arxiv.org/abs/astro-ph/9901122}{{\ttfamily
  astro-ph/9901122}}].

\bibitem{1992ApJ...385L...5H}
A.~J.~S. {Hamilton}, \emph{{Measuring Omega and the real correlation function
  from the redshift correlation function}}, .

\bibitem{Hamilton:1997zq}
A.~J.~S. Hamilton, \emph{{Linear redshift distortions: A Review}},  in
  \emph{{Ringberg Workshop on Large Scale Structure Ringberg, Germany,
  September 23-28, 1996}}, 1997,
  \href{https://arxiv.org/abs/astro-ph/9708102}{{\ttfamily astro-ph/9708102}},
  \href{https://doi.org/10.1007/978-94-011-4960-0_17}{DOI}.

\bibitem{10.1111/j.1365-2966.2011.20169.x}
L.~Samushia, W.~J. Percival and A.~Raccanelli, \emph{{Interpreting large-scale
  redshift-space distortion measurements}},
  \href{https://doi.org/10.1111/j.1365-2966.2011.20169.x}{\emph{Monthly Notices
  of the Royal Astronomical Society} {\bfseries 420} (2012) 2102}
  [\href{https://arxiv.org/abs/http://oup.prod.sis.lan/mnras/article-pdf/420/3/2102/3011995/mnras0420-2102.pdf}{{\ttfamily
  http://oup.prod.sis.lan/mnras/article-pdf/420/3/2102/3011995/mnras0420-2102.pdf}}].

\bibitem{Yoo:2013zga}
J.~Yoo and U.~Seljak, \emph{{Wide Angle Effects in Future Galaxy Surveys}},
  \href{https://doi.org/10.1093/mnras/stu2491}{\emph{Mon. Not. Roy. Astron.
  Soc.} {\bfseries 447} (2015) 1789}
  [\href{https://arxiv.org/abs/1308.1093}{{\ttfamily 1308.1093}}].

\bibitem{Vlah:2018ygt}
Z.~Vlah and M.~White, \emph{{Exploring redshift-space distortions in
  large-scale structure}},
  \href{https://doi.org/10.1088/1475-7516/2019/03/007}{\emph{JCAP} {\bfseries
  1903} (2019) 007} [\href{https://arxiv.org/abs/1812.02775}{{\ttfamily
  1812.02775}}].

\bibitem{White:2014gfa}
M.~White, \emph{{The Zel'dovich approximation}},
  \href{https://doi.org/10.1093/mnras/stu209}{\emph{Mon. Not. Roy. Astron.
  Soc.} {\bfseries 439} (2014) 3630}
  [\href{https://arxiv.org/abs/1401.5466}{{\ttfamily 1401.5466}}].

\bibitem{Bianchi:2014kba}
D.~Bianchi, M.~Chiesa and L.~Guzzo, \emph{{Improving the modelling of
  redshift-space distortions ? I. A bivariate Gaussian description for the
  galaxy pairwise velocity...}},
  \href{https://doi.org/10.1093/mnras/stu2080}{\emph{Mon. Not. Roy. Astron.
  Soc.} {\bfseries 446} (2015) 75}
  [\href{https://arxiv.org/abs/1407.4753}{{\ttfamily 1407.4753}}].

\bibitem{2012MNRAS.426.2719R}
B.~A. {Reid}, L.~{Samushia}, M.~{White}, W.~J. {Percival}, M.~{Manera},
  N.~{Padmanabhan} et~al., \emph{{The clustering of galaxies in the SDSS-III
  Baryon Oscillation Spectroscopic Survey: measurements of the growth of
  structure and expansion rate at z = 0.57 from anisotropic clustering}},
  \href{https://doi.org/10.1111/j.1365-2966.2012.21779.x}{\emph{\mnras}
  {\bfseries 426} (2012) 2719}
  [\href{https://arxiv.org/abs/1203.6641}{{\ttfamily 1203.6641}}].

\bibitem{2013MNRAS.429.1514S}
L.~{Samushia}, B.~A. {Reid}, M.~{White}, W.~J. {Percival}, A.~J. {Cuesta},
  L.~{Lombriser} et~al., \emph{{The clustering of galaxies in the SDSS-III DR9
  Baryon Oscillation Spectroscopic Survey: testing deviations from
  {\ensuremath{\Lambda}} and general relativity using anisotropic clustering of
  galaxies}}, \href{https://doi.org/10.1093/mnras/sts443}{\emph{\mnras}
  {\bfseries 429} (2013) 1514}
  [\href{https://arxiv.org/abs/1206.5309}{{\ttfamily 1206.5309}}].

\bibitem{2014MNRAS.439.3504S}
L.~{Samushia}, B.~A. {Reid}, M.~{White}, W.~J. {Percival}, A.~J. {Cuesta},
  G.-B. {Zhao} et~al., \emph{{The clustering of galaxies in the SDSS-III Baryon
  Oscillation Spectroscopic Survey: measuring growth rate and geometry with
  anisotropic clustering}},
  \href{https://doi.org/10.1093/mnras/stu197}{\emph{\mnras} {\bfseries 439}
  (2014) 3504} [\href{https://arxiv.org/abs/1312.4899}{{\ttfamily 1312.4899}}].

\bibitem{Alam:2016hwk}
{\scshape BOSS} collaboration, \emph{{The clustering of galaxies in the
  completed SDSS-III Baryon Oscillation Spectroscopic Survey: cosmological
  analysis of the DR12 galaxy sample}},
  \href{https://doi.org/10.1093/mnras/stx721}{\emph{Mon. Not. Roy. Astron.
  Soc.} {\bfseries 470} (2017) 2617}
  [\href{https://arxiv.org/abs/1607.03155}{{\ttfamily 1607.03155}}].

\bibitem{Zarrouk:2018vwy}
P.~Zarrouk et~al., \emph{{The clustering of the SDSS-IV extended Baryon
  Oscillation Spectroscopic Survey DR14 quasar sample: measurement of the
  growth rate of structure from the anisotropic correlation function between
  redshift 0.8 and 2.2}},
  \href{https://doi.org/10.1093/mnras/sty506}{\emph{Mon. Not. Roy. Astron.
  Soc.} {\bfseries 477} (2018) 1639}
  [\href{https://arxiv.org/abs/1801.03062}{{\ttfamily 1801.03062}}].

\bibitem{Lewis:1999bs}
A.~Lewis, A.~Challinor and A.~Lasenby, \emph{{Efficient computation of CMB
  anisotropies in closed FRW models}},
  \href{https://doi.org/10.1086/309179}{\emph{Astrophys. J.} {\bfseries 538}
  (2000) 473} [\href{https://arxiv.org/abs/astro-ph/9911177}{{\ttfamily
  astro-ph/9911177}}].

\bibitem{Icaza-Lizaola:2019zgk}
M.~Icaza-Lizaola et~al., \emph{{The clustering of the SDSS-IV extended Baryon
  Oscillation Spectroscopic Survey DR14 LRG sample: structure growth rate
  measurement from the anisotropic LRG correlation function in the redshift
  range 0.6 < z < 1.0}},  \href{https://arxiv.org/abs/1909.07742}{{\ttfamily
  1909.07742}}.

\bibitem{Vlah:2014nta}
Z.~Vlah, U.~Seljak and T.~Baldauf, \emph{{Lagrangian perturbation theory at one
  loop order: successes, failures, and improvements}},
  \href{https://doi.org/10.1103/PhysRevD.91.023508}{\emph{Phys. Rev.}
  {\bfseries D91} (2015) 023508}
  [\href{https://arxiv.org/abs/1410.1617}{{\ttfamily 1410.1617}}].

\bibitem{White:2016yhs}
M.~White, \emph{{A marked correlation function for constraining modified
  gravity models}},
  \href{https://doi.org/10.1088/1475-7516/2016/11/057}{\emph{JCAP} {\bfseries
  1611} (2016) 057} [\href{https://arxiv.org/abs/1609.08632}{{\ttfamily
  1609.08632}}].

\bibitem{Valogiannis:2017yxm}
G.~Valogiannis and R.~Bean, \emph{{Beyond $\delta$: Tailoring marked statistics
  to reveal modified gravity}},
  \href{https://doi.org/10.1103/PhysRevD.97.023535}{\emph{Phys. Rev.}
  {\bfseries D97} (2018) 023535}
  [\href{https://arxiv.org/abs/1708.05652}{{\ttfamily 1708.05652}}].

\bibitem{Hernandez-Aguayo:2018yrp}
C.~Hern~$\'{a}$ndez Aguayo, C.~M. Baugh and B.~Li, \emph{{Marked clustering
  statistics in $f(R)$ gravity cosmologies}},
  \href{https://doi.org/10.1093/mnras/sty1822}{\emph{Mon. Not. Roy. Astron.
  Soc.} {\bfseries 479} (2018) 4824}
  [\href{https://arxiv.org/abs/1801.08880}{{\ttfamily 1801.08880}}].

\bibitem{Armijo:2018urs}
J.~Armijo, Y.-C. Cai, N.~Padilla, B.~Li and J.~A. Peacock, \emph{{Testing
  modified gravity using a marked correlation function}},
  \href{https://doi.org/10.1093/mnras/sty1335}{\emph{Mon. Not. Roy. Astron.
  Soc.} {\bfseries 478} (2018) 3627}
  [\href{https://arxiv.org/abs/1801.08975}{{\ttfamily 1801.08975}}].

\bibitem{Satpathy:2019nvo}
S.~Satpathy, R.~A.~C. Croft, S.~Ho and B.~Li, \emph{{Measurement of marked
  correlation functions in SDSS-III Baryon Oscillation Spectroscopic Survey
  using LOWZ galaxies in Data Release 12}},
  \href{https://doi.org/10.1093/mnras/stz009}{\emph{Mon. Not. Roy. Astron.
  Soc.} {\bfseries 484} (2019) 2148}
  [\href{https://arxiv.org/abs/1901.01447}{{\ttfamily 1901.01447}}].

\end{thebibliography}






\end{document}